\documentclass[aps,prd,showpacs,floatfix,twocolumn,nofootinbib,superscriptaddress]{revtex4-1} %
\usepackage{dcolumn}

\usepackage{amsmath,amsfonts,amssymb,mathrsfs,times,bm,color}
\usepackage{extarrows}
\usepackage{graphicx}
\usepackage{float}
\usepackage{booktabs}
\usepackage{graphicx,epstopdf}
\usepackage{dcolumn}
\usepackage{exscale}
\usepackage{relsize}
\usepackage{latexsym}
\usepackage{longtable}
\usepackage[colorlinks=true,breaklinks=true,linkcolor=blue,citecolor=blue,urlcolor=blue]{hyperref}

\newcommand{\nn}{\nonumber}

\newcommand{\orcid}[1]{\href{https://orcid.org/#1}{\includegraphics[scale=0.035]{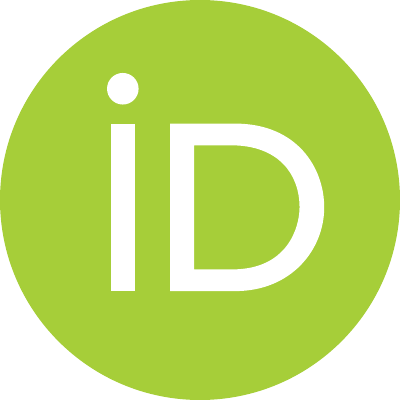}}}

\begin{document}

\title{Strong field gravitational lensing of particles by a black-bounce-Schwarzschild black hole}

\author{Guansheng He\hspace*{0.6pt}\orcid{0000-0002-6145-0449}\hspace*{0.8pt}}
\affiliation{School of Mathematics and Physics, University of South China, Hengyang 421001, China}
\author{Jiaxu Huang}
\affiliation{School of Mathematics and Physics, University of South China, Hengyang 421001, China}
\author{Zhongwen Feng}
\affiliation{Physics and Space Science College, China West Normal University, Nanchong 637009, China}
\author{Ghulam Mustafa\hspace*{0.6pt}\orcid{0000-0003-1409-2009}\hspace*{0.8pt}}
\affiliation{Department of Physics, Zhejiang Normal University, Jinhua 321004, China}
\author{Wenbin Lin\hspace*{0.6pt}\orcid{0000-0002-4282-066X}\hspace*{0.8pt}}
\email{lwb@usc.edu.cn}
\affiliation{School of Mathematics and Physics, University of South China, Hengyang 421001, China}
\affiliation{School of Physical Science and Technology, Southwest Jiaotong University, Chengdu 610031, China}

\date{\today}

\begin{abstract}
The gravitational lensing of relativistic and nonrelativistic neutral massive particles in the black-bounce-Schwarzschild black hole spacetime is investigated in the strong deflection limit. Beginning with the explicit equations of motion of a massive particle in the regular spacetime, we achieve the equation of the particle sphere and thus the radius of the unstable timelike circular orbit. It is interesting to find that the particle sphere equation can reduce to the well-known photon sphere equation, when the particle's initial velocity is equal to the speed of light. We adopt the strong field limit approach to calculate the black-bounce-Schwarzschild deflection angle of the particle subsequently, and obtain the strong-deflection lensing observables of the relativistic images of a pointlike particle source. The observables mainly include the apparent angular particle sphere radius, the angular separation between the outermost relativistic image and the other ones which are packed together, and the ratio between the particle-flux magnification of the outermost image and that of the packed ones (or equivalently, their resulted magnitudelike difference). The velocity effects induced by the deviation of the initial velocity of the particle from light speed on the corresponding strong-field lensing observables of the images of a pointlike light source in the regular geometry, along with these on the strong deflection limit coefficients and the critical impact parameter of the lightlike case, are then formulated. The influence of the spacetime bounce on the Schwarzschild lensing properties of the images of a massive-particle-emission source in the strong field limit is also considered. Serving as an application of the results, we finally concentrate on evaluating the astronomical detectability of the velocity- and bounce-induced effects on the lensing observables and analyzing their dependence on the parameters, by modeling the supermassive black hole at the galactic center (i.e., Sgr A$^{\ast}$) and the one at the center of galactic M87 (i.e., M87$^{\ast}$) as the lens, respectively.

\end{abstract}


\maketitle

\section{Introduction}
Owing to its fruitful astronomical applications (see, for instance,\,\cite{Refsdal1964,BN1992,VE2000,BS2001,KP2005,ZLBD2007,CJ2009,Reyes2010,NKYS2013,ZX2016,FLBPZ2017,LWLM2017,Collett2018,LMZL2018,JS2019,CLCS2020,HDE2025}), gravitational lensing (GL) has developed into a booming part and one of the most powerful tools in modern astrophysics. Compared with the traditional and widely explored gravitational lens effects in various non-regular spacetimes, the weak- or strong-field GL phenomena in a regular geometry (e.g., a regular black hole or wormhole spacetime) have attracted more and more attention from the relativity and astrophysics communities in the past three decades (see, e.g.,\,\cite{ES2011,ES2013,SS2015,GN2016,JOSVG2018,ZJ2018,LMWX2019,Ovgun2019,CX2021,GB2022,DJCC2023,ZJPH2023,XZSLD2024,TGA2025}), due to its special significance. And the reasons responsible for an increasing interest on the GL caused by a regular black hole lie mainly in two aspects. First, we know that in many cases, it is hard to practically distinguish the shadow of a Kerr black hole from these of some non-Kerr astrophysical black hole candidates (e.g., rotating regular black holes) within the current observational uncertainties\,\cite{Bambi2013,LB2014,LGPV2018,KKG2020,Torres2022}, despite some constraints placed on the parameters of the latter black holes. This leaves a window of chances to further test (regular) black holes, wormholes, and naked singularities via other effective avenues including the gravitational lensing effects of photonic or non-photonic messengers. A second, deeper, reason is that regular black holes, corresponding to a class of singularity-free solutions of the field equations\,\cite{We1972}, are intuitively attractive because of their nonsingular nature, and have been viewing as one of the main potential routes to overcome a series of long standing difficulties resulted from the spacetime singularity\,\cite{We1972,FLWJ2024,FJWY2025}. A variety of theoretical proposals for the construction of these regular gravitational sources were thus made (see\,\cite{Ba1968,RB1983,Hayward2006,BF2006,BM2013,FW2016,HY2019,BCMV2022,OCG2023,Ovalle2024,BCHM2025}, and references therein), along with vigorous exploration of them performed from the phenomenological or other perspectives\,\cite{CDLV2018,LM2022,Bargue2020,GD2020,ZTX2020,GD2021,RS2023}. Recently, a novel regular spacetime, the so-called black-bounce or black-bounce-Schwarzschild spacetime, was proposed by Simpson and Visser\,\cite{SV2019}. This geometry is especially interesting, since it not only represents minimal violence to the standard Schwarzschild spacetime in some sense, but also can describe a Schwarzschild black hole, a regular black hole with a one-way spacelike throat, a one-way wormhole with a null throat, a canonical traversable wormhole, or an Ellis-Bronnikov wormhole, by changing its spacetime parameters. Since the proposal of the black-bounce geometry, loads of efforts have been invested in the weak- or strong-field GL effects of electromagnetic signals in this spacetime or in its generalized versions\,\cite{NPPS2020,Ovgun2020,Tsuka2021a,Tsuka2021b,CX2021,IKG2021,GB2022,ZX2022,Tsuka2022,GLM2022,CGB2023,Vagnozzi2023,GL2023,Shaikh2023,JAPO2023,SAAK2023,PSSV2025}. For example, Nascimento \emph{et al.}\,\cite{NPPS2020} adopted the strong field limit analysis developed in\,\cite{BCIS2001,Bozza2002} to discuss the strong-deflection lensing characteristics of the relativistic images of the light source in the black-bounce-Schwarzschild geometry, together with a calculation of the weak-field black-bounce deflection angle of light. Islam \emph{et al.}\,\cite{IKG2021} considered the strong GL of light due to a rotating Simpson-Visser black hole, and examined the dependence of the observables on the bounce parameter and the angular momentum of the lens. Compared with the situation in which a photon acts as the test particle, there were only few works devoted to the considerations of the propagation and the related gravitational effects of timelike particles in the black-bounce spacetime or in its variations\,\cite{ZX2020b,ZX2022b,VRSA2023,MBRAS2024,HXJL2024,NZKR2024,DR2024,BO2025,JRT2025}. Very recently, the gravitational lensing effects of relativistic massive particles induced by a black-bounce-Schwarzschild black hole in the third post-Minkowskian approximation were studied\,\cite{HXJL2024}, where the so-called velocity effect\,\cite{WS2004,HL2014,LYJ2016}, induced by the deviation of the initial velocity of the timelike particle at infinity from the speed of light, on the observable weak-field lensing properties was explored in detail. However, to our knowledge, the strong-deflection GL phenomena, accompanied by other strong-field relativistic effects, of various timelike particles in the black-bounce geometry or in any of its generalized versions have not been discussed up to now.

Actually, we can now capture multiple cosmic messengers\,\cite{MFHM2019,Huerta2019}, including cosmic rays\,\cite{Ande1932,Rossi1941}, neutrinos\,\cite{BV1992}, and gravitational waves\,\cite{Abbott2016a,Abbott2017a},
although only electromagnetic radiation is used to explore the Universe in multiwavelength astronomy traditionally. With the coming of the era of multimessenger astronomy, a full theoretical probe for the strong-deflection gravitational lensing effects of massive particles in the Simpson-Visser spacetime and in other regular spacetimes deserves our consideration, for which three aspects are responsible. Firstly, similar to the weak-field case\,\cite{WS2004,Silverman1980,AR2002,AR2003,HL2016a,HL2017b,LZLH2019,HZFMWPL2020,LHZ2020,HL2022,WLMH2025}, the strong-field-limit deflection angle of a test particle for a given gravitational system increases with the decrease of the particle's initial velocity\,\cite{PJ2019,JH2021}, ending up with a significant result that some of the strong-deflection lensing properties of the images of the particle-emission source may be more evident than their optical counterparts respectively under the same conditions. It may thus bring more potential chances for observing gravitational lensing events, which has in turn inspired a number of attempts to discuss
the weak-field\,\cite{AP2004,BSN2007,PNH2014,CG2018,J2018,CGV2019,CGJ2019,JBGA2019,JL2019,LLJ2021,HC2023,HCL2024} and strong-field\,\cite{Tsupko2014,BT2017,HSC2023} GL effects of massive particles in a non-regular spacetime and to probe the mentioned velocity effects on the gravitational deflection angle as well as on the lensing observables\,\cite{WS2004,PJ2019,JL2019,HZFMWPL2020,HL2022}. It also motivates a further consideration of the strong-deflection GL of massive particles in a regular spacetime (including the black-bounce spacetime) and of the velocity effects on the strong-field lensing observables of the light-source images, which would be interesting and of theoretical significance. A second aspect lies in that some unique insights into our physical Universe and valuable information about the lens and the source, along with complementary constraints on the spacetime parameters, may be provided by exploring the strong-field GL phenomena of non-photonic messengers in various regular geometries. Additionally, it is also possible to accelerate the progression of mono-messenger or synergic multimessenger astronomical observations\,\cite{MFHM2019,Keivani2018,IceCube2018} by this exploration, since all of the messengers emitted by an astrophysical source may experience different geometrical and physical processes before reaching their receivers. Finally, we know that great achievements in techniques and instruments of high-accuracy astronomical observations have been acquired over the last decades (see, e.g.,\,\cite{Perryman2001,SN2009,Reid2009,Trippe2010,Malbet2012,ZRMZBDX2013,RH2014,Malbet2014,Prusti2016,Murphy2018,RD2020,Brown2021,LXLWBLYHL2022,LXBLLLH2022}). The astrometric precision in current surveys and forthcoming new telescope networks for multiwavelength observations is at the level of $1\!\sim\!10$ microarcseconds ($\mu$as) or better\,\cite{SN2009,Reid2009,Malbet2014,Murphy2018}. For instance, the Square Kilometre Array (SKA)\,\cite{BBGKW2015,LXLWBLYHL2022} and other next-generation radio observatories\,\cite{Murphy2018,RD2020} aim at an angular accuracy of $1\mu$as. It should be mentioned, however, that current astronomical detectors or instruments for measuring timelike particles or multimessengers have much lower angular resolutions of about one degree or better\,\cite{Aab2014,Bartoli2019,Albert2020}. For instance, the angular resolution of the ANTARES neutrino telescope is approximately $0.59^{\circ}\pm0.10^{\circ}$ for downward-going muons\,\cite{Albert2020}. Moreover, the current photometric precision is at the level of about $10\,\mu$mag\,\cite{Koch2010,BK2018,Kurtz2005}. Especially, the Kepler Mission reached an unprecedented photometric precision of a few $\mu$mag\,\cite{Koch2010,BK2018}, although it ended prematurely and was renamed as the K2 mission with a lower photometric precision\,\cite{VJ2014,AHILR2015,Huber2016}. Considering the rapid and continual improvement in astronomical measurements and the increase in (joint) multimessenger observations\,\cite{BT2017,IceCube2018,QJFZZ2021}, a detailed theoretical treatment of the strong field GL phenomena of various non-photonic messengers in regular spacetimes becomes more and more important.

In the present work, we investigate the strong deflection gravitational lensing of a neutral massive particle caused by a regular black-bounce-Schwarzschild black hole, which serves as a natural extension of its weak-field counterpart\,\cite{HXJL2024} and of the strong-field lensing case of light signals\,\cite{NPPS2020,Tsuka2021a}. After a review of the background metric and the timelike geodesic equation of the regular spacetime, we start our discussion with the derivation of the equation and radius of the particle sphere. The strong field limit approach\,\cite{BCIS2001,Bozza2002,FBT2025} is applied subsequently to the calculation of the black-bounce-Schwarzschild deflection angle of the particle. The strong-deflection lensing observables of the relativistic images of the pointlike particle source are then obtained. They mainly consist of the angular radius of the particle sphere, the angular separation between the outermost image and the packed other ones, and the ratio between their magnifications (or the magnitudelike difference converted from their fluxes). We further formulate the velocity effects, which originates from the difference between the particle's initial velocity at infinity and light speed, on the corresponding strong-deflection observables of the relativistic images of a pointlike light source and on the strong deflection limit coefficients as well as the critical impact parameter of the lightlike situation in the regular spacetime. The effects due to the spacetime bounce on the strong-field lensing observables of the particle-source images and on the two strong field limit coefficients of the Schwarzschild case are also considered. Finally, we model the Galactic supermassive black hole, Sgr A$^{\ast}$, and the central black hole of galactic M87 (i.e., M87$^{\ast}$) as the lens respectively, and estimate the detectability of the velocity- and bounce-induced effects on the observables, accompanied by a discussion of their dependence on the parameters.

This paper is organized as follows. Section\,\ref{sect2} gives the notation and assumptions used in this paper. In Sec.\,\ref{sect3}, we review the spacetime metric of a black-bounce-Schwarzschild black hole and the equations of motion of a massive particle, derive the equation and radius of the particle sphere, and then calculate the gravitational deflection of the particle in the strong field limit. The main strong-deflection lensing observables of the images of the particle source are obtained in Sec.\,\ref{sect4}, followed by the formulations of the velocity effects and the spacetime bounce effects in Sec.\,\ref{sect5}. Section\,\ref{sect6} presents an application of our formulae by assuming Sgr A$^{\ast}$ and M87$^{\ast}$ as the lens respectively, along with an analysis of the detectability of the velocity- and bounce-induced effects on the observables. The conclusions and a short discussion of the results are given in Sec.\,\ref{sect7}. Throughout this work, geometrized units $(G=c=1)$ and the metric signature $(+,~-,~-,~-)$ are adopted. Unless indicated otherwise, Greek indices run over $0,~1,~2$, and $3$ generally.

\section{Notation and assumptions} \label{sect2}
\begin{figure}[t]
\centering
\includegraphics[width=\linewidth]{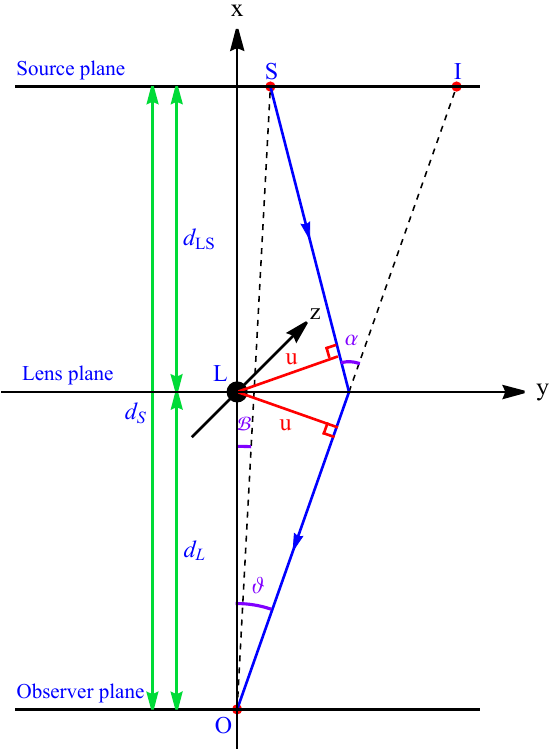}
\caption{Geometrical configuration of lensing of the massive particle in an asymptotically flat black-bounce-Schwarzschild black hole spacetime, adopted from~\cite{HXJL2024}.  }   \label{Figure1}
\end{figure}

We assume $w$ to be the initial velocity of a massive test particle, which is emitted by a pointlike source, strongly deflected by the black-bounce-Schwarzschild black hole, and received by a distant observer. The geometrical configuration for the lensing of the particle is shown in Fig.\,\ref{Figure1}. The barycenter of the black hole is located at the origin of a three-dimensional Cartesian coordinate system $(x,~y,~z)$, and the $x$ axis acts as the reference line joining the lens and the observer. We denote the source, observer, lens, and image by $S,~O,~L$, and $I$, respectively. All of them are situated in the equatorial plane (i.e., $x$-$y$ plane) of the lens, with $S$ and $O$ being situated in the asymptotically flat spacetime region. The observer-lens, observer-source, and lens-source angular diameter distances are represented by $d_{L}$, $d_{S}$, and $d_{LS}$ respectively. $\mathcal{B}$ and $\vartheta$ are used to denote the angular source and image positions, respectively, and $u~(=d_L\sin\vartheta)$ is the impact parameter. The blue lines stand for the asymptotes of the propagating trajectory of the test particle. In the strong-deflection case, the particle, which travels along an unbound orbit, may loop around the lens once or $i~(i\geq2,~i\in N^{+})$ times, or not go around it at all, before emerging. Hence, similar to the lightlike counterpart\,\cite{VE2000,ERT2002,Virbha2009,Virbhadra2022}, two infinite sets of relativistic images on both sides of the reference line, respectively, along with the positive-parity primary image and negative-parity secondary image of the particle source, are considered in our scenario. $\alpha$ is adopted to represent the gravitational deflection angle which is defined by the difference between the directions of particle propagation at the emission and reception points. For the convenience of discussion, we assume $\alpha$ for each of the images is non-negative\,\cite{Bozza2002,WS2004}, and this is a little different from the assumption used in\,\cite{ERT2002}. The primary and secondary images thus appear when $\alpha$ is smaller than $2\pi$, while the relativistic images, which are absent in the weak-deflection GL scenario, with clockwise or counterclockwise winding around the lens are formed when the condition $\alpha>2\pi$ is satisfied. Furthermore, the angular position $\vartheta$ of a lensed image is assumed to be always positive, as done in\,\cite{Bozza2002,KP2005,KP2006a,KP2006b}. It means that the source position $\mathcal{B}$ is positive when the image is on the same side of the lens (or the reference line) as the source, and is negative when the image is on the opposite side.

\section{Black-bounce-Schwarzschild deflection of massive particles in strong field limit}  \label{sect3}
In this section, we begin with reviewing the background metric of a black-bounce-Schwarzschild black hole spacetime and the equations of motion of massive particles in this geometry, and then apply the strong field limit approach\,\cite{BCIS2001,Bozza2002,FBT2025} to the analysis of the gravitational deflection of the particles through calculating the radius of the particle sphere.

\subsection{Spacetime and timelike geodesic equation}
The line element for the geometry of a static spherically symmetric regular black-bounce-Schwarzschild black hole in standard coordinates $x^\mu\equiv(t,~r,~\theta,~\varphi)$ reads\,\cite{SV2019,SMV2019,LRSSV2021}:
\begin{eqnarray}
ds^2=A(r)dt^2-B(r)dr^2-C(r)\left(d\theta^2+\sin^2\theta d\varphi^2\right)~,~~~  \label{Metric}
\end{eqnarray}
where
\begin{eqnarray}
&&A(r)=1-\frac{2M}{\sqrt{r^2+\eta^2}}~,                   \label{Ar} \\
&&B(r)=\left(1-\frac{2M}{\sqrt{r^2+\eta^2}}\right)^{-1}~, \label{Br} \\
&&C(r)=r^2+\eta^2~.                                       \label{Cr}
\end{eqnarray}
Here, $M$ is the rest mass of the gravitational body, and $\eta$ denotes the bounce parameter which is responsible for the spacetime regularization and has the dimension of length, with $0<\eta<2M$.

The geodesic equation of a test particle for a given gravitational system is equivalent to the Euler-Lagrangian equation $\frac{\partial \mathscr{L}}{\partial x^\mu}-\frac{d}{d\tau}\frac{\partial \mathscr{L}}{\partial \dot{x}^\mu}=0$ with the Lagrangian $\mathscr{L}\!=\!\frac{1}{2}g_{\rho\sigma}\dot{x}^\rho\dot{x}^\sigma$\cite{St1984,WS2004}, where $g_{\rho\sigma}$ is the metric tensor and a dot denotes differentiation with respect to the proper time $\tau$. For the equatorial propagation ($\theta=\pi/2$) of a massive particle in the black-bounce-Schwarzschild spacetime, the explicit equations of motion are thus expressed with the help of the timelike orbit $\mathscr{L}=1/2$ as follows\,\cite{HXJL2024}:
\begin{eqnarray}
&&\dot{t}=\frac{\mathcal{E}}{1-\frac{2M}{\sqrt{r^2+\eta^2}}}~, \label{GE1}  \\
&&\dot{\varphi}=\frac{\mathcal{L}}{r^2+\eta^2}~,               \label{GE2}  \\
&&\dot{r}^2=\mathcal{E}^2-U^2~,                                \label{GE3}
\end{eqnarray}
where the constants $\mathcal{E}$ and $\mathcal{L}$ denote the conserved orbital energy and angular momentum per unit mass of the particle at infinity, respectively, and they are related to the initial velocity of the particle by $\mathcal{E}=1/\sqrt{1-w^2}$ and $\mathcal{L}=wu\,\mathcal{E}$\,\cite{AR2002,AR2003,Tsupko2014,2023IR}. And $U$ stands for the effective potential\,\cite{MTW1973} in the spacetime of the regular black hole and is given by
\begin{eqnarray}
U(r)=\sqrt{\left(1-\frac{2M}{\sqrt{r^2+\eta^2}}\right)\left(1+\frac{\mathcal{L}^2}{r^2+\eta^2}\right)}~.   \label{EffP}
\end{eqnarray}

\subsection{Equation and radius of a particle sphere}
For a neutral timelike particle propagating in a black hole spacetime, its orbits can be unbound or bound. The bound orbits of the particle,
including the well-explored precessing and periodic orbits\,\cite{GRAVITY2020,Deng2020,LDE2022,LDE2023,LMMJ2023,LDA2023,HDE2024,HDP2024,LD2025P} which play a very important role in testing different theories of gravity, are located between its marginally bound unstable circular orbit and the particle's innermost stable circular orbit\,\cite{MTW1973}. The former circular orbit corresponds to the so-called particle sphere, whose concept has been discussed several times in the literature (see, e.g.,\,\cite{Zakh1994,LYJ2016,Zakh2018}). Similar to the photon sphere in the light-bending case ($w=1$), the particle sphere is also regarded as the starting point for the strong field expansion of the gravitational deflection of a massive particle in any given curved geometry.

A circular motion with a given radius $r_c$ of the particle indicates that the conditions $\left.\dot{r}\right|_{r=r_c}=0$ and $\left.\ddot{r}\right|_{r=r_c}=0$, which are equivalent to $U(r_c)=\mathcal{E}$ and $\left.\frac{dU}{dr}\right|_{r=r_c}=0$, respectively, are satisfied simultaneously\,\cite{CMBWZ2009,Tsupko2014,JTB2015,FBT2025}. It follows from these requirements that the equation for a particle sphere with $r_c$ in our scenario can be obtained as
\begin{eqnarray}
\frac{C'(r_c)}{C(r_c)}-\frac{A'(r_c)}{A(r_c)}\frac{1}{1-\frac{A(r_c)}{\mathcal{E}^2}}=0~,   \label{PS}
\end{eqnarray}
where a prime denotes differentiation with respect to $r$. Actually, Eq.\,\eqref{PS} also applies to other static, spherically symmetric, and asymptotically flat spacetimes, although the expressions for the metric coefficients are different. It is interesting to find that Eq.\,\eqref{PS}, which is slightly different from that in\,\cite{FBT2025}, can reduce to the photon sphere equation for a general static and spherically symmetric spacetime\,\cite{Bozza2002} in the limit $w\rightarrow1$ (or equivalently, $\mathcal{E}\rightarrow+\infty$).

The largest solution of Eq.\,\eqref{PS} then represents the radius of the particle sphere, and is expressed in the form in terms of $\mathcal{E}$ via defining $\hat{\eta}\equiv\eta/M$
\begin{eqnarray}
r_c=M\sqrt{\frac{\left(3\mathcal{E}^2-4+\mathcal{E}\sqrt{9\mathcal{E}^2-8}\hspace*{1pt}\right)^2}{4\left(\mathcal{E}^2-1\right)^2}-\hat{\eta}^2}~,~~~~   \label{RPSE}
\end{eqnarray}
or in terms of $w$,
\begin{eqnarray}
r_c=M\sqrt{\frac{\left(4w^2-1+\sqrt{1+8w^2}\hspace*{1pt}\right)^2}{4w^4}-\hat{\eta}^2}~,~~~   \label{RPSW}
\end{eqnarray}
which is consistent with the photon sphere radius $r_c=\sqrt{9-\hat{\eta}^2}M$ in the black-bounce-Schwarzschild black hole spacetime\,\cite{NPPS2020,ZX2022} in the limit $w\rightarrow1$, and matches well with the particle sphere radius in Schwarzschild geometry given in\,\cite{Tsupko2014,JTB2015,LYJ2016,FBT2025}, if the bounce parameter $\eta$ disappears. Additionally, for avoiding the scenario where the test massive particle is swallowed by the black hole, the condition that $r_c$ is not larger than the impact parameter $u$ of its trajectory should hold, where $u$ is related to the distance $r_0$ of closest approach to the central body by\,\cite{HXJL2024}
\begin{eqnarray}
&&\nn u=\frac{1}{w}\sqrt{\frac{C_0\left[1-\left(1-w^2\right)A_0\right]}{A_0}} \\
&&\hspace*{8pt}=\sqrt{\frac{\left(r_0^2+\eta^2\right)\left[\sqrt{r_0^2+\eta^2}+2\left(\frac{1}{w^2}-1\right)M\right]}{\sqrt{r_0^2+\eta^2}-2M}}~,~~~~  \label{b-r0}
\end{eqnarray}
which can be easily obtained from Eq.\,\eqref{GE3} via imposing $\left.\dot{r}\right|_{r=r_0}=0$ alone. Here and thereafter, we use the subscript $0$ to denote that a quantity is evaluated at $r=r_0$.

\subsection{Strong-field gravitational deflection angle of a massive particle}
We now turn to the calculation of the black-bounce-Schwarzschild deflection angle of the massive particle in the strong field limit. The exact expression of the deflection angle of the particle, which is emitted by the source ($r\rightarrow +\infty$) and travels along the mentioned unbound orbit with a distance $r_0$ of closest approach to the lens, can be written via Eqs.~\eqref{GE2} - \eqref{GE3} as\,\cite{We1972,FBT2025}

\begin{eqnarray}
&&\nn\alpha=2\int_{r_0}^{+\infty}\left|\frac{d\varphi}{dr}\right|dr-\pi  \\
&&\hspace*{9pt}=2\int_{r_0}^{+\infty}\!\frac{\sqrt{B}}{\sqrt{C}\sqrt{\frac{C A_0 H^2}{C_0 A H_0^2}-1}}\,dr-\pi~,  \label{alpha-exact-1}
\end{eqnarray}
with $H=\sqrt{1-\left(1-w^2\right)A(r)}$\,.

In order to obtain the expansion of the bending angle in the strong-deflection limit via the procedure proposed in\,\cite{BCIS2001,Bozza2002}, we define
\begin{eqnarray}
z=\frac{A(r)-A_0}{1-A_0}~~~~~\left(0\leq z<1\right)~,  \label{zdefine}
\end{eqnarray}
which indicates in turn
\begin{eqnarray}
r=\frac{\sqrt{r_0^2+z(2-z)\eta^2}}{1-z}~~~~~\left(r_0\leq r<+\infty\right)~.~  \label{rdefine}
\end{eqnarray}
The integral term in Eq.~\eqref{alpha-exact-1}, denoted by $I\left(r_0,\,w\right)$, thus can be expressed as
\begin{eqnarray}
I\left(r_0,\,w\right)=\int_0^1 R\left(z,\,r_0,\,w\right)f\left(z,\,r_0,\,w\right)\,dz~,~~~  \label{Idefine}
\end{eqnarray}
where
\begin{eqnarray}
&&R\left(z,\,r_0,\,w\right)=2\sqrt{\!\left[\!w^2\!+\!\frac{2\left(1\!-\!w^2\right)\!M}{\sqrt{r_0^2+\eta^2}}\!\right]\!\!\frac{r_0^2+\eta^2}{r_0^2\!+\!\left(2\!-\!z\right)\!z\eta^2}}~,~~~~~~~  \label{Rdefine}  \\
&&f\left(z,\,r_0,\,w\right)=\frac{1}{\sqrt{\rho_1 z+\rho_2 z^2-\rho_3 z^3}}~,~  \label{fdefine}
\end{eqnarray}
with
\begin{eqnarray}
&&\rho_1=2\left[w^2-\frac{\left(4w^2-1\right)M}{\sqrt{r_0^2+\eta^2}}-\frac{4\left(1-w^2\right)M^2}{r_0^2+\eta^2} \right]~,~~~~       \label{alpha}  \\
&&\rho_2=-w^2+\frac{2\left(4w^2-1\right)M}{\sqrt{r_0^2+\eta^2}}+\frac{12\left(1-w^2\right)M^2}{r_0^2+\eta^2}~,~~                     \label{beta}   \\
&&\rho_3=\frac{2M}{\sqrt{r_0^2+\eta^2}}\!\left[w^2+\frac{2\left(1-w^2\right)M}{\sqrt{r_0^2+\eta^2}}\right]~.~                          \label{gamma}
\end{eqnarray}
It should be mentioned that Eqs.~\eqref{Rdefine} - \eqref{fdefine} are in accord with Eqs. (64) - (65) of\,\cite{FBT2025} when the black-bounce parameter is ignored.

Since $f\left(z,\,r_0,\,w\right)$ diverges in the limit $z\rightarrow 0$, for a convenient calculation we split $I\left(r_0,\,w\right)$ into two parts, containing a divergent part $I_D\left(r_0,\,w\right)$ and a regular part $I_R\left(r_0,\,w\right)$, in the form
\begin{eqnarray}
&&I\left(r_0,\,w\right)=I_D\left(r_0,\,w\right)+I_R\left(r_0,\,w\right)~,                             \label{Idefine2}  \\
&&I_D\left(r_0,\,w\right)=\int_0^1 R\left(0,\,r_c,\,w\right)f_0\left(z,\,r_0,\,w\right)\,dz~, ~~~~    \label{ID}        \\
&&I_R\left(r_0,\,w\right)=\int_0^1 F\left(z,\,r_0,\,w\right)\,dz~,                                    \label{IR}
\end{eqnarray}
where
\begin{eqnarray}
&&f_0\left(z,\,r_0,\,w\right)\equiv\frac{1}{\sqrt{\rho_1 z+\rho_2 z^2}}~,                 \label{f0}  \\
&&\nn F\left(z,\,r_0,\,w\right)\equiv R\left(z,\,r_0,\,w\right)f\left(z,\,r_0,\,w\right)   \\
&&\hspace*{64pt}-R\left(0,\,r_c,\,w\right)f_0\left(z,\,r_0,\,w\right)~. ~~~~              \label{Fdefine}
\end{eqnarray}
The order of the divergence of $I\left(r_0,\,w\right)$ can be easily evaluated from the approximate version $f_0\left(z,\,r_0,\,w\right)$ of the function $f\left(z,\,r_0,\,w\right)$\,\cite{Bozza2002}. The leading order of the divergence in $f_0\left(z,\,r_0,\,w\right)$ is $z^{-1/2}$ for a nonzero value of the coefficient $\rho_1$, and it becomes $z^{-1}$ when $\rho_1$ is absent in the case $r_0=r_c$, which makes the integral diverge logarithmically. The divergent term on the right hand side of Eq.~\eqref{Idefine2} is then expanded via a convenient parameter $\xi\equiv\sqrt{1+8w^2}~\left(1<\xi\leq3\right)$ as
\begin{eqnarray}
&&\nn I_D\left(r_0,\,w\right)=-\,a\ln\left(\frac{r_0}{r_c}-1\right)+b_D  \\
&&\hspace*{2cm}+\,\mathcal{O}\left[\left(r_0-r_c\right)\ln\left(r_0-r_c\right)\right]~,~~~~    \label{ID2}
\end{eqnarray}
where
\begin{eqnarray}
&&a\left(w,\,\hat{\eta}\right)=\sqrt{\frac{\left(1+\xi\right)\left(4w^2-1+\xi\right)}{2w^2\xi\!\left[1-\frac{4w^4\hat{\eta}^2}{\left(4w^2-1+\xi\right)^2}\right]}}~,~   \label{aDefine}   \\
&&b_D\left(w,\,\hat{\eta}\right)=a\ln\!\left[\frac{2}{1-\frac{4w^4\hat{\eta}^2}{\left(4w^2-1+\xi\right)^2}}\right]~.~~~~    \label{bDDefine}
\end{eqnarray}
Moreover, the expansion of $F\left(z,\,r_0,\,w\right)$ in powers of $r_0-r_c$ yields
\begin{eqnarray}
I_R\left(r_0,\,w\right)=\sum_{j=0}^{+\infty}\frac{1}{j!}\left(r_0-r_c\right)^j\int_0^1\left.\frac{\partial^j F}{\partial r_0^j} \right|_{r_0=r_c} dz~. ~~~~    \label{IRSeries1}
\end{eqnarray}
We retain only the leading term with $j=0$ in Eq.~\eqref{IRSeries1} and denote it by
\begin{eqnarray}
b_R\left(w,\,\hat{\eta}\right)=\int_0^1 F\left(z,\,r_c,\,w\right)dz~,~~~~    \label{bR}
\end{eqnarray}
with the integrand taking the form
\begin{eqnarray}
&&\nn F\left(z,\,r_c,\,w\right)=\frac{\sqrt{3+\xi}}{z\sqrt{2\xi-\left(1+\xi\right)z}} \\
&&\hspace*{11pt}\times\!\left\{\!\frac{2}{\sqrt{1\!-\!\frac{\left(1+\xi\right)^2\left(1-z\right)^2\hat{\eta}^2}{4\left(3+\xi\right)^2}}}
-\sqrt{\frac{2\left[2\xi-\left(1+\xi\right)z\right]}{\xi\!\left[1-\frac{\left(1+\xi\right)^2\hat{\eta}^2}{4\left(3+\xi\right)^2}\right]}}\right\} ~.~~~~~~~    \label{FrC}
\end{eqnarray}
It should be pointed out that we have to evaluate $b_R\left(\hat{\eta},\,w\right)$ numerically, since the integration in Eq.\,\eqref{bR} cannot be calculated analytically, similar to the lightlike scenario\,\cite{NPPS2020}. Additionally, it is interesting to find that Eqs.~\eqref{aDefine}, \eqref{bDDefine} and \eqref{bR} for the case of no bounce effect $(\eta=0)$ are coincident with the strong-field results for the Schwarzschild deflection of the particle in the literature\,\cite{FBT2025}.

\begin{figure*}
\centering
\begin{minipage}[b]{16.cm}
\includegraphics[width=16.cm]{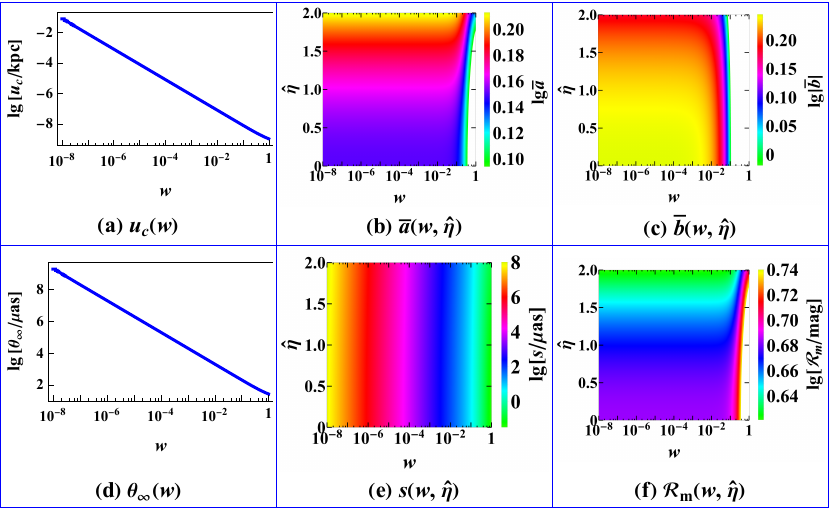}
\caption{The critical impact parameter $u_c$ and the color-indexed strong field limit coefficients $\bar{a}$ and $\bar{b}$, along with the particle sphere radius $\vartheta_{\infty}$ and the color-indexed observables $s$ and $\mathcal{R}_m$, plotted as the functions of $w$ (and $\hat{\eta}$). As an example, here we assume the black-bounce-Schwarzschild lens to be the galactic supermassive black hole which has a mass $M=4.2\times10^6M_{\odot}$\,\cite{BG2016,Parsa2017} and a distance $d_L=8.2$\,kpc\,\cite{BG2016} from us, with $M_{\odot}$ ($=4.8\times10^{-17}$\,kpc) being the rest mass of the Sun.}   \label{Figure2}
\end{minipage}
\end{figure*}

With the definition $b\equiv b_R+b_D-\pi$\,\cite{Bozza2002}, the black-bounce-Schwarzschild deflection angle of the massive particle in the strong-field limit can be expressed
in terms of $r_0$ and $w$ as
\begin{eqnarray}
&&\nn\alpha\left(r_0,\,w\right)=-\hspace*{1pt}a\ln\!\left(\frac{r_0}{r_c}-1\right)+b  \\
&&\hspace*{1.82cm}+\,\mathcal{O}\left[\left(r_0-r_c\right)\ln\left(r_0-r_c\right)\right]~.~~~~~   \label{alpha-Appro-1}
\end{eqnarray}
With the consideration that the impact parameter is related to $r_0$ and the angular image position $\vartheta$ by Eq.\,\eqref{b-r0} and $u=d_L\sin\vartheta\approx d_L\vartheta$, respectively, the deflection angle can be rewritten in terms of $\vartheta$ and $w$ in the form
\begin{eqnarray}
&&\nn\alpha\left(\vartheta,\,w\right)=-\hspace*{1pt}\bar{a}\ln\!\left(\frac{d_L\vartheta}{u_c}-1\right)+\bar{b} \\
&&\hspace*{1.7cm}+\,\mathcal{O}\left[\left(u-u_c\right)\ln\left(u-u_c\right)\right]~.~~~~~~  \label{alpha-Appro-2}
\end{eqnarray}
Here, $\bar{a}$ and $\bar{b}$ are the so-called strong deflection limit coefficients, while $u_c$ denotes the critical (or minimum) impact parameter which is found to be independent on $\eta$, similar to the lensing cases of lightlike signals in the black-bounce-Schwarzschild\,\cite{NPPS2020} or black-bounce-Reissner-Nordstr\"{o}m\,\cite{ZX2022} geometry. They are respectively given by
\begin{eqnarray}
&&\bar{a}\left(w,\,\hat{\eta}\right)=\frac{a}{2}~,~   \label{baraDefine}   \\
&&\bar{b}\left(w,\,\hat{\eta}\right)=-\bar{a}\ln\!\left[\frac{1}{2\xi}\sqrt{1-\frac{\left(1+\xi\right)^2\hat{\eta}^2}{4\left(3+\xi\right)^2}}\,\right]+b_R-\pi~, ~~~~  \label{barbDefine}  \\
&&u_c\left(w\right)=\sqrt{\frac{C_c\left[1-\left(1-w^2\right)A_c\right]}{w^2A_c}}=\frac{2\left(3+\xi\right)^{\frac{3}{2}}M}{\left(1+\xi\right)\sqrt{\xi-1}} ~,~~~~~~  \label{ucDefine}
\end{eqnarray}
where the subscript $c$ is adopted to denote that a quantity is evaluated at $r_0=r_c$. When the bounce parameter is absent, Eq.\,\eqref{alpha-Appro-2} matches well with the Schwarzschild deflection angle of the particle in the strong field limit\,\cite{FBT2025}. In the limit $\eta\rightarrow 0$ and $w\rightarrow 1$, we have $\bar{a}=1$, $\bar{b}=\ln\left[216(7-4\sqrt{3})\right]-\pi$, and $u_c=3\sqrt{3}M$, indicating a consistency of Eq.\,\eqref{alpha-Appro-2} with the strong-field-limit result for the Schwarzschild deflection of light\,\cite{Bozza2002}.

Finally, the critical impact parameter $u_c$ and the color-indexed strong deflection limit coefficients $\bar{a}$ and $\bar{b}$ are shown as the functions of the initial velocity of the particle (and the scaled bounce parameter) on the top plane of Fig.\,\ref{Figure2}. With the consideration of its independence of $\eta$, $u_c$ showed in Fig.\,\ref{Figure2}\,(a) in our scenario increases with the decrease of the particle's initial velocity monotonically and diverges in the limit $w\rightarrow 0$, which is the same as the behavior of the critical impact parameter in the Schwarzschild lensing case of massive particles~\cite{FBT2025}. A similar increase tendency except the divergence also applies to the coefficient $\bar{a}$ presented in Fig.\,\ref{Figure2}\,(b) for a given $\hat{\eta}$ on the domain $\left(0,~2\right)$, and to $\bar{b}$ given in Fig.\,\ref{Figure2}\,(c) for a fixed scaled bounce parameter with $0<\hat{\eta}\lesssim1$. The coefficient $\bar{b}$ first experiences a tiny decrease and then shows a monotonically increasing tendency with the decrease of $w$, when $\hat{\eta}$ takes a value on the range $1\lesssim\hat{\eta}<2$. Furthermore, for a given massive particle with an initial velocity $w\in(0,~1)$, $\bar{a}$ increases with increasing the scaled bounce parameter. Differently, $\bar{b}$ displays a diverse behavior. With the increase of $\hat{\eta}$, it first increases slightly and then decreases to a minimum value for a given test particle with $0.94\lesssim w<1$. For a given particle with $0<w\lesssim 0.94$, $\bar{b}$ exhibits an increasing tendency when decreasing $\hat{\eta}$ monotonically. In addition, due to the correction of the bounce parameter, we note that the value ranges of the strong field limit coefficients ($1<\bar{a}\lesssim 1.63$ and $-0.54\lesssim\bar{b}<1.76$) in our scenario are evidently broader than those (namely, $1<\bar{a}\lesssim 1.41$ and $-0.40\lesssim\bar{b}<1.76$~\cite{FBT2025}) in the Schwarzschild lensing case of massive particles, respectively.

\section{Lensing observables in strong field limit} \label{sect4}
We then consider the black-bounce-Schwarzschild lensing observables according to the properties of the lensed relativistic images of the particle source in the strong-deflection-limit approximation. Under the scenario in which the source, lens, and observer are almost aligned exactly, the deflection angle $\alpha$ is very close to a multiple of $2\pi$, and it is convenient to express it by $\alpha=2n\pi+\Delta\alpha_n~(n\in N)$\,\cite{VE2000,BCIS2001}. Here, $\Delta\alpha_n~\left(0<\Delta\alpha_n\ll1\right)$ denotes the offset of the deflection angle or the so-called effective deflection angle, with $n$ being the number of circles that the particle loops around the black hole. Hence, the Virbhadra-Ellis lens equation\,\cite{VE2000} for this special case takes the form\,\cite{BCIS2001}
\begin{eqnarray}
\mathcal{B}=\vartheta-\frac{d_{LS}}{d_S}\Delta\alpha_n~,  \label{LensE}
\end{eqnarray}
which holds for both the case where the source is on the same side of the reference line as the image $(\beta>0)$ and the case when the source is on the opposite side $(\beta<0)$.

Similar to the procedure for the case of light\,\cite{Bozza2002}, we can obtain the angular positions of the relativistic images of the particle source and thus the image-flux magnifications of the source of the massive particle signals\,\cite{PNHZ2014,BT2017}. By introducing an image position $\vartheta_n^0$ which satisfies the condition $\alpha\left(\vartheta_n^0,\,w\right)=2n\pi$ ($n\geq1$), using the power series expansion of $\alpha\left(\vartheta_n,\,w\right)$ around $\vartheta_n=\vartheta_n^0$, and substituting this expansion into Eq.\,\eqref{LensE}, we express approximately the position of the $n$-th relativistic image as
\begin{eqnarray}
\vartheta_n\left(w,\,\hat{\eta}\right)=\vartheta_n^0+\frac{d_S\,u_c\,e_n\left(\mathcal{B}-\vartheta_n^0\right)}{d_L d_{LS}\,\bar{a}}~,~~~~  \label{nIP}
\end{eqnarray}
with
\begin{eqnarray}
&&\vartheta_n^0\left(w,\,\hat{\eta}\right)=\frac{u_c\left(1+e_n\right)}{d_L}~,~~~~  \label{thetaN0} \\
&&e_n\left(w,\,\hat{\eta}\right)=e^{\frac{\bar{b}-2n\pi}{\bar{a}}}~.~~~~  \label{eN}
\end{eqnarray}
In the limit $n\rightarrow \infty$, Eq.\,\eqref{nIP} yields the apparent angular radius $\vartheta_{\infty}$ of the particle sphere, which is related to the minimum impact parameter via $u_c=d_L\vartheta_{\infty}$. Moreover, the magnification of the $n$-th lensed relativistic image, which relates the observable flux $\mathcal{F}_{n}$ of the $n$-th image to the flux $\mathcal{F}_S$ of the unlensed particle source
via $\mathcal{F}_{n}=|\mu_n|\mathcal{F}_S$\,\cite{WP2007,PNHZ2014,PJ2019}, is of the approximate form
\begin{eqnarray}
\mu_n\left(w,\,\hat{\eta}\right)=\frac{d_S u_c^2\left(1+e_n\right)e_n}{d_{LS}d_L^2\,\bar{a}\,\mathcal{B}}~.~~~~  \label{muN}
\end{eqnarray}
It is worth noting that magnitude is adopted to show conventionally the brightness of a source of electromagnetic waves and relates the received image flux to the source flux by the Pogson formula\,\cite{IMKTP2018}. Considering the existence of the magnification of massive particle fluxes, the concept of magnitude may be extended properly for discussing the image properties in lensing phenomena of timelike signals\,\cite{HXJL2024,HL2022}.

Under the elementary situation where only the outermost image ($n=1$) is resolved and the remaining relativistic images ($n\geq2$) are packed together at $\vartheta_{\infty}$, the main lensing observables in the strong deflection limit read\,\cite{Bozza2002}:
\begin{eqnarray}
&&\vartheta_{\infty}\left(w\right)=\frac{u_c\left(w\right)}{d_L}~,  \label{LO1}  \\
&&s\left(w,\,\hat{\eta}\right)=\vartheta_{\infty}\left(w\right)e^{\frac{\bar{b}\left(w,\,\hat{\eta}\right)-2\pi}{\bar{a}\left(w,\,\hat{\eta}\right)}}~,~~~~~  \label{LO2}  \\
&&\mathcal{R}_m\left(w,\,\hat{\eta}\right)=\frac{5\pi}{\ln10}\frac{1}{\bar{a}\left(w,\,\hat{\eta}\right)}~,  \label{LO3}
\end{eqnarray}
where $s\equiv\vartheta_1-\vartheta_{\infty}$ is the angular separation between the first relativistic image and the packed ones of the particle emission source, 
and $\mathcal{R}_m \equiv 2.5\lg \left(\frac{\mu_1}{\Sigma_{n=2}^{\infty}\mu_n}\right)$ denotes their magnitudelike difference\,\cite{HXJL2024} resulted from the conversion of the ratio of their energy fluxes (or magnifications). Eqs.\,\eqref{LO1} - \eqref{LO3} indicate a close connection between the observables and the strong deflection limit coefficients and the critical impact parameter for both the case of light signals\,\cite{Bozza2002} and the case of massive particle signals. It should be noted that the particle sphere radius presented in Eqs.\,\eqref{LO1} - \eqref{LO2} exhibits no divergence, since the assumed quasi-alignment condition for the lens equation with $0<\Delta\alpha_n\ll1$ guarantees a convergent upper limit of $u_c$ and thus a tiny but nonzero lower limit (denoted by $w_{min}$) of $w$ in Eqs.\,\eqref{LO1} - \eqref{LO3}, where the value of $w_{min}$ depends on several parameters and could be as small as $10^{-8}$ (or even smaller). By way of example, we adopt $w_{min}=10^{-8}$ for later convenience of illustration.

It is necessary to mention that the angular particle-sphere radius $\vartheta_{\infty}$, the angular separation $s$ of the particle-source images, and the magnitudelike difference $\mathcal{R}_m$ originated from the energy flux ratio could be viewed as direct observables in future potential astronomical observations for gravitational lensing phenomena of massive particles by high-accuracy particle emission detectors or telescopes. For example, the apparent angular radius $\vartheta_{\infty}$ of a neutrino sphere (``shadow" formed by neutrino emission), relative to the reference line defined above, may be measured directly in related gravitational lensing observations by future neutrino emission detectors with high enough angular resolutions. Additionally, with the consideration of the dependence of the lensing quantities $\vartheta_{\infty}$, $s$, and $\mathcal{R}_m$ on $M$, $\eta$, and $w$ shown in Eqs.\,\eqref{LO1} - \eqref{LO3}, we can determine or constrain in turn the spacetime parameters $M$ and $\eta$ of the central body and the particle's initial velocity $w$ by astronomically measuring $\vartheta_{\infty}$, $s$, and $\mathcal{R}_m$, when modeling a proper astrophysical black hole as the black-bounce-Schwarzschild lens. In this process, the strong field limit parameters $\bar{a}$, $\bar{b}$, and $u_c$ act as the link between the strong-deflection observables in related astrophysical observations for gravitational lensing of massive particles and the quantities $M$, $\eta$, and $w$.

The bottom plane of Fig.\,\ref{Figure2} shows the angular particle sphere radius $\vartheta_{\infty}$ and the color-indexed lensing observables $s$ and $\mathcal{R}_m$ as the functions of the particle's initial velocity (and the scaled bounce parameter) for the case of the galactic supermassive black hole being the lens. Figure\,\ref{Figure2}\,(d) indicates that the observable $\vartheta_{\infty}$ increases monotonically with decreasing $w$ on a limited range $[10^{-8},~1)$, when a value of the scaled bounce parameter is given on the range $\left(0,~2\right)$. Moreover, as shown in Fig.\,\ref{Figure2}\,(e)\,-\,(f), the angular image separation $s$ and the magnitudelike difference $\mathcal{R}_m$ of the images for a given $\hat{\eta}$ increases and decreases in different rhythms, respectively, with the decrease of $w$ from 1 to its small lower limit $w_{min}=10^{-8}$. These two observables with a fixed initial velocity of the particle on the range $[10^{-8},~1)$ exhibit tendencies to increase and decrease relatively slowly, respectively, when increasing $\hat{\eta}$.

\section{Velocity- and bounce-induced effects}  \label{sect5} 
\subsection{Velocity effects on lensing observables and strong field limit parameters of lightlike case}  \label{sect5.1}
The deviation of the initial velocity $w$ of a massive particle from the speed of light leads to a crucial difference between the gravitational lensing of massive particles and that of electromagnetic waves in any background spacetime. And it is important to analyze this velocity effect mentioned above, when studying the gravitational lensing phenomena of timelike signals in a given curved geometry\,\cite{WS2004,LYJ2016,PJ2019,HL2022}.

For the convenience of the discussion in the following, we base on the idea of \cite{HXJL2024} and Eqs.\,\eqref{LO1} - \eqref{LO3} to define the velocity effects on the angular photon sphere radius, on the angular separation between the first image and the packed relativistic ones of a pointlike light source, and on their magnitude (or brightness) difference in the black-bounce-Schwarzschild spacetime, respectively, as follows:
\begin{eqnarray}
&&\Delta\vartheta_{\infty}\left(w\right)=\frac{u_c\left(w\right)-u_c\left(1\right)}{d_L}~,~   \label{Vvartheta}   \\
&&\Delta s\left(w,\,\hat{\eta}\right)=\vartheta_{\infty}\left(w\right)e^{\frac{\bar{b}\left(w,\,\hat{\eta}\right)-2\pi}{\bar{a}\left(w,\,\hat{\eta}\right)}}
-\vartheta_{\infty}\left(1\right)e^{\frac{\bar{b}\left(1,\,\hat{\eta}\right)-2\pi}{\bar{a}\left(1,\,\hat{\eta}\right)}}~,~~~~~  \label{Vr}  \\
&&\Delta\mathcal{R}_m\left(w,\,\hat{\eta}\right)=\frac{5\pi}{\ln10}\left[\frac{1}{\bar{a}\left(w,\,\hat{\eta}\right)}-\frac{1}{\bar{a}\left(1,\,\hat{\eta}\right)}\right]~,~~~~~  \label{Vs}
\end{eqnarray}
in which $w\in[10^{-8},~1)$, as assumed in Sec.\,\ref{sect4}. Incidentally, the velocity effects on the strong deflection limit coefficients $\bar{a}$ and $\bar{b}$, and on the critical impact parameter $u_c$, of the lightlike case in the regular spacetime can also be written respectively on the basis of Eqs.\,\eqref{baraDefine} - \eqref{ucDefine} in the forms
\begin{eqnarray}
&&\Delta\bar{a}\left(w,\,\hat{\eta}\right)=\bar{a}\left(w,\,\hat{\eta}\right)-\bar{a}\left(1,\,\hat{\eta}\right)~,~   \label{Vbara}   \\
&&\Delta\bar{b}\left(w,\,\hat{\eta}\right)=\bar{b}\left(w,\,\hat{\eta}\right)-\bar{b}\left(1,\,\hat{\eta}\right)~, ~~~~  \label{Vbarb}  \\
&&\Delta u_c\left(w\right)=u_c\left(w\right)-u_c\left(1\right)~,~~~~~  \label{Vuc}
\end{eqnarray}
where $w$ takes values on the range $(0,~1)$, as done in Eqs.\,\eqref{baraDefine} - \eqref{ucDefine}. The velocity effects presented in Eqs.\,\eqref{Vbara} - \eqref{Vuc} are helpful for us to supplementarily analyze and evaluate the corrections caused by the deviation of the particle's initial velocity from light speed to the lightlike strong-deflection observables in related gravitational lensing observations.

\begin{figure*}
\centering
\begin{minipage}[b]{16.cm}
\includegraphics[width=16.cm]{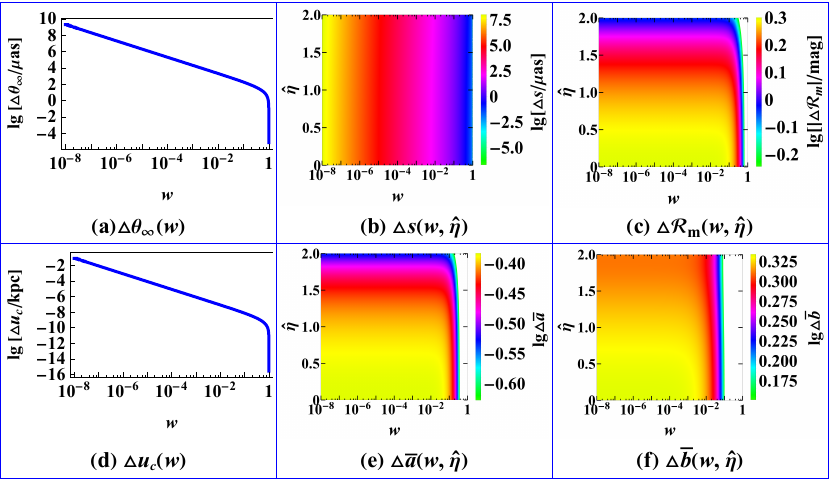}
\caption{The velocity effects $\Delta\vartheta_{\infty}$ and $\Delta u_c$, together with the color-indexed velocity effects $\Delta s$, $\Delta\mathcal{R}_m$, $\Delta\bar{a}$, and $\Delta\bar{b}$, plotted as the functions of $w$ (and $\hat{\eta}$) in the lensing scenario of Sgr A$^{\ast}$. Our attention is focused on the absolute values of these velocity effects in the discussion of their detectability. }   \label{Figure3}
\end{minipage}
\end{figure*}

\subsection{Bounce effects on lensing observables and strong field limit parameters of timelike case}
Besides the velocity effects defined above, it is also worth considering the influence of the spacetime bounce on the strong-field properties of the relativistic lensed images. Since the bounce-induced effects on the measurable properties of the primary, secondary, and relativistic images of a light source have been probed in detail in the literature (see, e.g., \cite{NPPS2020,CX2021,Tsukamoto2021,IKG2021,ZX2022}), the focus of this work in this aspect is on the impact of the bounce parameter on the strong-field lensing observables of the images of an emission source of massive particle signals, with the consideration of the bounce effects on the Schwarzschild lensing properties of the particle-source images analyzed in the weak-field limit\,\cite{HXJL2024}. In a similar way in Sec.\,\ref{sect5.1}, the correctional effects induced by the spacetime bounce on the angular separation between the first relativistic image and the packed ones of the particle source, as well as on their magnitudelike difference, of the Schwarzschild geometry are defined respectively by
\begin{eqnarray}
&&\delta s\left(w,\,\hat{\eta}\right)=\vartheta_{\infty}\left(w\right)\!\left[e^{\frac{\bar{b}\left(w,\,\hat{\eta}\right)-2\pi}{\bar{a}\left(w,\,\hat{\eta}\right)}}
-e^{\frac{\bar{b}\left(w,\,0\right)-2\pi}{\bar{a}\left(w,\,0\right)}}\right]~,~~~~~  \label{Bouncer}  \\
&&\delta\mathcal{R}_m\left(w,\,\hat{\eta}\right)=\frac{5\pi}{\ln10}\!\left[\frac{1}{\bar{a}\left(w,\,\hat{\eta}\right)}-\frac{1}{\bar{a}\left(w,\,0\right)}\right]~.~~~~~  \label{Bounces}
\end{eqnarray}
And the definitions of the bounce effects on the strong field limit coefficients $\bar{a}$ and $\bar{b}$ of the case where a massive particle, deflected by a Schwarzschild lens, acts as the test particle take the form
\begin{eqnarray}
&&\delta\bar{a}\left(w,\,\hat{\eta}\right)=\bar{a}\left(w,\,\hat{\eta}\right)-\bar{a}\left(w,\,0\right)~,~   \label{Bouncebara}   \\
&&\delta\bar{b}\left(w,\,\hat{\eta}\right)=\bar{b}\left(w,\,\hat{\eta}\right)-\bar{b}\left(w,\,0\right)~.~~~~  \label{Bouncebarb}
\end{eqnarray}

\section{Application to typical supermassive black holes}  \label{sect6} 

\subsection{Basics}
As a typical application of the results, in this section we model the supermassive black hole at the galactic center (i.e., Sgr A$^{\ast}$) and the one at the center of galactic M87
(i.e., M87$^{\ast}$) as the black-bounce-Schwarzschild lens, respectively. Our focus for potential applications mainly lies in the velocity effects on the strong-deflection black-bounce-Schwarzschild lensing properties\,\cite{NPPS2020,Tsuka2021a} of the relativistic images of the light source, the spacetime bounce effects on the strong-field lensing observables of the images of a massive-particle emission source, and the possibilities of their astronomical measurements, given that those correctional effects may also be detectable in future multimessenger synergic observations. However, it should be admitted that the experimental challenges, resulted from the extension of traditional multiwavelength astronomical observations to the observations with non-photonic messengers\,\cite{MFHM2019,Frost2023}, lead to a foreseeable situation that current multimessenger observatories or massive particle detectors are relatively rare and have lower (or even much lower) angular and flux resolutions (see, for instance,\,\cite{IceCube2006,Aab2014,Albert2017,KAZP2019,DEGG2021,Sharma2024}). Here we just perform a sketchy discussion of the potential observations of the velocity- and bounce-induced effects, in the light of the capability of current multiwavelength astronomical instruments\,\cite{RD2020,MAGIC:2022fww,Lico2023} and particle detectors, as illustrated in\,\cite{HXJL2024}.

The related parameters are given as follows. As mentioned above, the mass of Sgr A$^{\ast}$ and its distance from us are about $M=4.2\!\times\!10^6M_{\odot}$ and $d_L\!=\!8.2$\,kpc\,\cite{BG2016,Parsa2017}, respectively.
A mass of $M\!=\!6.5\!\times\!10^9 M_{\odot}$ and a distance of $d_L\!=\!16.9$\,Mpc\,\cite{2019ApJ...875L...6E} are adopted for the supermassive black hole M87$^{\ast}$.
Since the massive test particle in the strong-deflection lensing is not restricted to be relativistic\,\cite{FBT2025}, the range $10^{-8}\leq\,w<1$ is assumed, with the consideration of the mentioned lower limit of the particle's initial velocity originated from the quasi-alignment condition\,\cite{BCIS2001} of the lens equation in Eq.\,\eqref{LensE}. In addition, we set $0<\hat{\eta}<2$ for the scaled bounce parameter to indicate the regular black hole spacetime. And we use the following domains for performing the evaluation of the velocity effects:
\begin{eqnarray}
&& \mathcal{D}\equiv\left\{(w,~\hat{\eta})\,|\,10^{-8}\leq\,w<1,~0<\hat{\eta}<2\right\}~,            \label{D-1}  \\
&& \mathcal{D}_{\hat{\eta}1}\equiv\left\{(w,~\hat{\eta})\,|\,10^{-6}\leq w\leq0.999999,~\hat{\eta}=1\right\}~,~~~~~                   \label{D-2}  \\
&& \mathcal{D}_{\hat{\eta}2}\equiv\left\{(w,~\hat{\eta})\,|\,10^{-6}\leq w\leq0.999999,~\hat{\eta}=0\right\}~,                        \label{D-3}  \\
&& \mathcal{D}_{w}\equiv\left\{(w,~\hat{\eta})\,|\,w=0.1,~10^{-4}\leq\hat{\eta}\leq1.9999\right\}~, ~~~                               \label{D-4}
\end{eqnarray}
where $\mathcal{D}_{\hat{\eta}1}$, $\mathcal{D}_{\hat{\eta}2}$, and $\mathcal{D}_{w}$ act as three representative (or special) subdomains of the full domain $\mathcal{D}$, the order of magnitude ($|w-1|\!\sim\!10^{-6}$\,\cite{Agafon2012,Adam2012,Adamson2015}) of the velocities of ordinary massive neutrinos has been considered, and some possible constraints on the bounce parameter imposed by the detected precession of the star S2 around Sgr A$^{\ast}$\,\cite{ZTX2020} and by the Event Horizon Telescope observations for the angular diameter of the shadow of M87$^*$\,\cite{2019ApJ875L1E,Gralla2021} have been noticed.

\subsection{Examples of Sgr A$^{\ast}$ and M87$^{\ast}$}  \label{sect6.2}

\subsubsection{Discussion of the velocity effects}
\begin{figure*}
\centering
\begin{minipage}[b]{16cm}
\includegraphics[width=10.67cm]{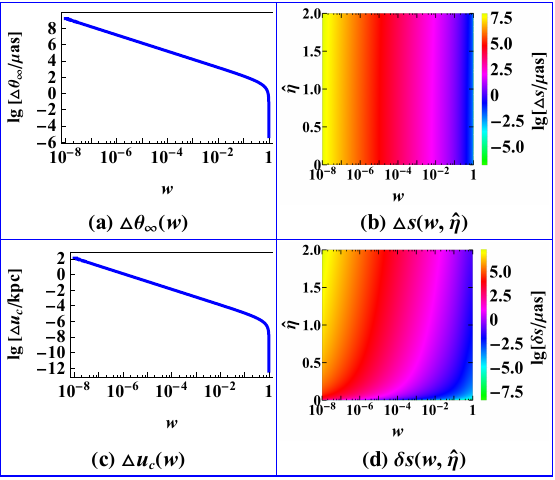}
\caption{The velocity effects $\Delta\vartheta_{\infty}$, $\Delta s$, and $\Delta u_c$, along with the bounce effect $\delta s$ in color-indexed form,
plotted for different $w$ (and $\hat{\eta}$) in the lensing case of M87$^{\ast}$. Notice that the results of $\Delta\mathcal{R}_m$, $\Delta\bar{a}$, and $\Delta\bar{b}$ for M87$^{\ast}$ are
the same as the corresponding ones in Fig.\,\ref{Figure3} for Sgr A$^{\ast}$. }   \label{Figure4}
\end{minipage}
\end{figure*}

The top plane of Fig.\,\ref{Figure3} presents the values of the velocity effects $\Delta\vartheta_{\infty}\left(w\right)$, $\Delta s\left(w,\,\hat{\eta}\right)$, and $\Delta\mathcal{R}_m\left(w,\,\hat{\eta}\right)$ on the strong-deflection lensing observables of the light-source images for the case of Sgr A$^{\ast}$ being the black-bounce-Schwarzschild lens, according to which two aspects should be mentioned. On the one hand, we find that the velocity effects on the angular photon sphere radius, and on the angular separation between the first relativistic image and the packed ones of the light source, increase monotonically with the decrease of the initial velocity of the particle for any given scaled bounce parameter on the domain $\mathcal{D}$, as shown in Fig.\,\ref{Figure3}\,(a)\,-\,(b). This is true when the supermassive black hole M87$^{\ast}$ is modeled as the lens (see Fig.\,\ref{Figure4}\,(a)\,-\,(b) for details). Conversely, the velocity effect on the magnitude difference (or equivalently, the difference in brightness) of the first image and the packed ones decreases with decreasing $w$ for a given $\hat{\eta}\in\mathcal{D}$. On a representative subdomain $\mathcal{D}_{\hat{\eta}1}$, $\Delta s$ ranges from about $9.1\times10^{-8}$ to $8.9\times10^{5}\mu$as in the lensing case of Sgr A$^{\ast}$ and varies from $6.8\times10^{-8}$ to $6.6\times10^{5}\mu$as for the case of M87$^{\ast}$, while $|\Delta\mathcal{R}_m|$ varies from $1.3\times10^{-6}$ to $1.8$ mag. On $\mathcal{D}_{\hat{\eta}2}$, $\Delta\vartheta_{\infty}$ changes from $1.8\times10^{-5}\mu$as to $2.0\times10^{7}\mu$as for Sgr A$^{\ast}$, and this velocity effect for M87$^{\ast}$ ranges from $1.3\times10^{-5}$ to $1.5\times10^{7}\mu$as, which is about two orders of magnitude smaller than the angular resolutions (at the level of $\lesssim 1^\circ$\,\cite{Aab2014,Bartoli2019,Albert2020}) of current instruments for observing massive particles or multimessengers. Moreover, for a given $w$ on $\mathcal{D}$, both $\Delta s$ shown in Fig.\,\ref{Figure3}\,(b) for the case of Sgr A$^{\ast}$ and in Fig.\,\ref{Figure4}\,(b) for M87$^{\ast}$ and $\Delta\mathcal{R}_m$ shown in Fig.\,\ref{Figure3}\,(c) display a monotonic and relatively slow increase with the increase of the scaled bounce parameter. On the subdomain $\mathcal{D}_{w}$, the minimum and maximum values of $\Delta s$ are respectively $4.6\mu$as and $6.6\mu$as for Sgr A$^{\ast}$, and they become $3.5\mu$as and $4.9\mu$as respectively for the case of M87$^{\ast}$, accompanied by a range of $|\Delta\mathcal{R}_m|$ from 0.9 to $1.9$ mag. On the other hand, we base on the results given in Fig.\,\ref{Figure3} and Fig.\,\ref{Figure4} to analyze the possibilities of the detection of these velocity effects briefly. In the light of Fig.\,\ref{Figure3}\,(a) and Fig.\,\ref{Figure4}\,(a), it is found that there is a relatively large possibility to detect the velocity effect on the photon sphere radius, within the capability of (near) future high-accuracy joint multimessenger detectors or observatories. Even for a relativistic massive particle with a high initial velocity $w=0.92$, such as a cosmic-ray neutron or proton with a corresponding energy of about $2.4$\,GeV, $\Delta\vartheta_{\infty}$ can reach about $1.3\mu$as for Sgr A$^{\ast}$ and about $1.1\mu$as for M87$^{\ast}$, both of which are still larger than the planned angular resolution ($\sim\!1\mu$as) of the SKA\,\cite{BBGKW2015,LXLWBLYHL2022} or other next-generation radio observatories\,\cite{Murphy2018,RD2020}. It indicates that for most of the relativistic particles and all of the nonrelativistic particles, the velocity effect $\Delta\vartheta_{\infty}$ may be observed by future multimessenger detectors whose astrometric precisions are nearly equal to or better than that of the SKA mentioned above. Furthermore, Fig.\,\ref{Figure3}\,(b) and Fig.\,\ref{Figure4}\,(b) show that it is also possible to observe the velocity effect $\Delta s$ on the angular image separation within the capability of these SKA-level multimessenger detectors, although a tighter restriction on the particle's initial velocity ($w\lesssim0.23$ for Sgr A$^{\ast}$, and $w\lesssim0.2$ for M87$^{\ast}$) has to be imposed for a given $\hat{\eta}$ on $\mathcal{D}$. For example, in the cases $\{w=0.1,~\hat{\eta}=1\}$ and $\{w=0.001,~\hat{\eta}=0.01\}$, the approximate values of $\Delta s$ for Sgr A$^{\ast}$ are respectively $5.1\mu$as and $820.9\mu$as, both of which are evidently larger than the SKA's intended accuracy. In addition, from Fig.\,\ref{Figure3}\,(c) we note that the potential possibility to measure the velocity effect on the brightness difference of the first image and the packed ones of the light source is not small on almost all of the domain $\mathcal{D}$, via future multimessenger emission detectors whose photometric and flux resolutions are approximately equal to or better than these of the nominal Kepler mission (with a photometric resolution of a few $\mu$mag\,\cite{Koch2010,BK2018}). It may be feasible to detect $\Delta\mathcal{R}_m$ by these Kepler-level multimessenger emission detectors, even in the situations where the particle has an ultrarelativistic initial velocity and the lens possesses a very large bounce parameter. For instance, $|\Delta\mathcal{R}_m|$ can reach about $243.6\mu$mag for the case in which $w=0.999$ and $\hat{\eta}=1.99$. Finally, we display the values of the velocity effects $\Delta\bar{a}$, $\Delta\bar{b}$, and $\Delta u_c$ on the strong field limit parameters of the lightlike case for various initial velocities of the particle (and different scaled bounce parameters) in the lensing scenario of Sgr A$^{\ast}$ on the bottom plane of Fig.\,\ref{Figure3} incidentally, and the values of $\Delta u_c$ for the case of M87$^{\ast}$ are presented in Fig.\,\ref{Figure4}\,(c). An increasing tendency of $\Delta u_c$ and $\Delta\bar{a}$ for a given $\hat{\eta}\in\left(0,~2\right)$ is shown in Fig.\,\ref{Figure3}\,(d)\,-\,(e) and Fig.\,\ref{Figure4}\,(c), when decreasing $w$ on the domain $\mathcal{D}$. This is true for the velocity effect on the strong field limit coefficient $\bar{b}$, provided that the scaled bounce parameter takes a value in the range $0<\hat{\eta}\lesssim 0.99$. We note that for a fixed spacetime bounce on the range $0.99\lesssim\hat{\eta}<2$, $\Delta\bar{b}$ first decreases slightly to a negative minimum value and then increase monotonically with the decrease of $w$. Furthermore, for a given $w\in\mathcal{D}$, we note that $\Delta\bar{a}$ increases with the decrease of $\hat{\eta}$, while $\Delta\bar{b}$ behaviors in a more complex manner by exhibiting a small decrease firstly and then a monotonous increase, with decreasing the scaled bounce parameter on $\mathcal{D}$. The maximum deviations of the strong field limit parameters $\bar{a}\left(w,\,\hat{\eta}\right)$, $\bar{b}\left(w,\,\hat{\eta}\right)$, and $u_c\left(w\right)$ of the timelike case from those of the lightlike case can respectively reach about $\Delta\bar{a}=0.4$, $\Delta\bar{b}=2.1$, and $\Delta u_c=8.0\times10^{-4}\,$kpc on the domain $\mathcal{D}_{\hat{\eta}1}$ for the lensing scenario of Sgr A$^{\ast}$. And the maximum value of $\Delta u_c$ on $\mathcal{D}_{\hat{\eta}1}$ becomes larger and approaches $1.2\,$kpc for the case of M87$^{\ast}$.

\begin{figure*}
\centering
\begin{minipage}[b]{16cm}
\includegraphics[width=10.67cm]{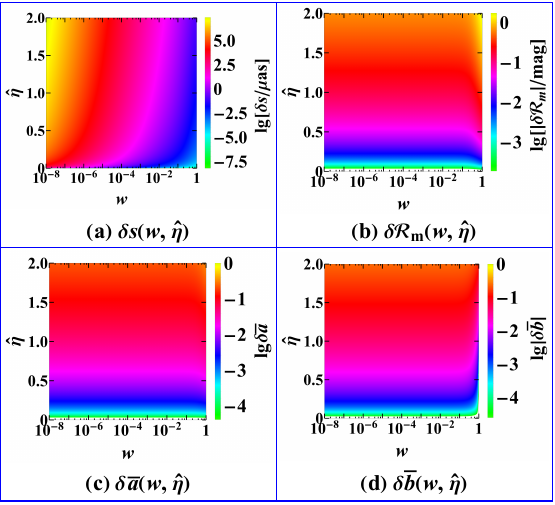}
\caption{The color-indexed bounce-induced effects $\delta s$, $\delta\mathcal{R}_m$, $\delta\bar{a}$, and $\delta\bar{b}$ plotted as the functions of $w$ and $\hat{\eta}$ in the lensing scenario of Sgr A$^{\ast}$.  }   \label{Figure5}
\end{minipage}
\end{figure*}

\subsubsection{Discussion of the bounce effects}
The values of the bounce effects on the strong-field Schwarzschild lensing observables of the particle-source images for the case where Sgr A$^{\ast}$ serves as the black-bounce-Schwarzschild lens are shown on the top plane of Fig.\,\ref{Figure5}. It can be seen from Fig.\,\ref{Figure5}\,(a) that the bounce effect $\delta s\left(w,\,\hat{\eta}\right)$ on the angular separation between the first relativistic particle-source image and the packed ones of the Schwarzschild lensing case exhibits an increase tendency with decreasing the particle's initial velocity and increasing the scaled bounce parameter on the domain $\mathcal{D}$, similar to the behavior of the strong field limit coefficient $\bar{a}\left(w,\,\hat{\eta}\right)$. This tendency of $\delta s$ also applies when the lens is modeled by another astrophysical black hole such as M87$^{\ast}$ (see Fig.\,\ref{Figure4}\,(d) for details). Differently, Fig.\,\ref{Figure5}\,(b) indicates that the bounce effect $\delta\mathcal{R}_m\left(w,\,\hat{\eta}\right)$ on the magnitudelike difference between the first image and the packed ones of the particle source decreases monotonically, when both $w$ ($\in\mathcal{D}$) and $\hat{\eta}$ (on the range $0.02\lesssim \hat{\eta}<2$) are increased. For a fixed $\hat{\eta}$ within the range $0<\hat{\eta}\lesssim 0.02$, $\delta\mathcal{R}_m$ first increases and then decreases to a negative minimum value, with the increase of $w$. The maximum value of $\delta s$ and the minimum value of $\delta\mathcal{R}_m$ are respectively about $6.2\times10^{4}\mu$as and $-0.39$\,mag on the subdomain $\mathcal{D}_{\hat{\eta}1}$, and reduce to $2.1\mu$as and $-0.67$\,mag on $\mathcal{D}_{w}$, respectively, when considering the lensing case of Sgr A$^{\ast}$. In addition, we note that for a given massive particle with $10^{-8}\leq w\lesssim0.94$, the bounce effects $\delta\bar{a}\left(w,\,\hat{\eta}\right)$ and $\delta\bar{b}\left(w,\,\hat{\eta}\right)$ on the strong deflection limit coefficients of the Schwarzschild lensing case of the particle increases and decreases respectively with the increase of $\hat{\eta}$, which can be revealed qualitatively by Fig.\,\ref{Figure5}\,(c)\,-\,(d). For a massive particle with $0.94\lesssim w<1$, $\delta\bar{a}$ increases monotonically, while $\delta\bar{b}$ shows a tendency to first increase and then decrease,
as $\hat{\eta}$ increases. When the scaled bounce parameter is fixed, $\delta\bar{a}$ for the case $0.03\lesssim\hat{\eta}<2$ and $\delta\bar{b}$ for the case $0.01\lesssim\hat{\eta}<2$ are found to increase monotonically with increasing $w$. They show a slight decrease followed by an increase, or other diverse behaviors, provided that the scaled bounce parameter takes a value on $0<\hat{\eta}\lesssim0.03$ for $\delta\bar{a}$ and on $0<\hat{\eta}\lesssim0.01$ for $\delta\bar{b}$. Regarding the detectability of the effects of the spacetime bounce, we find that it is possible to measure the bounce effect on the angular image separation within the capability of future high-accuracy joint multimessenger observatories, considering that most of the values of $\delta s$ on the subdomain $\left\{(w,\,\hat{\eta})\,|\,10^{-8}\!\leq w\lesssim0.17,~4.2\!\times\!10^{-4}\!\lesssim\hat{\eta}<\!2\right\}$ for Sgr A$^{\ast}$ and on $\left\{(w,~\hat{\eta})\,|\,10^{-8}\!\leq w\lesssim0.14,~4.9\!\times\!10^{-4}\!\lesssim\hat{\eta}<2\right\}$ for M87$^{\ast}$ are larger than the planned angular resolution of the SKA. For instance, in the case $\{w\!=\!0.001,~\hat{\eta}\!=\!1.3\}$, $\delta s$ can reach about $106.6\mu$as for Sgr A$^{\ast}$ and $79.8\mu$as for M87$^{\ast}$, both of which evidently exceed the astrometric precisions of the mentioned SKA-level multimessenger detectors. Furthermore, although smaller than that for the velocity effect $\Delta\mathcal{R}_m$, the possibility of detecting the bounce effect on the magnitudelike difference of the particle-source images by the mentioned Kepler-level multimessenger emission detectors is relatively large, since we note that the values of $|\delta\mathcal{R}_m|$ on the subdomain
$\left\{(w,~\hat{\eta})\,|\,10^{-8}\!\leq w<\!1,~8.2\times10^{-3}\!\lesssim\hat{\eta}<\!2\right\}$ are not smaller than $10\mu$mag. For example, $|\delta\mathcal{R}_m|$ can reach about $0.038$mag and is within the capability of these future high-accuracy multimessenger emission detectors, when $w=10^{-6}$ and $\hat{\eta}=0.5$ are assumed.

\section{Conclusions and discussion} \label{sect7}
In this work, we have studied the strong deflection gravitational lensing of neutral massive particles induced by a black-bounce-Schwarzschild black hole. On the basis of the equations of motion of a timelike particle, we have obtained the expressions for the equation and radius of the particle sphere. This particle sphere equation is found to be consistent with the equation of the photon sphere of the regular spacetime, in the special case where the initial velocity $w$ of the particle reaches the speed of light. A subsequent derivation of the black-bounce-Schwarzschild deflection angle of a massive particle in the strong field limit has been performed. The main strong-deflection observables of the lensed relativistic images of a pointlike particle source, including the angular radius of the particle sphere, the angular separation between the first image and the packed other ones of the particle source, and the magnitude difference resulted from the conversion of the ratio of their magnifications (or fluxes), have also been achieved. Additionally, we have formulated the velocity effects, resulted from the deviation of $w$ from light speed, on the strong-field lensing observables of the relativistic light-source images and on the strong field limit parameters of the lightlike scenario in the regular geometry. The impact of the spacetime bounce on the Schwarzschild strong-deflection lensing observables of the images of a massive-particle emission source, as well as on the strong deflection limit coefficients of the timelike case, has also been considered.

As typical examples, the galactic supermassive black hole and the central black hole in M87 have been modeled as the lens respectively to estimate the astronomical observability of the velocity- and bounce-induced effects on the observables, together with an analysis of their dependence on the parameters. We find that the velocity effects on the angular photon sphere radius and on the angular separation between the first image and the packed ones of a pointlike light source, along with the absolute value of the velocity effect on their magnitude difference, display a tendency to increase monotonically with the decrease of the initial velocity of the particle for a given scaled bounce parameter $\hat{\eta}$. It theoretically indicates that under the same conditions, the main strong-deflection black-bounce-Schwarzschild lensing observables of the relativistic images of a pointlike particle source are more evident than those of the light-source images, respectively, which implies a window of opportunity for detecting these velocity effects. We notice that the bounce effect on the angular separation between the first relativistic image and the packed ones of a massive-particle source in the Schwarzschild lensing case exhibits an increase tendency with decreasing $w$ and increasing $\hat{\eta}$, and that the bounce effect on their magnitudelike difference decreases when both $w$ and $\hat{\eta}$ increase. Though not accessible with current limited techniques, our results suggest a relatively large possibility to detect the velocity effect on the angular photon sphere radius, in the capability of future high-accuracy particle and multimessenger detectors whose astrometric accuracies are similar to that of the Square Kilometre Array. Within the resolution capability of these detectors, it is also likely to observe the velocity effect on the angular separation between the first image and the packed ones of the light source, as well as the bounce effect on the angular separation of the particle-source images, with a tighter limit on $w$ and $\hat{\eta}$. As to the velocity effect on the brightness difference of the light-source images, it is found that the possibility of its detection is not small via future multimessenger emission detectors whose photometric and flux resolutions are similar to these of the Kepler Mission. Additionally, we note that it is possible to detect the bounce effect on the magnitudelike difference of the particle-source images by those future emission detectors.

Compared with those on the practical weak-field black-bounce-Schwarzschild lensing observables of the primary and secondary images of a pointlike light source discussed in\,\cite{HXJL2024}, the velocity effects discussed in this work are on the strong-deflection lensing properties of the relativistic light-source images of the same regular geometry, and apply to various test massive particles with relativistic or nonrelativistic initial velocities (not limiting to the cases of relativistic particles). It indicates that there could be new complementary and unique opportunities to probe a series of regular black hole spacetimes, including the one considered in present paper, via the strong-field gravitational lensing phenomena of massive particles in current multimessenger era.
In comparison with the bounce-induced effects on the weak- and strong-field Schwarzschild\,\cite{NPPS2020,CX2021} or Reissner-Nordstr\"{o}m\,\cite{ZX2022} lensing observables of the light-source images,
as well as on the observable weak-field Schwarzschild lensing properties of the particle-source images\,\cite{HXJL2024}, each of the bounce effects on the strong-deflection Schwarzschild observables of the relativistic images of the massive-particle source shown in Sec.\,\ref{sect6.2} has a wider range with a higher upper limit, due to a broader choice of massive test particles with various initial velocities.
This diversity of test particles (including relativistic and nonrelativistic particles) would effectively result in more chances and better detectability for the observational verifications of the bounce effects on the strong-field observables than for those of the bounce effects on the weak-field lensing observables. 

Finally, it is worth mentioning that our consideration of the strong deflection gravitational lensing of relativistic and nonrelativistic particles due to the regular black hole could be extended to other static spherically symmetric or stationary axisymmetric spacetimes. In addition, we note that there could be other creative approaches to derive analytically the gravitational deflection of massive particles in the regular spacetime in the strong-deflection case, such as by adopting the analogy\,\cite{KL1992,Tsupko2014} between the motion of a photon with some frequency in an unmagnetized cold homogeneous plasma and the timelike geodesic motion of a massive test particle with the initial velocity $w$ in the same geometry. Furthermore, it should be admitted that we have to confront some severe experimental challenges and limitations in current and near future observations of the gravitational lensing of massive particles. However, we could expect that the extensive significant achievements in the theoretical and experimental studies of various gravitational lensing phenomena over the last decades\,\cite{VK2008,WLFY2012,CJ2015J,Liu2016,TG2017,HL2017a,ZX2017E,CX2018,TG2018,MLMHBN2019,JBO2019,LX2019E,ZX2020a,KIG2020a,LX2021,HLL2021a,JRPO2022,GX2022,SVP2023,GYYH2023,
PPOD2023,aaaa2023,SVP2024,ZX2024,SCHL2024,SGM2025,Igata2026,BH2026}, together with the rapid progress made in the joint photonic and non-photonic messenger observations\,\cite{MFHM2019,QJFZZ2021,Tambo2025}, will give birth to a good prospect on this aspect with unprecedented opportunities in multimessenger astronomy.

\begin{acknowledgments}
This work was supported in part by the National Natural Science Foundation of China (Grant Nos. 12475057 and 12205139), the Natural Science Foundation of Hunan Province (Grant No. 2026JJ50351), and the Scientific Research Foundation of Hunan Provincial Education Department (Grant No. 25B0373).
\end{acknowledgments}


\begin{thebibliography}{222}%
\makeatletter
\providecommand \@ifxundefined [1]{%
 \@ifx{#1\undefined}
}%
\providecommand \@ifnum [1]{%
 \ifnum #1\expandafter \@firstoftwo
 \else \expandafter \@secondoftwo
 \fi
}%
\providecommand \@ifx [1]{%
 \ifx #1\expandafter \@firstoftwo
 \else \expandafter \@secondoftwo
 \fi
}%
\providecommand \natexlab [1]{#1}%
\providecommand \enquote  [1]{``#1''}%
\providecommand \bibnamefont  [1]{#1}%
\providecommand \bibfnamefont [1]{#1}%
\providecommand \citenamefont [1]{#1}%
\providecommand \href@noop [0]{\@secondoftwo}%
\providecommand \href [0]{\begingroup \@sanitize@url \@href}%
\providecommand \@href[1]{\@@startlink{#1}\@@href}%
\providecommand \@@href[1]{\endgroup#1\@@endlink}%
\providecommand \@sanitize@url [0]{\catcode `\\12\catcode `\$12\catcode
  `\&12\catcode `\#12\catcode `\^12\catcode `\_12\catcode `\%12\relax}%
\providecommand \@@startlink[1]{}%
\providecommand \@@endlink[0]{}%
\providecommand \url  [0]{\begingroup\@sanitize@url \@url }%
\providecommand \@url [1]{\endgroup\@href {#1}{\urlprefix }}%
\providecommand \urlprefix  [0]{URL }%
\providecommand \Eprint [0]{\href }%
\providecommand \doibase [0]{http://dx.doi.org/}%
\providecommand \selectlanguage [0]{\@gobble}%
\providecommand \bibinfo  [0]{\@secondoftwo}%
\providecommand \bibfield  [0]{\@secondoftwo}%
\providecommand \translation [1]{[#1]}%
\providecommand \BibitemOpen [0]{}%
\providecommand \bibitemStop [0]{}%
\providecommand \bibitemNoStop [0]{.\EOS\space}%
\providecommand \EOS [0]{\spacefactor3000\relax}%
\providecommand \BibitemShut  [1]{\csname bibitem#1\endcsname}%
\let\auto@bib@innerbib\@empty
\bibitem [{\citenamefont {{Refsdal}}(1964)}]{Refsdal1964}%
  \BibitemOpen
  \bibfield  {author} {\bibinfo {author} {\bibfnamefont {S.}~\bibnamefont
  {{Refsdal}}},\ }\href {\doibase 10.1093/mnras/128.4.307} {\bibfield
  {journal} {\bibinfo  {journal} {Mon. Not. R. Astron. Soc.}\ }\textbf
  {\bibinfo {volume} {128}},\ \bibinfo {pages} {307} (\bibinfo {year}
  {1964})}\BibitemShut {NoStop}%
\bibitem [{\citenamefont {{Blandford}}\ and\ \citenamefont
  {{Narayan}}(1992)}]{BN1992}%
  \BibitemOpen
  \bibfield  {author} {\bibinfo {author} {\bibfnamefont {R.~D.}\ \bibnamefont
  {{Blandford}}}\ and\ \bibinfo {author} {\bibfnamefont {R.}~\bibnamefont
  {{Narayan}}},\ }\href {\doibase 10.1146/annurev.astro.30.1.311} {\bibfield
  {journal} {\bibinfo  {journal} {Annu. Rev. Astron. Astrophys.}\ }\textbf
  {\bibinfo {volume} {30}},\ \bibinfo {pages} {311} (\bibinfo {year}
  {1992})}\BibitemShut {NoStop}%
\bibitem [{\citenamefont {{Virbhadra}}\ and\ \citenamefont
  {{Ellis}}(2000)}]{VE2000}%
  \BibitemOpen
  \bibfield  {author} {\bibinfo {author} {\bibfnamefont {K.~S.}\ \bibnamefont
  {{Virbhadra}}}\ and\ \bibinfo {author} {\bibfnamefont {G.~F.~R.}\
  \bibnamefont {{Ellis}}},\ }\href {\doibase 10.1103/PhysRevD.62.084003}
  {\bibfield  {journal} {\bibinfo  {journal} {\prd}\ }\textbf {\bibinfo
  {volume} {62}},\ \bibinfo {eid} {084003} (\bibinfo {year} {2000})},\ \Eprint
  {http://arxiv.org/abs/astro-ph/9904193} {arXiv:astro-ph/9904193 [astro-ph]}
  \BibitemShut {NoStop}%
\bibitem [{\citenamefont {{Bartelmann}}\ and\ \citenamefont
  {{Schneider}}(2001)}]{BS2001}%
  \BibitemOpen
  \bibfield  {author} {\bibinfo {author} {\bibfnamefont {M.}~\bibnamefont
  {{Bartelmann}}}\ and\ \bibinfo {author} {\bibfnamefont {P.}~\bibnamefont
  {{Schneider}}},\ }\href {\doibase 10.1016/S0370-1573(00)00082-X} {\bibfield
  {journal} {\bibinfo  {journal} {Phys. Rep.}\ }\textbf {\bibinfo {volume}
  {340}},\ \bibinfo {pages} {291} (\bibinfo {year} {2001})},\ \Eprint
  {http://arxiv.org/abs/astro-ph/9912508} {arXiv:astro-ph/9912508 [astro-ph]}
  \BibitemShut {NoStop}%
\bibitem [{\citenamefont {{Keeton}}\ and\ \citenamefont
  {{Petters}}(2005)}]{KP2005}%
  \BibitemOpen
  \bibfield  {author} {\bibinfo {author} {\bibfnamefont {C.~R.}\ \bibnamefont
  {{Keeton}}}\ and\ \bibinfo {author} {\bibfnamefont {A.~O.}\ \bibnamefont
  {{Petters}}},\ }\href {\doibase 10.1103/PhysRevD.72.104006} {\bibfield
  {journal} {\bibinfo  {journal} {\prd}\ }\textbf {\bibinfo {volume} {72}},\
  \bibinfo {eid} {104006} (\bibinfo {year} {2005})},\ \Eprint
  {http://arxiv.org/abs/gr-qc/0511019} {arXiv:gr-qc/0511019 [gr-qc]}
  \BibitemShut {NoStop}%
\bibitem [{\citenamefont {{Zhang}}\ \emph {et~al.}(2007)\citenamefont
  {{Zhang}}, \citenamefont {{Liguori}}, \citenamefont {{Bean}},\ and\
  \citenamefont {{Dodelson}}}]{ZLBD2007}%
  \BibitemOpen
  \bibfield  {author} {\bibinfo {author} {\bibfnamefont {P.}~\bibnamefont
  {{Zhang}}}, \bibinfo {author} {\bibfnamefont {M.}~\bibnamefont {{Liguori}}},
  \bibinfo {author} {\bibfnamefont {R.}~\bibnamefont {{Bean}}}, \ and\ \bibinfo
  {author} {\bibfnamefont {S.}~\bibnamefont {{Dodelson}}},\ }\href {\doibase
  10.1103/PhysRevLett.99.141302} {\bibfield  {journal} {\bibinfo  {journal}
  {\prl}\ }\textbf {\bibinfo {volume} {99}},\ \bibinfo {eid} {141302} (\bibinfo
  {year} {2007})},\ \Eprint {http://arxiv.org/abs/0704.1932} {arXiv:0704.1932
  [astro-ph]} \BibitemShut {NoStop}%
\bibitem [{\citenamefont {{Chen}}\ and\ \citenamefont {{Jing}}(2009)}]{CJ2009}%
  \BibitemOpen
  \bibfield  {author} {\bibinfo {author} {\bibfnamefont {S.}~\bibnamefont
  {{Chen}}}\ and\ \bibinfo {author} {\bibfnamefont {J.}~\bibnamefont
  {{Jing}}},\ }\href {\doibase 10.1103/PhysRevD.80.024036} {\bibfield
  {journal} {\bibinfo  {journal} {\prd}\ }\textbf {\bibinfo {volume} {80}},\
  \bibinfo {eid} {024036} (\bibinfo {year} {2009})},\ \Eprint
  {http://arxiv.org/abs/0905.2055} {arXiv:0905.2055 [gr-qc]} \BibitemShut
  {NoStop}%
\bibitem [{\citenamefont {{Reyes}}\ \emph {et~al.}(2010)\citenamefont
  {{Reyes}}, \citenamefont {{Mandelbaum}}, \citenamefont {{Seljak}},
  \citenamefont {{Baldauf}}, \citenamefont {{Gunn}}, \citenamefont
  {{Lombriser}},\ and\ \citenamefont {{Smith}}}]{Reyes2010}%
  \BibitemOpen
  \bibfield  {author} {\bibinfo {author} {\bibfnamefont {R.}~\bibnamefont
  {{Reyes}}}, \bibinfo {author} {\bibfnamefont {R.}~\bibnamefont
  {{Mandelbaum}}}, \bibinfo {author} {\bibfnamefont {U.}~\bibnamefont
  {{Seljak}}}, \bibinfo {author} {\bibfnamefont {T.}~\bibnamefont {{Baldauf}}},
  \bibinfo {author} {\bibfnamefont {J.~E.}\ \bibnamefont {{Gunn}}}, \bibinfo
  {author} {\bibfnamefont {L.}~\bibnamefont {{Lombriser}}}, \ and\ \bibinfo
  {author} {\bibfnamefont {R.~E.}\ \bibnamefont {{Smith}}},\ }\href {\doibase
  10.1038/nature08857} {\bibfield  {journal} {\bibinfo  {journal} {\nat}\
  }\textbf {\bibinfo {volume} {464}},\ \bibinfo {pages} {256} (\bibinfo {year}
  {2010})},\ \Eprint {http://arxiv.org/abs/1003.2185} {arXiv:1003.2185
  [astro-ph.CO]} \BibitemShut {NoStop}%
\bibitem [{\citenamefont {{Narikawa}}\ \emph {et~al.}(2013)\citenamefont
  {{Narikawa}}, \citenamefont {{Kobayashi}}, \citenamefont {{Yamauchi}},\ and\
  \citenamefont {{Saito}}}]{NKYS2013}%
  \BibitemOpen
  \bibfield  {author} {\bibinfo {author} {\bibfnamefont {T.}~\bibnamefont
  {{Narikawa}}}, \bibinfo {author} {\bibfnamefont {T.}~\bibnamefont
  {{Kobayashi}}}, \bibinfo {author} {\bibfnamefont {D.}~\bibnamefont
  {{Yamauchi}}}, \ and\ \bibinfo {author} {\bibfnamefont {R.}~\bibnamefont
  {{Saito}}},\ }\href {\doibase 10.1103/PhysRevD.87.124006} {\bibfield
  {journal} {\bibinfo  {journal} {\prd}\ }\textbf {\bibinfo {volume} {87}},\
  \bibinfo {eid} {124006} (\bibinfo {year} {2013})},\ \Eprint
  {http://arxiv.org/abs/1302.2311} {arXiv:1302.2311 [astro-ph.CO]} \BibitemShut
  {NoStop}%
\bibitem [{\citenamefont {{Zhao}}\ and\ \citenamefont {{Xie}}(2016)}]{ZX2016}%
  \BibitemOpen
  \bibfield  {author} {\bibinfo {author} {\bibfnamefont {S.-S.}\ \bibnamefont
  {{Zhao}}}\ and\ \bibinfo {author} {\bibfnamefont {Y.}~\bibnamefont {{Xie}}},\
  }\href {\doibase 10.1088/1475-7516/2016/07/007} {\bibfield  {journal}
  {\bibinfo  {journal} {J. Cosmol. Astropart. Phys.}\ }\textbf {\bibinfo
  {volume} {2016}},\ \bibinfo {eid} {007} (\bibinfo {year} {2016})},\ \Eprint
  {http://arxiv.org/abs/1603.00637} {arXiv:1603.00637 [gr-qc]} \BibitemShut
  {NoStop}%
\bibitem [{\citenamefont {{Fan}}\ \emph {et~al.}(2017)\citenamefont {{Fan}},
  \citenamefont {{Liao}}, \citenamefont {{Biesiada}}, \citenamefont
  {{Pi{\'o}rkowska-Kurpas}},\ and\ \citenamefont {{Zhu}}}]{FLBPZ2017}%
  \BibitemOpen
  \bibfield  {author} {\bibinfo {author} {\bibfnamefont {X.-L.}\ \bibnamefont
  {{Fan}}}, \bibinfo {author} {\bibfnamefont {K.}~\bibnamefont {{Liao}}},
  \bibinfo {author} {\bibfnamefont {M.}~\bibnamefont {{Biesiada}}}, \bibinfo
  {author} {\bibfnamefont {A.}~\bibnamefont {{Pi{\'o}rkowska-Kurpas}}}, \ and\
  \bibinfo {author} {\bibfnamefont {Z.-H.}\ \bibnamefont {{Zhu}}},\ }\href
  {\doibase 10.1103/PhysRevLett.118.091102} {\bibfield  {journal} {\bibinfo
  {journal} {\prl}\ }\textbf {\bibinfo {volume} {118}},\ \bibinfo {eid}
  {091102} (\bibinfo {year} {2017})},\ \Eprint
  {http://arxiv.org/abs/1612.04095} {arXiv:1612.04095 [gr-qc]} \BibitemShut
  {NoStop}%
\bibitem [{\citenamefont {{Liu}}\ \emph {et~al.}(2017)\citenamefont {{Liu}},
  \citenamefont {{Wang}}, \citenamefont {{Li}},\ and\ \citenamefont
  {{Ma}}}]{LWLM2017}%
  \BibitemOpen
  \bibfield  {author} {\bibinfo {author} {\bibfnamefont {H.}~\bibnamefont
  {{Liu}}}, \bibinfo {author} {\bibfnamefont {X.}~\bibnamefont {{Wang}}},
  \bibinfo {author} {\bibfnamefont {H.}~\bibnamefont {{Li}}}, \ and\ \bibinfo
  {author} {\bibfnamefont {Y.}~\bibnamefont {{Ma}}},\ }\href {\doibase
  10.1140/epjc/s10052-017-5308-5} {\bibfield  {journal} {\bibinfo  {journal}
  {Eur. Phys. J. C}\ }\textbf {\bibinfo {volume} {77}},\ \bibinfo {eid} {723}
  (\bibinfo {year} {2017})},\ \Eprint {http://arxiv.org/abs/1508.02647}
  {arXiv:1508.02647 [gr-qc]} \BibitemShut {NoStop}%
\bibitem [{\citenamefont {{Collett}}\ \emph {et~al.}(2018)\citenamefont
  {{Collett}} \emph {et~al.}}]{Collett2018}%
  \BibitemOpen
  \bibfield  {author} {\bibinfo {author} {\bibfnamefont {T.~E.}\ \bibnamefont
  {{Collett}}} \emph {et~al.},\ }\href {\doibase 10.1126/science.aao2469}
  {\bibfield  {journal} {\bibinfo  {journal} {Science}\ }\textbf {\bibinfo
  {volume} {360}},\ \bibinfo {pages} {1342} (\bibinfo {year} {2018})},\ \Eprint
  {http://arxiv.org/abs/1806.08300} {arXiv:1806.08300 [astro-ph.CO]}
  \BibitemShut {NoStop}%
\bibitem [{\citenamefont {{Li}}\ \emph {et~al.}(2018)\citenamefont {{Li}},
  \citenamefont {{Mao}}, \citenamefont {{Zhao}},\ and\ \citenamefont
  {{Lu}}}]{LMZL2018}%
  \BibitemOpen
  \bibfield  {author} {\bibinfo {author} {\bibfnamefont {S.-S.}\ \bibnamefont
  {{Li}}}, \bibinfo {author} {\bibfnamefont {S.}~\bibnamefont {{Mao}}},
  \bibinfo {author} {\bibfnamefont {Y.}~\bibnamefont {{Zhao}}}, \ and\ \bibinfo
  {author} {\bibfnamefont {Y.}~\bibnamefont {{Lu}}},\ }\href {\doibase
  10.1093/mnras/sty411} {\bibfield  {journal} {\bibinfo  {journal} {Mon. Not.
  R. Astron. Soc.}\ }\textbf {\bibinfo {volume} {476}},\ \bibinfo {pages}
  {2220} (\bibinfo {year} {2018})},\ \Eprint {http://arxiv.org/abs/1802.05089}
  {arXiv:1802.05089 [astro-ph.CO]} \BibitemShut {NoStop}%
\bibitem [{\citenamefont {{Jung}}\ and\ \citenamefont {{Shin}}(2019)}]{JS2019}%
  \BibitemOpen
  \bibfield  {author} {\bibinfo {author} {\bibfnamefont {S.}~\bibnamefont
  {{Jung}}}\ and\ \bibinfo {author} {\bibfnamefont {C.~S.}\ \bibnamefont
  {{Shin}}},\ }\href {\doibase 10.1103/PhysRevLett.122.041103} {\bibfield
  {journal} {\bibinfo  {journal} {\prl}\ }\textbf {\bibinfo {volume} {122}},\
  \bibinfo {eid} {041103} (\bibinfo {year} {2019})},\ \Eprint
  {http://arxiv.org/abs/1712.01396} {arXiv:1712.01396 [astro-ph.CO]}
  \BibitemShut {NoStop}%
\bibitem [{\citenamefont {{Chen}}\ \emph {et~al.}(2020)\citenamefont {{Chen}},
  \citenamefont {{Luo}}, \citenamefont {{Cai}},\ and\ \citenamefont
  {{Saridakis}}}]{CLCS2020}%
  \BibitemOpen
  \bibfield  {author} {\bibinfo {author} {\bibfnamefont {Z.}~\bibnamefont
  {{Chen}}}, \bibinfo {author} {\bibfnamefont {W.}~\bibnamefont {{Luo}}},
  \bibinfo {author} {\bibfnamefont {Y.-F.}\ \bibnamefont {{Cai}}}, \ and\
  \bibinfo {author} {\bibfnamefont {E.~N.}\ \bibnamefont {{Saridakis}}},\
  }\href {\doibase 10.1103/PhysRevD.102.104044} {\bibfield  {journal} {\bibinfo
   {journal} {\prd}\ }\textbf {\bibinfo {volume} {102}},\ \bibinfo {eid}
  {104044} (\bibinfo {year} {2020})},\ \Eprint
  {http://arxiv.org/abs/1907.12225} {arXiv:1907.12225 [astro-ph.CO]}
  \BibitemShut {NoStop}%
\bibitem [{\citenamefont {{Huang}}\ and\ \citenamefont
  {{Deng}}(2025)}]{HDE2025}%
  \BibitemOpen
  \bibfield  {author} {\bibinfo {author} {\bibfnamefont {G.-Y.}\ \bibnamefont
  {{Huang}}}\ and\ \bibinfo {author} {\bibfnamefont {X.-M.}\ \bibnamefont
  {{Deng}}},\ }\href {\doibase 10.1140/epjc/s10052-025-15148-z} {\bibfield
  {journal} {\bibinfo  {journal} {Eur. Phys. J. C}\ }\textbf {\bibinfo {volume}
  {85}},\ \bibinfo {eid} {1410} (\bibinfo {year} {2025})}\BibitemShut {NoStop}%
\bibitem [{\citenamefont {{Eiroa}}\ and\ \citenamefont
  {{Sendra}}(2011)}]{ES2011}%
  \BibitemOpen
  \bibfield  {author} {\bibinfo {author} {\bibfnamefont {E.~F.}\ \bibnamefont
  {{Eiroa}}}\ and\ \bibinfo {author} {\bibfnamefont {C.~M.}\ \bibnamefont
  {{Sendra}}},\ }\href {\doibase 10.1088/0264-9381/28/8/085008} {\bibfield
  {journal} {\bibinfo  {journal} {Classical and Quantum Gravity}\ }\textbf
  {\bibinfo {volume} {28}},\ \bibinfo {eid} {085008} (\bibinfo {year}
  {2011})},\ \Eprint {http://arxiv.org/abs/1011.2455} {arXiv:1011.2455 [gr-qc]}
  \BibitemShut {NoStop}%
\bibitem [{\citenamefont {{Eiroa}}\ and\ \citenamefont
  {{Sendra}}(2013)}]{ES2013}%
  \BibitemOpen
  \bibfield  {author} {\bibinfo {author} {\bibfnamefont {E.~F.}\ \bibnamefont
  {{Eiroa}}}\ and\ \bibinfo {author} {\bibfnamefont {C.~M.}\ \bibnamefont
  {{Sendra}}},\ }\href {\doibase 10.1103/PhysRevD.88.103007} {\bibfield
  {journal} {\bibinfo  {journal} {\prd}\ }\textbf {\bibinfo {volume} {88}},\
  \bibinfo {eid} {103007} (\bibinfo {year} {2013})},\ \Eprint
  {http://arxiv.org/abs/1308.5959} {arXiv:1308.5959 [gr-qc]} \BibitemShut
  {NoStop}%
\bibitem [{\citenamefont {{Schee}}\ and\ \citenamefont
  {{Stuchl{\'\i}k}}(2015)}]{SS2015}%
  \BibitemOpen
  \bibfield  {author} {\bibinfo {author} {\bibfnamefont {J.}~\bibnamefont
  {{Schee}}}\ and\ \bibinfo {author} {\bibfnamefont {Z.}~\bibnamefont
  {{Stuchl{\'\i}k}}},\ }\href {\doibase 10.1088/1475-7516/2015/06/048}
  {\bibfield  {journal} {\bibinfo  {journal} {J. Cosmol. Astropart. Phys.}\
  }\textbf {\bibinfo {volume} {2015}},\ \bibinfo {pages} {048} (\bibinfo {year}
  {2015})},\ \Eprint {http://arxiv.org/abs/1501.00835} {arXiv:1501.00835
  [astro-ph.HE]} \BibitemShut {NoStop}%
\bibitem [{\citenamefont {{Ghaffarnejad}}\ and\ \citenamefont
  {{niad}}(2016)}]{GN2016}%
  \BibitemOpen
  \bibfield  {author} {\bibinfo {author} {\bibfnamefont {H.}~\bibnamefont
  {{Ghaffarnejad}}}\ and\ \bibinfo {author} {\bibfnamefont {H.}~\bibnamefont
  {{niad}}},\ }\href {\doibase 10.1007/s10773-015-2787-8} {\bibfield  {journal}
  {\bibinfo  {journal} {Int. J. Theor. Phys.}\ }\textbf {\bibinfo {volume}
  {55}},\ \bibinfo {pages} {1492} (\bibinfo {year} {2016})},\ \Eprint
  {http://arxiv.org/abs/1411.7247} {arXiv:1411.7247 [gr-qc]} \BibitemShut
  {NoStop}%
\bibitem [{\citenamefont {{Jusufi}}\ \emph {et~al.}(2018)\citenamefont
  {{Jusufi}}, \citenamefont {{{\"O}vg{\"u}n}}, \citenamefont {{Saavedra}},
  \citenamefont {{V{\'a}squez}},\ and\ \citenamefont
  {{Gonz{\'a}lez}}}]{JOSVG2018}%
  \BibitemOpen
  \bibfield  {author} {\bibinfo {author} {\bibfnamefont {K.}~\bibnamefont
  {{Jusufi}}}, \bibinfo {author} {\bibfnamefont {A.}~\bibnamefont
  {{{\"O}vg{\"u}n}}}, \bibinfo {author} {\bibfnamefont {J.}~\bibnamefont
  {{Saavedra}}}, \bibinfo {author} {\bibfnamefont {Y.}~\bibnamefont
  {{V{\'a}squez}}}, \ and\ \bibinfo {author} {\bibfnamefont {P.~A.}\
  \bibnamefont {{Gonz{\'a}lez}}},\ }\href {\doibase 10.1103/PhysRevD.97.124024}
  {\bibfield  {journal} {\bibinfo  {journal} {\prd}\ }\textbf {\bibinfo
  {volume} {97}},\ \bibinfo {eid} {124024} (\bibinfo {year} {2018})},\ \Eprint
  {http://arxiv.org/abs/1804.00643} {arXiv:1804.00643 [gr-qc]} \BibitemShut
  {NoStop}%
\bibitem [{\citenamefont {{Zhang}}\ and\ \citenamefont
  {{Jing}}(2018)}]{ZJ2018}%
  \BibitemOpen
  \bibfield  {author} {\bibinfo {author} {\bibfnamefont {R.}~\bibnamefont
  {{Zhang}}}\ and\ \bibinfo {author} {\bibfnamefont {J.}~\bibnamefont
  {{Jing}}},\ }\href {\doibase 10.1140/epjc/s10052-018-6272-4} {\bibfield
  {journal} {\bibinfo  {journal} {Eur. Phys. J. C}\ }\textbf {\bibinfo {volume}
  {78}},\ \bibinfo {eid} {796} (\bibinfo {year} {2018})}\BibitemShut {NoStop}%
\bibitem [{\citenamefont {{Liu}}\ \emph {et~al.}(2019)\citenamefont {{Liu}},
  \citenamefont {{Mai}}, \citenamefont {{Wu}},\ and\ \citenamefont
  {{Xie}}}]{LMWX2019}%
  \BibitemOpen
  \bibfield  {author} {\bibinfo {author} {\bibfnamefont {F.-Y.}\ \bibnamefont
  {{Liu}}}, \bibinfo {author} {\bibfnamefont {Y.-F.}\ \bibnamefont {{Mai}}},
  \bibinfo {author} {\bibfnamefont {W.-Y.}\ \bibnamefont {{Wu}}}, \ and\
  \bibinfo {author} {\bibfnamefont {Y.}~\bibnamefont {{Xie}}},\ }\href
  {\doibase 10.1016/j.physletb.2019.06.052} {\bibfield  {journal} {\bibinfo
  {journal} {Phys. Lett. B}\ }\textbf {\bibinfo {volume} {795}},\ \bibinfo
  {pages} {475} (\bibinfo {year} {2019})}\BibitemShut {NoStop}%
\bibitem [{\citenamefont {{{\"O}vg{\"u}n}}(2019)}]{Ovgun2019}%
  \BibitemOpen
  \bibfield  {author} {\bibinfo {author} {\bibfnamefont {A.}~\bibnamefont
  {{{\"O}vg{\"u}n}}},\ }\href {\doibase 10.1103/PhysRevD.99.104075} {\bibfield
  {journal} {\bibinfo  {journal} {\prd}\ }\textbf {\bibinfo {volume} {99}},\
  \bibinfo {eid} {104075} (\bibinfo {year} {2019})},\ \Eprint
  {http://arxiv.org/abs/1902.04411} {arXiv:1902.04411 [gr-qc]} \BibitemShut
  {NoStop}%
\bibitem [{\citenamefont {{Cheng}}\ and\ \citenamefont {{Xie}}(2021)}]{CX2021}%
  \BibitemOpen
  \bibfield  {author} {\bibinfo {author} {\bibfnamefont {X.-T.}\ \bibnamefont
  {{Cheng}}}\ and\ \bibinfo {author} {\bibfnamefont {Y.}~\bibnamefont
  {{Xie}}},\ }\href {\doibase 10.1103/PhysRevD.103.064040} {\bibfield
  {journal} {\bibinfo  {journal} {\prd}\ }\textbf {\bibinfo {volume} {103}},\
  \bibinfo {eid} {064040} (\bibinfo {year} {2021})}\BibitemShut {NoStop}%
\bibitem [{\citenamefont {{Ghosh}}\ and\ \citenamefont
  {{Bhattacharyya}}(2022)}]{GB2022}%
  \BibitemOpen
  \bibfield  {author} {\bibinfo {author} {\bibfnamefont {S.}~\bibnamefont
  {{Ghosh}}}\ and\ \bibinfo {author} {\bibfnamefont {A.}~\bibnamefont
  {{Bhattacharyya}}},\ }\href {\doibase 10.1088/1475-7516/2022/11/006}
  {\bibfield  {journal} {\bibinfo  {journal} {J. Cosmol. Astropart. Phys.}\
  }\textbf {\bibinfo {volume} {2022}},\ \bibinfo {eid} {006} (\bibinfo {year}
  {2022})},\ \Eprint {http://arxiv.org/abs/2206.09954} {arXiv:2206.09954
  [gr-qc]} \BibitemShut {NoStop}%
\bibitem [{\citenamefont {{de Paula}}\ \emph {et~al.}(2023)\citenamefont {{de
  Paula}}, \citenamefont {{Junior}}, \citenamefont {{Cunha}},\ and\
  \citenamefont {{Crispino}}}]{DJCC2023}%
  \BibitemOpen
  \bibfield  {author} {\bibinfo {author} {\bibfnamefont {M.~A.~A.}\
  \bibnamefont {{de Paula}}}, \bibinfo {author} {\bibfnamefont {H.~C.~D.~L.}\
  \bibnamefont {{Junior}}}, \bibinfo {author} {\bibfnamefont {P.~V.~P.}\
  \bibnamefont {{Cunha}}}, \ and\ \bibinfo {author} {\bibfnamefont {L.~C.~B.}\
  \bibnamefont {{Crispino}}},\ }\href {\doibase 10.1103/PhysRevD.108.084029}
  {\bibfield  {journal} {\bibinfo  {journal} {\prd}\ }\textbf {\bibinfo
  {volume} {108}},\ \bibinfo {eid} {084029} (\bibinfo {year} {2023})},\ \Eprint
  {http://arxiv.org/abs/2305.04776} {arXiv:2305.04776 [gr-qc]} \BibitemShut
  {NoStop}%
\bibitem [{\citenamefont {{Zhang}}\ \emph {et~al.}(2023)\citenamefont
  {{Zhang}}, \citenamefont {{Jing}}, \citenamefont {{Peng}},\ and\
  \citenamefont {{Huang}}}]{ZJPH2023}%
  \BibitemOpen
  \bibfield  {author} {\bibinfo {author} {\bibfnamefont {R.}~\bibnamefont
  {{Zhang}}}, \bibinfo {author} {\bibfnamefont {J.}~\bibnamefont {{Jing}}},
  \bibinfo {author} {\bibfnamefont {Z.}~\bibnamefont {{Peng}}}, \ and\ \bibinfo
  {author} {\bibfnamefont {Q.}~\bibnamefont {{Huang}}},\ }\href {\doibase
  10.1088/1674-1137/acf489} {\bibfield  {journal} {\bibinfo  {journal} {Chin.
  Phys. C}\ }\textbf {\bibinfo {volume} {47}},\ \bibinfo {eid} {105105}
  (\bibinfo {year} {2023})},\ \Eprint {http://arxiv.org/abs/2304.13263}
  {arXiv:2304.13263 [gr-qc]} \BibitemShut {NoStop}%
\bibitem [{\citenamefont {{Xie}}\ \emph {et~al.}(2024)\citenamefont {{Xie}},
  \citenamefont {{Zhang}}, \citenamefont {{Sun}}, \citenamefont {{Li}},\ and\
  \citenamefont {{Duan}}}]{XZSLD2024}%
  \BibitemOpen
  \bibfield  {author} {\bibinfo {author} {\bibfnamefont {C.-H.}\ \bibnamefont
  {{Xie}}}, \bibinfo {author} {\bibfnamefont {Y.}~\bibnamefont {{Zhang}}},
  \bibinfo {author} {\bibfnamefont {Q.}~\bibnamefont {{Sun}}}, \bibinfo
  {author} {\bibfnamefont {Q.-Q.}\ \bibnamefont {{Li}}}, \ and\ \bibinfo
  {author} {\bibfnamefont {P.-F.}\ \bibnamefont {{Duan}}},\ }\href {\doibase
  10.1088/1475-7516/2024/05/121} {\bibfield  {journal} {\bibinfo  {journal} {J.
  Cosmol. Astropart. Phys.}\ }\textbf {\bibinfo {volume} {2024}},\ \bibinfo
  {eid} {121} (\bibinfo {year} {2024})},\ \Eprint
  {http://arxiv.org/abs/2401.05454} {arXiv:2401.05454 [gr-qc]} \BibitemShut
  {NoStop}%
\bibitem [{\citenamefont {{Turakhonov}}\ \emph {et~al.}(2025)\citenamefont
  {{Turakhonov}}, \citenamefont {{Atamurotov}}, \citenamefont {{Ghosh}},\ and\
  \citenamefont {{Abdujabbarov}}}]{TGA2025}%
  \BibitemOpen
  \bibfield  {author} {\bibinfo {author} {\bibfnamefont {Z.}~\bibnamefont
  {{Turakhonov}}}, \bibinfo {author} {\bibfnamefont {F.}~\bibnamefont
  {{Atamurotov}}}, \bibinfo {author} {\bibfnamefont {S.~G.}\ \bibnamefont
  {{Ghosh}}}, \ and\ \bibinfo {author} {\bibfnamefont {A.}~\bibnamefont
  {{Abdujabbarov}}},\ }\href {\doibase 10.1016/j.dark.2025.101880} {\bibfield
  {journal} {\bibinfo  {journal} {Phys. Dark Univ.}\ }\textbf {\bibinfo
  {volume} {48}},\ \bibinfo {eid} {101880} (\bibinfo {year}
  {2025})}\BibitemShut {NoStop}%
\bibitem [{\citenamefont {{Bambi}}(2013)}]{Bambi2013}%
  \BibitemOpen
  \bibfield  {author} {\bibinfo {author} {\bibfnamefont {C.}~\bibnamefont
  {{Bambi}}},\ }\href {\doibase 10.1088/1475-7516/2013/08/055} {\bibfield
  {journal} {\bibinfo  {journal} {J. Cosmol. Astropart. Phys.}\ }\textbf
  {\bibinfo {volume} {2013}},\ \bibinfo {eid} {055} (\bibinfo {year} {2013})},\
  \Eprint {http://arxiv.org/abs/1305.5409} {arXiv:1305.5409 [gr-qc]}
  \BibitemShut {NoStop}%
\bibitem [{\citenamefont {{Li}}\ and\ \citenamefont {{Bambi}}(2014)}]{LB2014}%
  \BibitemOpen
  \bibfield  {author} {\bibinfo {author} {\bibfnamefont {Z.}~\bibnamefont
  {{Li}}}\ and\ \bibinfo {author} {\bibfnamefont {C.}~\bibnamefont {{Bambi}}},\
  }\href {\doibase 10.1088/1475-7516/2014/01/041} {\bibfield  {journal}
  {\bibinfo  {journal} {J. Cosmol. Astropart. Phys.}\ }\textbf {\bibinfo
  {volume} {2014}},\ \bibinfo {eid} {041} (\bibinfo {year} {2014})},\ \Eprint
  {http://arxiv.org/abs/1309.1606} {arXiv:1309.1606 [gr-qc]} \BibitemShut
  {NoStop}%
\bibitem [{\citenamefont {{Lamy}}\ \emph {et~al.}(2018)\citenamefont {{Lamy}},
  \citenamefont {{Gourgoulhon}}, \citenamefont {{Paumard}},\ and\ \citenamefont
  {{Vincent}}}]{LGPV2018}%
  \BibitemOpen
  \bibfield  {author} {\bibinfo {author} {\bibfnamefont {F.}~\bibnamefont
  {{Lamy}}}, \bibinfo {author} {\bibfnamefont {E.}~\bibnamefont
  {{Gourgoulhon}}}, \bibinfo {author} {\bibfnamefont {T.}~\bibnamefont
  {{Paumard}}}, \ and\ \bibinfo {author} {\bibfnamefont {F.~H.}\ \bibnamefont
  {{Vincent}}},\ }\href {\doibase 10.1088/1361-6382/aabd97} {\bibfield
  {journal} {\bibinfo  {journal} {Classical and Quantum Gravity}\ }\textbf
  {\bibinfo {volume} {35}},\ \bibinfo {eid} {115009} (\bibinfo {year}
  {2018})},\ \Eprint {http://arxiv.org/abs/1802.01635} {arXiv:1802.01635
  [gr-qc]} \BibitemShut {NoStop}%
\bibitem [{\citenamefont {{Kumar}}\ \emph
  {et~al.}(2020{\natexlab{a}})\citenamefont {{Kumar}}, \citenamefont
  {{Kumar}},\ and\ \citenamefont {{Ghosh}}}]{KKG2020}%
  \BibitemOpen
  \bibfield  {author} {\bibinfo {author} {\bibfnamefont {R.}~\bibnamefont
  {{Kumar}}}, \bibinfo {author} {\bibfnamefont {A.}~\bibnamefont {{Kumar}}}, \
  and\ \bibinfo {author} {\bibfnamefont {S.~G.}\ \bibnamefont {{Ghosh}}},\
  }\href {\doibase 10.3847/1538-4357/ab8c4a} {\bibfield  {journal} {\bibinfo
  {journal} {Astrophys. J.}\ }\textbf {\bibinfo {volume} {896}},\ \bibinfo
  {eid} {89} (\bibinfo {year} {2020}{\natexlab{a}})},\ \Eprint
  {http://arxiv.org/abs/2006.09869} {arXiv:2006.09869 [gr-qc]} \BibitemShut
  {NoStop}%
\bibitem [{\citenamefont {{Torres}}(2022)}]{Torres2022}%
  \BibitemOpen
  \bibfield  {author} {\bibinfo {author} {\bibfnamefont {R.}~\bibnamefont
  {{Torres}}},\ }\href {\doibase 10.48550/arXiv.2208.12713} {\bibfield
  {journal} {\bibinfo  {journal} {arXiv e-prints}\ ,\ \bibinfo {eid}
  {arXiv:2208.12713}} (\bibinfo {year} {2022})},\ \Eprint
  {http://arxiv.org/abs/2208.12713} {arXiv:2208.12713 [gr-qc]} \BibitemShut
  {NoStop}%
\bibitem [{\citenamefont {{Weinberg}}(1972)}]{We1972}%
  \BibitemOpen
  \bibfield  {author} {\bibinfo {author} {\bibfnamefont {S.}~\bibnamefont
  {{Weinberg}}},\ }\href@noop {} {\emph {\bibinfo {title} {{Gravitation and
  Cosmology: Principles and Applications of the General Theory of
  Relativity}}}}\ (\bibinfo  {publisher} {Wiley},\ \bibinfo {address} {New
  York},\ \bibinfo {year} {1972})\BibitemShut {NoStop}%
\bibitem [{\citenamefont {{Feng}}\ \emph {et~al.}(2024)\citenamefont {{Feng}},
  \citenamefont {{Ling}}, \citenamefont {{Wu}},\ and\ \citenamefont
  {{Jiang}}}]{FLWJ2024}%
  \BibitemOpen
  \bibfield  {author} {\bibinfo {author} {\bibfnamefont {Z.}~\bibnamefont
  {{Feng}}}, \bibinfo {author} {\bibfnamefont {Y.}~\bibnamefont {{Ling}}},
  \bibinfo {author} {\bibfnamefont {X.}~\bibnamefont {{Wu}}}, \ and\ \bibinfo
  {author} {\bibfnamefont {Q.}~\bibnamefont {{Jiang}}},\ }\href {\doibase
  10.1007/s11433-023-2373-0} {\bibfield  {journal} {\bibinfo  {journal} {Sci.
  China-Phys. Mech. Astron.}\ }\textbf {\bibinfo {volume} {67}},\ \bibinfo
  {eid} {270412} (\bibinfo {year} {2024})},\ \Eprint
  {http://arxiv.org/abs/2308.15689} {arXiv:2308.15689 [gr-qc]} \BibitemShut
  {NoStop}%
\bibitem [{\citenamefont {{Feng}}\ \emph {et~al.}(2025)\citenamefont {{Feng}},
  \citenamefont {{Jiang}}, \citenamefont {{Ling}}, \citenamefont {{Wu}},\ and\
  \citenamefont {{Yu}}}]{FJWY2025}%
  \BibitemOpen
  \bibfield  {author} {\bibinfo {author} {\bibfnamefont {Z.}~\bibnamefont
  {{Feng}}}, \bibinfo {author} {\bibfnamefont {Q.}~\bibnamefont {{Jiang}}},
  \bibinfo {author} {\bibfnamefont {Y.}~\bibnamefont {{Ling}}}, \bibinfo
  {author} {\bibfnamefont {X.}~\bibnamefont {{Wu}}}, \ and\ \bibinfo {author}
  {\bibfnamefont {Z.}~\bibnamefont {{Yu}}},\ }\href {\doibase
  10.1007/s11433-024-2623-8} {\bibfield  {journal} {\bibinfo  {journal} {Sci.
  China-Phys. Mech. Astron.}\ }\textbf {\bibinfo {volume} {68}},\ \bibinfo
  {eid} {260412} (\bibinfo {year} {2025})},\ \Eprint
  {http://arxiv.org/abs/2408.01780} {arXiv:2408.01780 [gr-qc]} \BibitemShut
  {NoStop}%
\bibitem [{\citenamefont {{Bardeen}}(1968)}]{Ba1968}%
  \BibitemOpen
  \bibfield  {author} {\bibinfo {author} {\bibfnamefont {J.}~\bibnamefont
  {{Bardeen}}},\ }in\ \href@noop {} {\emph {\bibinfo {booktitle} {Proceedings
  of the 5th International Conference on Gravitation and the Theory of
  Relativity}}}\ (\bibinfo {year} {1968})\ p.~\bibinfo {pages} {87}\BibitemShut
  {NoStop}%
\bibitem [{\citenamefont {{Roman}}\ and\ \citenamefont
  {{Bergmann}}(1983)}]{RB1983}%
  \BibitemOpen
  \bibfield  {author} {\bibinfo {author} {\bibfnamefont {T.~A.}\ \bibnamefont
  {{Roman}}}\ and\ \bibinfo {author} {\bibfnamefont {P.~G.}\ \bibnamefont
  {{Bergmann}}},\ }\href {\doibase 10.1103/PhysRevD.28.1265} {\bibfield
  {journal} {\bibinfo  {journal} {\prd}\ }\textbf {\bibinfo {volume} {28}},\
  \bibinfo {pages} {1265} (\bibinfo {year} {1983})}\BibitemShut {NoStop}%
\bibitem [{\citenamefont {{Hayward}}(2006)}]{Hayward2006}%
  \BibitemOpen
  \bibfield  {author} {\bibinfo {author} {\bibfnamefont {S.~A.}\ \bibnamefont
  {{Hayward}}},\ }\href {\doibase 10.1103/PhysRevLett.96.031103} {\bibfield
  {journal} {\bibinfo  {journal} {\prl}\ }\textbf {\bibinfo {volume} {96}},\
  \bibinfo {eid} {031103} (\bibinfo {year} {2006})},\ \Eprint
  {http://arxiv.org/abs/gr-qc/0506126} {arXiv:gr-qc/0506126 [gr-qc]}
  \BibitemShut {NoStop}%
\bibitem [{\citenamefont {{Bronnikov}}\ and\ \citenamefont
  {{Fabris}}(2006)}]{BF2006}%
  \BibitemOpen
  \bibfield  {author} {\bibinfo {author} {\bibfnamefont {K.~A.}\ \bibnamefont
  {{Bronnikov}}}\ and\ \bibinfo {author} {\bibfnamefont {J.~C.}\ \bibnamefont
  {{Fabris}}},\ }\href {\doibase 10.1103/PhysRevLett.96.251101} {\bibfield
  {journal} {\bibinfo  {journal} {\prl}\ }\textbf {\bibinfo {volume} {96}},\
  \bibinfo {eid} {251101} (\bibinfo {year} {2006})},\ \Eprint
  {http://arxiv.org/abs/gr-qc/0511109} {arXiv:gr-qc/0511109 [gr-qc]}
  \BibitemShut {NoStop}%
\bibitem [{\citenamefont {{Bambi}}\ and\ \citenamefont
  {{Modesto}}(2013)}]{BM2013}%
  \BibitemOpen
  \bibfield  {author} {\bibinfo {author} {\bibfnamefont {C.}~\bibnamefont
  {{Bambi}}}\ and\ \bibinfo {author} {\bibfnamefont {L.}~\bibnamefont
  {{Modesto}}},\ }\href {\doibase 10.1016/j.physletb.2013.03.025} {\bibfield
  {journal} {\bibinfo  {journal} {Phys. Lett. B}\ }\textbf {\bibinfo {volume}
  {721}},\ \bibinfo {pages} {329} (\bibinfo {year} {2013})},\ \Eprint
  {http://arxiv.org/abs/1302.6075} {arXiv:1302.6075 [gr-qc]} \BibitemShut
  {NoStop}%
\bibitem [{\citenamefont {{Fan}}\ and\ \citenamefont {{Wang}}(2016)}]{FW2016}%
  \BibitemOpen
  \bibfield  {author} {\bibinfo {author} {\bibfnamefont {Z.-Y.}\ \bibnamefont
  {{Fan}}}\ and\ \bibinfo {author} {\bibfnamefont {X.}~\bibnamefont {{Wang}}},\
  }\href {\doibase 10.1103/PhysRevD.94.124027} {\bibfield  {journal} {\bibinfo
  {journal} {\prd}\ }\textbf {\bibinfo {volume} {94}},\ \bibinfo {eid} {124027}
  (\bibinfo {year} {2016})},\ \Eprint {http://arxiv.org/abs/1610.02636}
  {arXiv:1610.02636 [gr-qc]} \BibitemShut {NoStop}%
\bibitem [{\citenamefont {{Huang}}\ and\ \citenamefont
  {{Yang}}(2019)}]{HY2019}%
  \BibitemOpen
  \bibfield  {author} {\bibinfo {author} {\bibfnamefont {H.}~\bibnamefont
  {{Huang}}}\ and\ \bibinfo {author} {\bibfnamefont {J.}~\bibnamefont
  {{Yang}}},\ }\href {\doibase 10.1103/PhysRevD.100.124063} {\bibfield
  {journal} {\bibinfo  {journal} {\prd}\ }\textbf {\bibinfo {volume} {100}},\
  \bibinfo {eid} {124063} (\bibinfo {year} {2019})},\ \Eprint
  {http://arxiv.org/abs/1909.04603} {arXiv:1909.04603 [gr-qc]} \BibitemShut
  {NoStop}%
\bibitem [{\citenamefont {{Barrientos}}\ \emph {et~al.}(2022)\citenamefont
  {{Barrientos}}, \citenamefont {{Cisterna}}, \citenamefont {{Mora}},\ and\
  \citenamefont {{Vigan{\`o}}}}]{BCMV2022}%
  \BibitemOpen
  \bibfield  {author} {\bibinfo {author} {\bibfnamefont {J.}~\bibnamefont
  {{Barrientos}}}, \bibinfo {author} {\bibfnamefont {A.}~\bibnamefont
  {{Cisterna}}}, \bibinfo {author} {\bibfnamefont {N.}~\bibnamefont {{Mora}}},
  \ and\ \bibinfo {author} {\bibfnamefont {A.}~\bibnamefont {{Vigan{\`o}}}},\
  }\href {\doibase 10.1103/PhysRevD.106.024038} {\bibfield  {journal} {\bibinfo
   {journal} {\prd}\ }\textbf {\bibinfo {volume} {106}},\ \bibinfo {eid}
  {024038} (\bibinfo {year} {2022})},\ \Eprint
  {http://arxiv.org/abs/2202.06706} {arXiv:2202.06706 [hep-th]} \BibitemShut
  {NoStop}%
\bibitem [{\citenamefont {{Ovalle}}\ \emph {et~al.}(2023)\citenamefont
  {{Ovalle}}, \citenamefont {{Casadio}},\ and\ \citenamefont
  {{Giusti}}}]{OCG2023}%
  \BibitemOpen
  \bibfield  {author} {\bibinfo {author} {\bibfnamefont {J.}~\bibnamefont
  {{Ovalle}}}, \bibinfo {author} {\bibfnamefont {R.}~\bibnamefont {{Casadio}}},
  \ and\ \bibinfo {author} {\bibfnamefont {A.}~\bibnamefont {{Giusti}}},\
  }\href {\doibase 10.1016/j.physletb.2023.138085} {\bibfield  {journal}
  {\bibinfo  {journal} {Phys. Lett. B}\ }\textbf {\bibinfo {volume} {844}},\
  \bibinfo {eid} {138085} (\bibinfo {year} {2023})},\ \Eprint
  {http://arxiv.org/abs/2304.03263} {arXiv:2304.03263 [gr-qc]} \BibitemShut
  {NoStop}%
\bibitem [{\citenamefont {{Ovalle}}(2024)}]{Ovalle2024}%
  \BibitemOpen
  \bibfield  {author} {\bibinfo {author} {\bibfnamefont {J.}~\bibnamefont
  {{Ovalle}}},\ }\href {\doibase 10.1103/PhysRevD.109.104032} {\bibfield
  {journal} {\bibinfo  {journal} {\prd}\ }\textbf {\bibinfo {volume} {109}},\
  \bibinfo {eid} {104032} (\bibinfo {year} {2024})},\ \Eprint
  {http://arxiv.org/abs/2405.06731} {arXiv:2405.06731 [gr-qc]} \BibitemShut
  {NoStop}%
\bibitem [{\citenamefont {{Bueno}}\ \emph {et~al.}(2025)\citenamefont
  {{Bueno}}, \citenamefont {{Cano}}, \citenamefont {{Hennigar}},\ and\
  \citenamefont {{Murcia}}}]{BCHM2025}%
  \BibitemOpen
  \bibfield  {author} {\bibinfo {author} {\bibfnamefont {P.}~\bibnamefont
  {{Bueno}}}, \bibinfo {author} {\bibfnamefont {P.~A.}\ \bibnamefont {{Cano}}},
  \bibinfo {author} {\bibfnamefont {R.~A.}\ \bibnamefont {{Hennigar}}}, \ and\
  \bibinfo {author} {\bibfnamefont {{\'A}.~J.}\ \bibnamefont {{Murcia}}},\
  }\href {\doibase 10.1103/PhysRevLett.134.181401} {\bibfield  {journal}
  {\bibinfo  {journal} {\prl}\ }\textbf {\bibinfo {volume} {134}},\ \bibinfo
  {eid} {181401} (\bibinfo {year} {2025})},\ \Eprint
  {http://arxiv.org/abs/2412.02742} {arXiv:2412.02742 [gr-qc]} \BibitemShut
  {NoStop}%
\bibitem [{\citenamefont {{Carballo-Rubio}}\ \emph {et~al.}(2018)\citenamefont
  {{Carballo-Rubio}}, \citenamefont {{Di Filippo}}, \citenamefont
  {{Liberati}},\ and\ \citenamefont {{Visser}}}]{CDLV2018}%
  \BibitemOpen
  \bibfield  {author} {\bibinfo {author} {\bibfnamefont {R.}~\bibnamefont
  {{Carballo-Rubio}}}, \bibinfo {author} {\bibfnamefont {F.}~\bibnamefont {{Di
  Filippo}}}, \bibinfo {author} {\bibfnamefont {S.}~\bibnamefont {{Liberati}}},
  \ and\ \bibinfo {author} {\bibfnamefont {M.}~\bibnamefont {{Visser}}},\
  }\href {\doibase 10.1103/PhysRevD.98.124009} {\bibfield  {journal} {\bibinfo
  {journal} {\prd}\ }\textbf {\bibinfo {volume} {98}},\ \bibinfo {eid} {124009}
  (\bibinfo {year} {2018})},\ \Eprint {http://arxiv.org/abs/1809.08238}
  {arXiv:1809.08238 [gr-qc]} \BibitemShut {NoStop}%
\bibitem [{\citenamefont {{Li}}\ and\ \citenamefont {{Miao}}(2022)}]{LM2022}%
  \BibitemOpen
  \bibfield  {author} {\bibinfo {author} {\bibfnamefont {Y.}~\bibnamefont
  {{Li}}}\ and\ \bibinfo {author} {\bibfnamefont {Y.-G.}\ \bibnamefont
  {{Miao}}},\ }\href {\doibase 10.1140/epjc/s10052-022-10458-y} {\bibfield
  {journal} {\bibinfo  {journal} {Eur. Phys. J. C}\ }\textbf {\bibinfo {volume}
  {82}},\ \bibinfo {eid} {503} (\bibinfo {year} {2022})},\ \Eprint
  {http://arxiv.org/abs/2110.14201} {arXiv:2110.14201 [hep-th]} \BibitemShut
  {NoStop}%
\bibitem [{\citenamefont {{Bargue{\~n}o}}(2020)}]{Bargue2020}%
  \BibitemOpen
  \bibfield  {author} {\bibinfo {author} {\bibfnamefont {P.}~\bibnamefont
  {{Bargue{\~n}o}}},\ }\href {\doibase 10.1103/PhysRevD.102.104028} {\bibfield
  {journal} {\bibinfo  {journal} {\prd}\ }\textbf {\bibinfo {volume} {102}},\
  \bibinfo {eid} {104028} (\bibinfo {year} {2020})},\ \Eprint
  {http://arxiv.org/abs/2008.02680} {arXiv:2008.02680 [gr-qc]} \BibitemShut
  {NoStop}%
\bibitem [{\citenamefont {{Gao}}\ and\ \citenamefont {{Deng}}(2020)}]{GD2020}%
  \BibitemOpen
  \bibfield  {author} {\bibinfo {author} {\bibfnamefont {B.}~\bibnamefont
  {{Gao}}}\ and\ \bibinfo {author} {\bibfnamefont {X.-M.}\ \bibnamefont
  {{Deng}}},\ }\href {\doibase 10.1016/j.aop.2020.168194} {\bibfield  {journal}
  {\bibinfo  {journal} {Ann. Phys.}\ }\textbf {\bibinfo {volume} {418}},\
  \bibinfo {eid} {168194} (\bibinfo {year} {2020})}\BibitemShut {NoStop}%
\bibitem [{\citenamefont {{Zhou}}\ and\ \citenamefont
  {{Xie}}(2020{\natexlab{a}})}]{ZTX2020}%
  \BibitemOpen
  \bibfield  {author} {\bibinfo {author} {\bibfnamefont {T.-Y.}\ \bibnamefont
  {{Zhou}}}\ and\ \bibinfo {author} {\bibfnamefont {Y.}~\bibnamefont {{Xie}}},\
  }\href {\doibase 10.1140/epjc/s10052-020-08661-w} {\bibfield  {journal}
  {\bibinfo  {journal} {Eur. Phys. J. C}\ }\textbf {\bibinfo {volume} {80}},\
  \bibinfo {eid} {1070} (\bibinfo {year} {2020}{\natexlab{a}})}\BibitemShut
  {NoStop}%
\bibitem [{\citenamefont {{Gao}}\ and\ \citenamefont {{Deng}}(2021)}]{GD2021}%
  \BibitemOpen
  \bibfield  {author} {\bibinfo {author} {\bibfnamefont {B.}~\bibnamefont
  {{Gao}}}\ and\ \bibinfo {author} {\bibfnamefont {X.-M.}\ \bibnamefont
  {{Deng}}},\ }\href {\doibase 10.1142/S0217732321502370} {\bibfield  {journal}
  {\bibinfo  {journal} {Mod. Phys. Lett. A}\ }\textbf {\bibinfo {volume}
  {36}},\ \bibinfo {eid} {2150237} (\bibinfo {year} {2021})}\BibitemShut
  {NoStop}%
\bibitem [{\citenamefont {{Rodrigues}}\ and\ \citenamefont
  {{Silva}}(2023)}]{RS2023}%
  \BibitemOpen
  \bibfield  {author} {\bibinfo {author} {\bibfnamefont {M.~E.}\ \bibnamefont
  {{Rodrigues}}}\ and\ \bibinfo {author} {\bibfnamefont {M.~V. d.~S.}\
  \bibnamefont {{Silva}}},\ }\href {\doibase 10.1103/PhysRevD.107.044064}
  {\bibfield  {journal} {\bibinfo  {journal} {\prd}\ }\textbf {\bibinfo
  {volume} {107}},\ \bibinfo {eid} {044064} (\bibinfo {year} {2023})},\ \Eprint
  {http://arxiv.org/abs/2302.10772} {arXiv:2302.10772 [gr-qc]} \BibitemShut
  {NoStop}%
\bibitem [{\citenamefont {{Simpson}}\ and\ \citenamefont
  {{Visser}}(2019)}]{SV2019}%
  \BibitemOpen
  \bibfield  {author} {\bibinfo {author} {\bibfnamefont {A.}~\bibnamefont
  {{Simpson}}}\ and\ \bibinfo {author} {\bibfnamefont {M.}~\bibnamefont
  {{Visser}}},\ }\href {\doibase 10.1088/1475-7516/2019/02/042} {\bibfield
  {journal} {\bibinfo  {journal} {J. Cosmol. Astropart. Phys.}\ }\textbf
  {\bibinfo {volume} {2019}},\ \bibinfo {eid} {042} (\bibinfo {year} {2019})},\
  \Eprint {http://arxiv.org/abs/1812.07114} {arXiv:1812.07114 [gr-qc]}
  \BibitemShut {NoStop}%
\bibitem [{\citenamefont {{Nascimento}}\ \emph {et~al.}(2020)\citenamefont
  {{Nascimento}}, \citenamefont {{Petrov}}, \citenamefont {{Porf{\'\i}rio}},\
  and\ \citenamefont {{Soares}}}]{NPPS2020}%
  \BibitemOpen
  \bibfield  {author} {\bibinfo {author} {\bibfnamefont {J.~R.}\ \bibnamefont
  {{Nascimento}}}, \bibinfo {author} {\bibfnamefont {A.~Y.}\ \bibnamefont
  {{Petrov}}}, \bibinfo {author} {\bibfnamefont {P.~J.}\ \bibnamefont
  {{Porf{\'\i}rio}}}, \ and\ \bibinfo {author} {\bibfnamefont {A.~R.}\
  \bibnamefont {{Soares}}},\ }\href {\doibase 10.1103/PhysRevD.102.044021}
  {\bibfield  {journal} {\bibinfo  {journal} {\prd}\ }\textbf {\bibinfo
  {volume} {102}},\ \bibinfo {eid} {044021} (\bibinfo {year}
  {2020})}\BibitemShut {NoStop}%
\bibitem [{\citenamefont {{{\"O}vg{\"u}n}}(2020)}]{Ovgun2020}%
  \BibitemOpen
  \bibfield  {author} {\bibinfo {author} {\bibfnamefont {A.}~\bibnamefont
  {{{\"O}vg{\"u}n}}},\ }\href {\doibase 10.3906/fiz-2008-11} {\bibfield
  {journal} {\bibinfo  {journal} {Turk. J. Phys.}\ }\textbf {\bibinfo {volume}
  {44}},\ \bibinfo {eid} {465} (\bibinfo {year} {2020})},\ \Eprint
  {http://arxiv.org/abs/2011.04423} {arXiv:2011.04423 [gr-qc]} \BibitemShut
  {NoStop}%
\bibitem [{\citenamefont {{Tsukamoto}}(2021{\natexlab{a}})}]{Tsuka2021a}%
  \BibitemOpen
  \bibfield  {author} {\bibinfo {author} {\bibfnamefont {N.}~\bibnamefont
  {{Tsukamoto}}},\ }\href {\doibase 10.1103/PhysRevD.103.024033} {\bibfield
  {journal} {\bibinfo  {journal} {\prd}\ }\textbf {\bibinfo {volume} {103}},\
  \bibinfo {eid} {024033} (\bibinfo {year} {2021}{\natexlab{a}})},\ \Eprint
  {http://arxiv.org/abs/2011.03932} {arXiv:2011.03932 [gr-qc]} \BibitemShut
  {NoStop}%
\bibitem [{\citenamefont {{Tsukamoto}}(2021{\natexlab{b}})}]{Tsuka2021b}%
  \BibitemOpen
  \bibfield  {author} {\bibinfo {author} {\bibfnamefont {N.}~\bibnamefont
  {{Tsukamoto}}},\ }\href {\doibase 10.1103/PhysRevD.104.064022} {\bibfield
  {journal} {\bibinfo  {journal} {\prd}\ }\textbf {\bibinfo {volume} {104}},\
  \bibinfo {eid} {064022} (\bibinfo {year} {2021}{\natexlab{b}})},\ \Eprint
  {http://arxiv.org/abs/2105.14336} {arXiv:2105.14336 [gr-qc]} \BibitemShut
  {NoStop}%
\bibitem [{\citenamefont {{Islam}}\ \emph {et~al.}(2021)\citenamefont
  {{Islam}}, \citenamefont {{Kumar}},\ and\ \citenamefont {{Ghosh}}}]{IKG2021}%
  \BibitemOpen
  \bibfield  {author} {\bibinfo {author} {\bibfnamefont {S.~U.}\ \bibnamefont
  {{Islam}}}, \bibinfo {author} {\bibfnamefont {J.}~\bibnamefont {{Kumar}}}, \
  and\ \bibinfo {author} {\bibfnamefont {S.~G.}\ \bibnamefont {{Ghosh}}},\
  }\href {\doibase 10.1088/1475-7516/2021/10/013} {\bibfield  {journal}
  {\bibinfo  {journal} {J. Cosmol. Astropart. Phys.}\ }\textbf {\bibinfo
  {volume} {2021}},\ \bibinfo {eid} {013} (\bibinfo {year} {2021})},\ \Eprint
  {http://arxiv.org/abs/2104.00696} {arXiv:2104.00696 [gr-qc]} \BibitemShut
  {NoStop}%
\bibitem [{\citenamefont {{Zhang}}\ and\ \citenamefont
  {{Xie}}(2022{\natexlab{a}})}]{ZX2022}%
  \BibitemOpen
  \bibfield  {author} {\bibinfo {author} {\bibfnamefont {J.}~\bibnamefont
  {{Zhang}}}\ and\ \bibinfo {author} {\bibfnamefont {Y.}~\bibnamefont
  {{Xie}}},\ }\href {\doibase 10.1140/epjc/s10052-022-10441-7} {\bibfield
  {journal} {\bibinfo  {journal} {Eur. Phys. J. C}\ }\textbf {\bibinfo {volume}
  {82}},\ \bibinfo {eid} {471} (\bibinfo {year}
  {2022}{\natexlab{a}})}\BibitemShut {NoStop}%
\bibitem [{\citenamefont {{Tsukamoto}}(2022)}]{Tsuka2022}%
  \BibitemOpen
  \bibfield  {author} {\bibinfo {author} {\bibfnamefont {N.}~\bibnamefont
  {{Tsukamoto}}},\ }\href {\doibase 10.1103/PhysRevD.105.084036} {\bibfield
  {journal} {\bibinfo  {journal} {\prd}\ }\textbf {\bibinfo {volume} {105}},\
  \bibinfo {eid} {084036} (\bibinfo {year} {2022})},\ \Eprint
  {http://arxiv.org/abs/2202.09641} {arXiv:2202.09641 [gr-qc]} \BibitemShut
  {NoStop}%
\bibitem [{\citenamefont {{Guo}}\ \emph {et~al.}(2022)\citenamefont {{Guo}},
  \citenamefont {{Lan}},\ and\ \citenamefont {{Miao}}}]{GLM2022}%
  \BibitemOpen
  \bibfield  {author} {\bibinfo {author} {\bibfnamefont {Y.}~\bibnamefont
  {{Guo}}}, \bibinfo {author} {\bibfnamefont {C.}~\bibnamefont {{Lan}}}, \ and\
  \bibinfo {author} {\bibfnamefont {Y.-G.}\ \bibnamefont {{Miao}}},\ }\href
  {\doibase 10.1103/PhysRevD.106.124052} {\bibfield  {journal} {\bibinfo
  {journal} {\prd}\ }\textbf {\bibinfo {volume} {106}},\ \bibinfo {eid}
  {124052} (\bibinfo {year} {2022})}\BibitemShut {NoStop}%
\bibitem [{\citenamefont {{Chowdhuri}}\ \emph {et~al.}(2023)\citenamefont
  {{Chowdhuri}}, \citenamefont {{Ghosh}},\ and\ \citenamefont
  {{Bhattacharyya}}}]{CGB2023}%
  \BibitemOpen
  \bibfield  {author} {\bibinfo {author} {\bibfnamefont {A.}~\bibnamefont
  {{Chowdhuri}}}, \bibinfo {author} {\bibfnamefont {S.}~\bibnamefont
  {{Ghosh}}}, \ and\ \bibinfo {author} {\bibfnamefont {A.}~\bibnamefont
  {{Bhattacharyya}}},\ }\href {\doibase 10.3389/fphy.2023.1113909} {\bibfield
  {journal} {\bibinfo  {journal} {Front. Phys.}\ }\textbf {\bibinfo {volume}
  {11}},\ \bibinfo {eid} {1113909} (\bibinfo {year} {2023})},\ \Eprint
  {http://arxiv.org/abs/2303.02069} {arXiv:2303.02069 [gr-qc]} \BibitemShut
  {NoStop}%
\bibitem [{\citenamefont {{Vagnozzi}}\ \emph {et~al.}(2023)\citenamefont
  {{Vagnozzi}} \emph {et~al.}}]{Vagnozzi2023}%
  \BibitemOpen
  \bibfield  {author} {\bibinfo {author} {\bibfnamefont {S.}~\bibnamefont
  {{Vagnozzi}}} \emph {et~al.},\ }\href {\doibase 10.1088/1361-6382/acd97b}
  {\bibfield  {journal} {\bibinfo  {journal} {Classical and Quantum Gravity}\
  }\textbf {\bibinfo {volume} {40}},\ \bibinfo {eid} {165007} (\bibinfo {year}
  {2023})},\ \Eprint {http://arxiv.org/abs/2205.07787} {arXiv:2205.07787
  [gr-qc]} \BibitemShut {NoStop}%
\bibitem [{\citenamefont {{Gao}}\ and\ \citenamefont {{Liu}}(2023)}]{GL2023}%
  \BibitemOpen
  \bibfield  {author} {\bibinfo {author} {\bibfnamefont {K.}~\bibnamefont
  {{Gao}}}\ and\ \bibinfo {author} {\bibfnamefont {L.-H.}\ \bibnamefont
  {{Liu}}},\ }\href {\doibase 10.48550/arXiv.2307.16627} {\bibfield  {journal}
  {\bibinfo  {journal} {arXiv e-prints}\ ,\ \bibinfo {eid} {arXiv:2307.16627}}
  (\bibinfo {year} {2023})},\ \Eprint {http://arxiv.org/abs/2307.16627}
  {arXiv:2307.16627 [gr-qc]} \BibitemShut {NoStop}%
\bibitem [{\citenamefont {{Shaikh}}(2023)}]{Shaikh2023}%
  \BibitemOpen
  \bibfield  {author} {\bibinfo {author} {\bibfnamefont {R.}~\bibnamefont
  {{Shaikh}}},\ }\href {\doibase 10.1093/mnras/stad1383} {\bibfield  {journal}
  {\bibinfo  {journal} {Mon. Not. R. Astron. Soc.}\ }\textbf {\bibinfo {volume}
  {523}},\ \bibinfo {pages} {375} (\bibinfo {year} {2023})},\ \Eprint
  {http://arxiv.org/abs/2208.01995} {arXiv:2208.01995 [gr-qc]} \BibitemShut
  {NoStop}%
\bibitem [{\citenamefont {{Javed}}\ \emph {et~al.}(2023)\citenamefont
  {{Javed}}, \citenamefont {{Atique}}, \citenamefont {{Pantig}},\ and\
  \citenamefont {{{\"O}vg{\"u}n}}}]{JAPO2023}%
  \BibitemOpen
  \bibfield  {author} {\bibinfo {author} {\bibfnamefont {W.}~\bibnamefont
  {{Javed}}}, \bibinfo {author} {\bibfnamefont {M.}~\bibnamefont {{Atique}}},
  \bibinfo {author} {\bibfnamefont {R.~C.}\ \bibnamefont {{Pantig}}}, \ and\
  \bibinfo {author} {\bibfnamefont {A.}~\bibnamefont {{{\"O}vg{\"u}n}}},\
  }\href {\doibase 10.3390/sym15010148} {\bibfield  {journal} {\bibinfo
  {journal} {Symmetry}\ }\textbf {\bibinfo {volume} {15}},\ \bibinfo {pages}
  {148} (\bibinfo {year} {2023})}\BibitemShut {NoStop}%
\bibitem [{\citenamefont {{Sarikulov}}\ \emph {et~al.}(2023)\citenamefont
  {{Sarikulov}}, \citenamefont {{Atamurotov}}, \citenamefont {{Abdujabbarov}},\
  and\ \citenamefont {{Khamidov}}}]{SAAK2023}%
  \BibitemOpen
  \bibfield  {author} {\bibinfo {author} {\bibfnamefont {F.}~\bibnamefont
  {{Sarikulov}}}, \bibinfo {author} {\bibfnamefont {F.}~\bibnamefont
  {{Atamurotov}}}, \bibinfo {author} {\bibfnamefont {A.}~\bibnamefont
  {{Abdujabbarov}}}, \ and\ \bibinfo {author} {\bibfnamefont {V.}~\bibnamefont
  {{Khamidov}}},\ }\href {\doibase 10.1088/1674-1137/acedf2} {\bibfield
  {journal} {\bibinfo  {journal} {Chinese Phys. C}\ }\textbf {\bibinfo {volume}
  {47}},\ \bibinfo {eid} {115101} (\bibinfo {year} {2023})}\BibitemShut
  {NoStop}%
\bibitem [{\citenamefont {{Pereira}}\ \emph {et~al.}(2025)\citenamefont
  {{Pereira}}, \citenamefont {{Soares}}, \citenamefont {{Silva}}, \citenamefont
  {{Vit{\'o}ria}},\ and\ \citenamefont {{Belich}}}]{PSSV2025}%
  \BibitemOpen
  \bibfield  {author} {\bibinfo {author} {\bibfnamefont {C.~F.~S.}\
  \bibnamefont {{Pereira}}}, \bibinfo {author} {\bibfnamefont {A.~R.}\
  \bibnamefont {{Soares}}}, \bibinfo {author} {\bibfnamefont {M.~V. d.~S.}\
  \bibnamefont {{Silva}}}, \bibinfo {author} {\bibfnamefont {R.~L.~L.}\
  \bibnamefont {{Vit{\'o}ria}}}, \ and\ \bibinfo {author} {\bibfnamefont
  {H.}~\bibnamefont {{Belich}}},\ }\href {\doibase 10.1103/fqvd-8nl7}
  {\bibfield  {journal} {\bibinfo  {journal} {\prd}\ }\textbf {\bibinfo
  {volume} {112}},\ \bibinfo {eid} {064012} (\bibinfo {year} {2025})},\ \Eprint
  {http://arxiv.org/abs/2505.12577} {arXiv:2505.12577 [gr-qc]} \BibitemShut
  {NoStop}%
\bibitem [{\citenamefont {{Bozza}}\ \emph {et~al.}(2001)\citenamefont
  {{Bozza}}, \citenamefont {{Capozziello}}, \citenamefont {{Iovane}},\ and\
  \citenamefont {{Scarpetta}}}]{BCIS2001}%
  \BibitemOpen
  \bibfield  {author} {\bibinfo {author} {\bibfnamefont {V.}~\bibnamefont
  {{Bozza}}}, \bibinfo {author} {\bibfnamefont {S.}~\bibnamefont
  {{Capozziello}}}, \bibinfo {author} {\bibfnamefont {G.}~\bibnamefont
  {{Iovane}}}, \ and\ \bibinfo {author} {\bibfnamefont {G.}~\bibnamefont
  {{Scarpetta}}},\ }\href {\doibase 10.1023/A:1012292927358} {\bibfield
  {journal} {\bibinfo  {journal} {Gen. Rel. Grav.}\ }\textbf {\bibinfo {volume}
  {33}},\ \bibinfo {pages} {1535} (\bibinfo {year} {2001})},\ \Eprint
  {http://arxiv.org/abs/gr-qc/0102068} {arXiv:gr-qc/0102068 [gr-qc]}
  \BibitemShut {NoStop}%
\bibitem [{\citenamefont {{Bozza}}(2002)}]{Bozza2002}%
  \BibitemOpen
  \bibfield  {author} {\bibinfo {author} {\bibfnamefont {V.}~\bibnamefont
  {{Bozza}}},\ }\href {\doibase 10.1103/PhysRevD.66.103001} {\bibfield
  {journal} {\bibinfo  {journal} {\prd}\ }\textbf {\bibinfo {volume} {66}},\
  \bibinfo {eid} {103001} (\bibinfo {year} {2002})},\ \Eprint
  {http://arxiv.org/abs/gr-qc/0208075} {arXiv:gr-qc/0208075 [gr-qc]}
  \BibitemShut {NoStop}%
\bibitem [{\citenamefont {{Zhou}}\ and\ \citenamefont
  {{Xie}}(2020{\natexlab{b}})}]{ZX2020b}%
  \BibitemOpen
  \bibfield  {author} {\bibinfo {author} {\bibfnamefont {T.-Y.}\ \bibnamefont
  {{Zhou}}}\ and\ \bibinfo {author} {\bibfnamefont {Y.}~\bibnamefont {{Xie}}},\
  }\href {\doibase 10.1140/epjc/s10052-020-08661-w} {\bibfield  {journal}
  {\bibinfo  {journal} {Eur. Phys. J. C}\ }\textbf {\bibinfo {volume} {80}},\
  \bibinfo {eid} {1070} (\bibinfo {year} {2020}{\natexlab{b}})}\BibitemShut
  {NoStop}%
\bibitem [{\citenamefont {{Zhang}}\ and\ \citenamefont
  {{Xie}}(2022{\natexlab{b}})}]{ZX2022b}%
  \BibitemOpen
  \bibfield  {author} {\bibinfo {author} {\bibfnamefont {J.}~\bibnamefont
  {{Zhang}}}\ and\ \bibinfo {author} {\bibfnamefont {Y.}~\bibnamefont
  {{Xie}}},\ }\href {\doibase 10.1140/epjc/s10052-022-10846-4} {\bibfield
  {journal} {\bibinfo  {journal} {Eur. Phys. J. C}\ }\textbf {\bibinfo {volume}
  {82}},\ \bibinfo {eid} {854} (\bibinfo {year}
  {2022}{\natexlab{b}})}\BibitemShut {NoStop}%
\bibitem [{\citenamefont {{Vrba}}\ \emph {et~al.}(2023)\citenamefont {{Vrba}},
  \citenamefont {{Rayimbaev}}, \citenamefont {{Stuchlik}},\ and\ \citenamefont
  {{Ahmedov}}}]{VRSA2023}%
  \BibitemOpen
  \bibfield  {author} {\bibinfo {author} {\bibfnamefont {J.}~\bibnamefont
  {{Vrba}}}, \bibinfo {author} {\bibfnamefont {J.}~\bibnamefont {{Rayimbaev}}},
  \bibinfo {author} {\bibfnamefont {Z.}~\bibnamefont {{Stuchlik}}}, \ and\
  \bibinfo {author} {\bibfnamefont {B.}~\bibnamefont {{Ahmedov}}},\ }\href
  {\doibase 10.1140/epjc/s10052-023-12023-7} {\bibfield  {journal} {\bibinfo
  {journal} {Eur. Phys. J. C}\ }\textbf {\bibinfo {volume} {83}},\ \bibinfo
  {eid} {854} (\bibinfo {year} {2023})}\BibitemShut {NoStop}%
\bibitem [{\citenamefont {{Murodov}}\ \emph {et~al.}(2024)\citenamefont
  {{Murodov}}, \citenamefont {{Badalov}}, \citenamefont {{Rayimbaev}},
  \citenamefont {{Ahmedov}},\ and\ \citenamefont {{Stuchlik}}}]{MBRAS2024}%
  \BibitemOpen
  \bibfield  {author} {\bibinfo {author} {\bibfnamefont {S.}~\bibnamefont
  {{Murodov}}}, \bibinfo {author} {\bibfnamefont {K.}~\bibnamefont
  {{Badalov}}}, \bibinfo {author} {\bibfnamefont {J.}~\bibnamefont
  {{Rayimbaev}}}, \bibinfo {author} {\bibfnamefont {B.}~\bibnamefont
  {{Ahmedov}}}, \ and\ \bibinfo {author} {\bibfnamefont {Z.}~\bibnamefont
  {{Stuchlik}}},\ }\href {\doibase 10.3390/sym16010109} {\bibfield  {journal}
  {\bibinfo  {journal} {Symmetry}\ }\textbf {\bibinfo {volume} {16}},\ \bibinfo
  {eid} {109} (\bibinfo {year} {2024})}\BibitemShut {NoStop}%
\bibitem [{\citenamefont {{He}}\ \emph {et~al.}(2024)\citenamefont {{He}},
  \citenamefont {{Xie}}, \citenamefont {{Jiang}},\ and\ \citenamefont
  {{Lin}}}]{HXJL2024}%
  \BibitemOpen
  \bibfield  {author} {\bibinfo {author} {\bibfnamefont {G.}~\bibnamefont
  {{He}}}, \bibinfo {author} {\bibfnamefont {Y.}~\bibnamefont {{Xie}}},
  \bibinfo {author} {\bibfnamefont {C.}~\bibnamefont {{Jiang}}}, \ and\
  \bibinfo {author} {\bibfnamefont {W.}~\bibnamefont {{Lin}}},\ }\href
  {\doibase 10.1103/PhysRevD.110.064008} {\bibfield  {journal} {\bibinfo
  {journal} {\prd}\ }\textbf {\bibinfo {volume} {110}},\ \bibinfo {eid}
  {064008} (\bibinfo {year} {2024})},\ \Eprint
  {http://arxiv.org/abs/2402.01548} {arXiv:2402.01548 [gr-qc]} \BibitemShut
  {NoStop}%
\bibitem [{\citenamefont {{Nishonov}}\ \emph {et~al.}(2024)\citenamefont
  {{Nishonov}}, \citenamefont {{Zahid}}, \citenamefont {{Khan}}, \citenamefont
  {{Rayimbaev}},\ and\ \citenamefont {{Abdujabbarov}}}]{NZKR2024}%
  \BibitemOpen
  \bibfield  {author} {\bibinfo {author} {\bibfnamefont {I.}~\bibnamefont
  {{Nishonov}}}, \bibinfo {author} {\bibfnamefont {M.}~\bibnamefont {{Zahid}}},
  \bibinfo {author} {\bibfnamefont {S.~U.}\ \bibnamefont {{Khan}}}, \bibinfo
  {author} {\bibfnamefont {J.}~\bibnamefont {{Rayimbaev}}}, \ and\ \bibinfo
  {author} {\bibfnamefont {A.}~\bibnamefont {{Abdujabbarov}}},\ }\href
  {\doibase 10.1140/epjc/s10052-024-13204-8} {\bibfield  {journal} {\bibinfo
  {journal} {Eur. Phys. J. C}\ }\textbf {\bibinfo {volume} {84}},\ \bibinfo
  {eid} {829} (\bibinfo {year} {2024})}\BibitemShut {NoStop}%
\bibitem [{\citenamefont {{de S. Silva}}\ and\ \citenamefont
  {{Rodrigues}}(2024)}]{DR2024}%
  \BibitemOpen
  \bibfield  {author} {\bibinfo {author} {\bibfnamefont {M.~V.}\ \bibnamefont
  {{de S. Silva}}}\ and\ \bibinfo {author} {\bibfnamefont {M.~E.}\ \bibnamefont
  {{Rodrigues}}},\ }\href {\doibase 10.1007/s10773-024-05644-5} {\bibfield
  {journal} {\bibinfo  {journal} {Int. J. Theor. Phys.}\ }\textbf {\bibinfo
  {volume} {63}},\ \bibinfo {eid} {101} (\bibinfo {year} {2024})},\ \Eprint
  {http://arxiv.org/abs/2404.15792} {arXiv:2404.15792 [gr-qc]} \BibitemShut
  {NoStop}%
\bibitem [{\citenamefont {{Bragado}}\ and\ \citenamefont
  {{Olmo}}(2025)}]{BO2025}%
  \BibitemOpen
  \bibfield  {author} {\bibinfo {author} {\bibfnamefont {A.}~\bibnamefont
  {{Bragado}}}\ and\ \bibinfo {author} {\bibfnamefont {G.~J.}\ \bibnamefont
  {{Olmo}}},\ }\href {\doibase 10.1088/1361-6382/adffde} {\bibfield  {journal}
  {\bibinfo  {journal} {Classical and Quantum Gravity}\ }\textbf {\bibinfo
  {volume} {42}},\ \bibinfo {eid} {185004} (\bibinfo {year} {2025})},\ \Eprint
  {http://arxiv.org/abs/2506.15600} {arXiv:2506.15600 [gr-qc]} \BibitemShut
  {NoStop}%
\bibitem [{\citenamefont {{Jumaniyozov}}\ \emph {et~al.}(2025)\citenamefont
  {{Jumaniyozov}}, \citenamefont {{Rayimbaev}},\ and\ \citenamefont
  {{Turaev}}}]{JRT2025}%
  \BibitemOpen
  \bibfield  {author} {\bibinfo {author} {\bibfnamefont {S.}~\bibnamefont
  {{Jumaniyozov}}}, \bibinfo {author} {\bibfnamefont {J.}~\bibnamefont
  {{Rayimbaev}}}, \ and\ \bibinfo {author} {\bibfnamefont {Y.}~\bibnamefont
  {{Turaev}}},\ }\href {\doibase 10.1140/epjc/s10052-025-14834-2} {\bibfield
  {journal} {\bibinfo  {journal} {Eur. Phys. J. C}\ }\textbf {\bibinfo {volume}
  {85}},\ \bibinfo {eid} {1247} (\bibinfo {year} {2025})}\BibitemShut {NoStop}%
\bibitem [{\citenamefont {{Wucknitz}}\ and\ \citenamefont
  {{Sperhake}}(2004)}]{WS2004}%
  \BibitemOpen
  \bibfield  {author} {\bibinfo {author} {\bibfnamefont {O.}~\bibnamefont
  {{Wucknitz}}}\ and\ \bibinfo {author} {\bibfnamefont {U.}~\bibnamefont
  {{Sperhake}}},\ }\href {\doibase 10.1103/PhysRevD.69.063001} {\bibfield
  {journal} {\bibinfo  {journal} {\prd}\ }\textbf {\bibinfo {volume} {69}},\
  \bibinfo {eid} {063001} (\bibinfo {year} {2004})},\ \Eprint
  {http://arxiv.org/abs/astro-ph/0401362} {arXiv:astro-ph/0401362 [astro-ph]}
  \BibitemShut {NoStop}%
\bibitem [{\citenamefont {{He}}\ and\ \citenamefont {{Lin}}(2014)}]{HL2014}%
  \BibitemOpen
  \bibfield  {author} {\bibinfo {author} {\bibfnamefont {G.}~\bibnamefont
  {{He}}}\ and\ \bibinfo {author} {\bibfnamefont {W.}~\bibnamefont {{Lin}}},\
  }\href {\doibase 10.1142/S021827181450031X} {\bibfield  {journal} {\bibinfo
  {journal} {Int. J. Mod. Phys. D}\ }\textbf {\bibinfo {volume} {23}},\
  \bibinfo {eid} {1450031} (\bibinfo {year} {2014})},\ \Eprint
  {http://arxiv.org/abs/2007.12317} {arXiv:2007.12317 [gr-qc]} \BibitemShut
  {NoStop}%
\bibitem [{\citenamefont {{Liu}}\ \emph
  {et~al.}(2016{\natexlab{a}})\citenamefont {{Liu}}, \citenamefont {{Yang}},\
  and\ \citenamefont {{Jia}}}]{LYJ2016}%
  \BibitemOpen
  \bibfield  {author} {\bibinfo {author} {\bibfnamefont {X.}~\bibnamefont
  {{Liu}}}, \bibinfo {author} {\bibfnamefont {N.}~\bibnamefont {{Yang}}}, \
  and\ \bibinfo {author} {\bibfnamefont {J.}~\bibnamefont {{Jia}}},\ }\href
  {\doibase 10.1088/0264-9381/33/17/175014} {\bibfield  {journal} {\bibinfo
  {journal} {Classical and Quantum Gravity}\ }\textbf {\bibinfo {volume}
  {33}},\ \bibinfo {eid} {175014} (\bibinfo {year} {2016}{\natexlab{a}})},\
  \Eprint {http://arxiv.org/abs/1512.04037} {arXiv:1512.04037 [gr-qc]}
  \BibitemShut {NoStop}%
\bibitem [{\citenamefont {{M{\'e}sz{\'a}ros}}\ \emph
  {et~al.}(2019)\citenamefont {{M{\'e}sz{\'a}ros}}, \citenamefont {{Fox}},
  \citenamefont {{Hanna}},\ and\ \citenamefont {{Murase}}}]{MFHM2019}%
  \BibitemOpen
  \bibfield  {author} {\bibinfo {author} {\bibfnamefont {P.}~\bibnamefont
  {{M{\'e}sz{\'a}ros}}}, \bibinfo {author} {\bibfnamefont {D.~B.}\ \bibnamefont
  {{Fox}}}, \bibinfo {author} {\bibfnamefont {C.}~\bibnamefont {{Hanna}}}, \
  and\ \bibinfo {author} {\bibfnamefont {K.}~\bibnamefont {{Murase}}},\ }\href
  {\doibase 10.1038/s42254-019-0101-z} {\bibfield  {journal} {\bibinfo
  {journal} {Nat. Rev. Phys.}\ }\textbf {\bibinfo {volume} {1}},\ \bibinfo
  {pages} {585} (\bibinfo {year} {2019})},\ \Eprint
  {http://arxiv.org/abs/1906.10212} {arXiv:1906.10212 [astro-ph.HE]}
  \BibitemShut {NoStop}%
\bibitem [{\citenamefont {{Huerta}}\ \emph {et~al.}(2019)\citenamefont
  {{Huerta}} \emph {et~al.}}]{Huerta2019}%
  \BibitemOpen
  \bibfield  {author} {\bibinfo {author} {\bibfnamefont {E.~A.}\ \bibnamefont
  {{Huerta}}} \emph {et~al.},\ }\href {\doibase 10.1038/s42254-019-0097-4}
  {\bibfield  {journal} {\bibinfo  {journal} {Nat. Rev. Phys.}\ }\textbf
  {\bibinfo {volume} {1}},\ \bibinfo {pages} {600} (\bibinfo {year} {2019})},\
  \Eprint {http://arxiv.org/abs/1911.11779} {arXiv:1911.11779 [gr-qc]}
  \BibitemShut {NoStop}%
\bibitem [{\citenamefont {{Anderson}}(1932)}]{Ande1932}%
  \BibitemOpen
  \bibfield  {author} {\bibinfo {author} {\bibfnamefont {C.~D.}\ \bibnamefont
  {{Anderson}}},\ }\href {\doibase 10.1103/PhysRev.41.405} {\bibfield
  {journal} {\bibinfo  {journal} {Phys. Rev.}\ }\textbf {\bibinfo {volume}
  {41}},\ \bibinfo {pages} {405} (\bibinfo {year} {1932})}\BibitemShut
  {NoStop}%
\bibitem [{\citenamefont {{Rossi}}\ and\ \citenamefont
  {{Greisen}}(1941)}]{Rossi1941}%
  \BibitemOpen
  \bibfield  {author} {\bibinfo {author} {\bibfnamefont {B.}~\bibnamefont
  {{Rossi}}}\ and\ \bibinfo {author} {\bibfnamefont {K.}~\bibnamefont
  {{Greisen}}},\ }\href {\doibase 10.1103/RevModPhys.13.240} {\bibfield
  {journal} {\bibinfo  {journal} {Rev. Mod. Phys.}\ }\textbf {\bibinfo {volume}
  {13}},\ \bibinfo {pages} {240} (\bibinfo {year} {1941})}\BibitemShut
  {NoStop}%
\bibitem [{\citenamefont {{Boehm}}\ and\ \citenamefont
  {{Vogel}}(1992)}]{BV1992}%
  \BibitemOpen
  \bibfield  {author} {\bibinfo {author} {\bibfnamefont {F.}~\bibnamefont
  {{Boehm}}}\ and\ \bibinfo {author} {\bibfnamefont {P.}~\bibnamefont
  {{Vogel}}},\ }\href@noop {} {\emph {\bibinfo {title} {{Physics of Massive
  Neutrinos}}}}\ (\bibinfo  {publisher} {Cambridge University Press},\ \bibinfo
  {address} {Cambridge},\ \bibinfo {year} {1992})\BibitemShut {NoStop}%
\bibitem [{\citenamefont {{Abbott}}\ \emph {et~al.}(2016)\citenamefont
  {{Abbott}} \emph {et~al.}}]{Abbott2016a}%
  \BibitemOpen
  \bibfield  {author} {\bibinfo {author} {\bibfnamefont {B.~P.}\ \bibnamefont
  {{Abbott}}} \emph {et~al.},\ }\href {\doibase 10.1103/PhysRevLett.116.061102}
  {\bibfield  {journal} {\bibinfo  {journal} {\prl}\ }\textbf {\bibinfo
  {volume} {116}},\ \bibinfo {eid} {061102} (\bibinfo {year} {2016})},\ \Eprint
  {http://arxiv.org/abs/1602.03837} {arXiv:1602.03837 [gr-qc]} \BibitemShut
  {NoStop}%
\bibitem [{\citenamefont {{Abbott}}\ \emph {et~al.}(2017)\citenamefont
  {{Abbott}} \emph {et~al.}}]{Abbott2017a}%
  \BibitemOpen
  \bibfield  {author} {\bibinfo {author} {\bibfnamefont {B.~P.}\ \bibnamefont
  {{Abbott}}} \emph {et~al.},\ }\href {\doibase 10.1103/PhysRevLett.119.161101}
  {\bibfield  {journal} {\bibinfo  {journal} {\prl}\ }\textbf {\bibinfo
  {volume} {119}},\ \bibinfo {eid} {161101} (\bibinfo {year} {2017})},\ \Eprint
  {http://arxiv.org/abs/1710.05832} {arXiv:1710.05832 [gr-qc]} \BibitemShut
  {NoStop}%
\bibitem [{\citenamefont {{Silverman}}(1980)}]{Silverman1980}%
  \BibitemOpen
  \bibfield  {author} {\bibinfo {author} {\bibfnamefont {M.~P.}\ \bibnamefont
  {{Silverman}}},\ }\href {\doibase 10.1119/1.12268} {\bibfield  {journal}
  {\bibinfo  {journal} {Am. J. Phys.}\ }\textbf {\bibinfo {volume} {48}},\
  \bibinfo {pages} {72} (\bibinfo {year} {1980})}\BibitemShut {NoStop}%
\bibitem [{\citenamefont {{Accioly}}\ and\ \citenamefont
  {{Ragusa}}(2002)}]{AR2002}%
  \BibitemOpen
  \bibfield  {author} {\bibinfo {author} {\bibfnamefont {A.}~\bibnamefont
  {{Accioly}}}\ and\ \bibinfo {author} {\bibfnamefont {S.}~\bibnamefont
  {{Ragusa}}},\ }\href {\doibase 10.1088/0264-9381/19/21/308} {\bibfield
  {journal} {\bibinfo  {journal} {Classical and Quantum Gravity}\ }\textbf
  {\bibinfo {volume} {19}},\ \bibinfo {pages} {5429} (\bibinfo {year}
  {2002})}\BibitemShut {NoStop}%
\bibitem [{\citenamefont {{Accioly}}\ and\ \citenamefont
  {{Ragusa}}(2003)}]{AR2003}%
  \BibitemOpen
  \bibfield  {author} {\bibinfo {author} {\bibfnamefont {A.}~\bibnamefont
  {{Accioly}}}\ and\ \bibinfo {author} {\bibfnamefont {S.}~\bibnamefont
  {{Ragusa}}},\ }\href {\doibase 10.1088/0264-9381/20/22/C01} {\bibfield
  {journal} {\bibinfo  {journal} {Classical and Quantum Gravity}\ }\textbf
  {\bibinfo {volume} {20}},\ \bibinfo {pages} {4963} (\bibinfo {year}
  {2003})}\BibitemShut {NoStop}%
\bibitem [{\citenamefont {{He}}\ and\ \citenamefont {{Lin}}(2016)}]{HL2016a}%
  \BibitemOpen
  \bibfield  {author} {\bibinfo {author} {\bibfnamefont {G.}~\bibnamefont
  {{He}}}\ and\ \bibinfo {author} {\bibfnamefont {W.}~\bibnamefont {{Lin}}},\
  }\href {\doibase 10.1088/0264-9381/33/9/095007} {\bibfield  {journal}
  {\bibinfo  {journal} {Classical and Quantum Gravity}\ }\textbf {\bibinfo
  {volume} {33}},\ \bibinfo {eid} {095007} (\bibinfo {year} {2016})},\ \Eprint
  {http://arxiv.org/abs/2007.08754} {arXiv:2007.08754 [gr-qc]} \BibitemShut
  {NoStop}%
\bibitem [{\citenamefont {{He}}\ and\ \citenamefont
  {{Lin}}(2017{\natexlab{a}})}]{HL2017b}%
  \BibitemOpen
  \bibfield  {author} {\bibinfo {author} {\bibfnamefont {G.}~\bibnamefont
  {{He}}}\ and\ \bibinfo {author} {\bibfnamefont {W.}~\bibnamefont {{Lin}}},\
  }\href {\doibase 10.1088/1361-6382/aa691d} {\bibfield  {journal} {\bibinfo
  {journal} {Classical and Quantum Gravity}\ }\textbf {\bibinfo {volume}
  {34}},\ \bibinfo {eid} {105006} (\bibinfo {year} {2017}{\natexlab{a}})},\
  \Eprint {http://arxiv.org/abs/2007.09831} {arXiv:2007.09831 [gr-qc]}
  \BibitemShut {NoStop}%
\bibitem [{\citenamefont {{Li}}\ \emph {et~al.}(2019)\citenamefont {{Li}},
  \citenamefont {{Zhou}}, \citenamefont {{Li}},\ and\ \citenamefont
  {{He}}}]{LZLH2019}%
  \BibitemOpen
  \bibfield  {author} {\bibinfo {author} {\bibfnamefont {Z.}~\bibnamefont
  {{Li}}}, \bibinfo {author} {\bibfnamefont {X.}~\bibnamefont {{Zhou}}},
  \bibinfo {author} {\bibfnamefont {W.}~\bibnamefont {{Li}}}, \ and\ \bibinfo
  {author} {\bibfnamefont {G.}~\bibnamefont {{He}}},\ }\href {\doibase
  10.1088/0253-6102/71/10/1219} {\bibfield  {journal} {\bibinfo  {journal}
  {Commun. Theor. Phys.}\ }\textbf {\bibinfo {volume} {71}},\ \bibinfo {eid}
  {1219} (\bibinfo {year} {2019})},\ \Eprint {http://arxiv.org/abs/1906.06470}
  {arXiv:1906.06470 [gr-qc]} \BibitemShut {NoStop}%
\bibitem [{\citenamefont {{He}}\ \emph {et~al.}(2020)\citenamefont {{He}},
  \citenamefont {{Zhou}}, \citenamefont {{Feng}}, \citenamefont {{Mu}},
  \citenamefont {{Wang}}, \citenamefont {{Li}}, \citenamefont {{Pan}},\ and\
  \citenamefont {{Lin}}}]{HZFMWPL2020}%
  \BibitemOpen
  \bibfield  {author} {\bibinfo {author} {\bibfnamefont {G.}~\bibnamefont
  {{He}}}, \bibinfo {author} {\bibfnamefont {X.}~\bibnamefont {{Zhou}}},
  \bibinfo {author} {\bibfnamefont {Z.}~\bibnamefont {{Feng}}}, \bibinfo
  {author} {\bibfnamefont {X.}~\bibnamefont {{Mu}}}, \bibinfo {author}
  {\bibfnamefont {H.}~\bibnamefont {{Wang}}}, \bibinfo {author} {\bibfnamefont
  {W.}~\bibnamefont {{Li}}}, \bibinfo {author} {\bibfnamefont {C.}~\bibnamefont
  {{Pan}}}, \ and\ \bibinfo {author} {\bibfnamefont {W.}~\bibnamefont
  {{Lin}}},\ }\href {\doibase 10.1140/epjc/s10052-020-8382-z} {\bibfield
  {journal} {\bibinfo  {journal} {Eur. Phys. J. C}\ }\textbf {\bibinfo {volume}
  {80}},\ \bibinfo {eid} {835} (\bibinfo {year} {2020})}\BibitemShut {NoStop}%
\bibitem [{\citenamefont {{Li}}\ \emph {et~al.}(2020)\citenamefont {{Li}},
  \citenamefont {{He}},\ and\ \citenamefont {{Zhou}}}]{LHZ2020}%
  \BibitemOpen
  \bibfield  {author} {\bibinfo {author} {\bibfnamefont {Z.}~\bibnamefont
  {{Li}}}, \bibinfo {author} {\bibfnamefont {G.}~\bibnamefont {{He}}}, \ and\
  \bibinfo {author} {\bibfnamefont {T.}~\bibnamefont {{Zhou}}},\ }\href
  {\doibase 10.1103/PhysRevD.101.044001} {\bibfield  {journal} {\bibinfo
  {journal} {\prd}\ }\textbf {\bibinfo {volume} {101}},\ \bibinfo {eid}
  {044001} (\bibinfo {year} {2020})},\ \Eprint
  {http://arxiv.org/abs/1908.01647} {arXiv:1908.01647 [gr-qc]} \BibitemShut
  {NoStop}%
\bibitem [{\citenamefont {{He}}\ and\ \citenamefont {{Lin}}(2022)}]{HL2022}%
  \BibitemOpen
  \bibfield  {author} {\bibinfo {author} {\bibfnamefont {G.}~\bibnamefont
  {{He}}}\ and\ \bibinfo {author} {\bibfnamefont {W.}~\bibnamefont {{Lin}}},\
  }\href {\doibase 10.1103/PhysRevD.105.104034} {\bibfield  {journal} {\bibinfo
   {journal} {\prd}\ }\textbf {\bibinfo {volume} {105}},\ \bibinfo {eid}
  {104034} (\bibinfo {year} {2022})},\ \Eprint
  {http://arxiv.org/abs/2112.08142} {arXiv:2112.08142 [gr-qc]} \BibitemShut
  {NoStop}%
\bibitem [{\citenamefont {{Wang}}\ \emph {et~al.}(2025)\citenamefont {{Wang}},
  \citenamefont {{Lin}}, \citenamefont {{Mustafa}},\ and\ \citenamefont
  {{He}}}]{WLMH2025}%
  \BibitemOpen
  \bibfield  {author} {\bibinfo {author} {\bibfnamefont {X.}~\bibnamefont
  {{Wang}}}, \bibinfo {author} {\bibfnamefont {W.}~\bibnamefont {{Lin}}},
  \bibinfo {author} {\bibfnamefont {G.}~\bibnamefont {{Mustafa}}}, \ and\
  \bibinfo {author} {\bibfnamefont {G.}~\bibnamefont {{He}}},\ }\href {\doibase
  10.1103/PhysRevD.111.084009} {\bibfield  {journal} {\bibinfo  {journal}
  {\prd}\ }\textbf {\bibinfo {volume} {111}},\ \bibinfo {eid} {084009}
  (\bibinfo {year} {2025})},\ \Eprint {http://arxiv.org/abs/2503.17164}
  {arXiv:2503.17164 [gr-qc]} \BibitemShut {NoStop}%
\bibitem [{\citenamefont {{Pang}}\ and\ \citenamefont {{Jia}}(2019)}]{PJ2019}%
  \BibitemOpen
  \bibfield  {author} {\bibinfo {author} {\bibfnamefont {X.}~\bibnamefont
  {{Pang}}}\ and\ \bibinfo {author} {\bibfnamefont {J.}~\bibnamefont {{Jia}}},\
  }\href {\doibase 10.1088/1361-6382/ab0512} {\bibfield  {journal} {\bibinfo
  {journal} {Classical and Quantum Gravity}\ }\textbf {\bibinfo {volume}
  {36}},\ \bibinfo {eid} {065012} (\bibinfo {year} {2019})},\ \Eprint
  {http://arxiv.org/abs/1806.04719} {arXiv:1806.04719 [gr-qc]} \BibitemShut
  {NoStop}%
\bibitem [{\citenamefont {{Jia}}\ and\ \citenamefont {{Huang}}(2021)}]{JH2021}%
  \BibitemOpen
  \bibfield  {author} {\bibinfo {author} {\bibfnamefont {J.}~\bibnamefont
  {{Jia}}}\ and\ \bibinfo {author} {\bibfnamefont {K.}~\bibnamefont
  {{Huang}}},\ }\href {\doibase 10.1140/epjc/s10052-021-09026-7} {\bibfield
  {journal} {\bibinfo  {journal} {Eur. Phys. J. C}\ }\textbf {\bibinfo {volume}
  {81}},\ \bibinfo {eid} {242} (\bibinfo {year} {2021})},\ \Eprint
  {http://arxiv.org/abs/2011.08084} {arXiv:2011.08084 [gr-qc]} \BibitemShut
  {NoStop}%
\bibitem [{\citenamefont {{Accioly}}\ and\ \citenamefont
  {{Paszko}}(2004)}]{AP2004}%
  \BibitemOpen
  \bibfield  {author} {\bibinfo {author} {\bibfnamefont {A.}~\bibnamefont
  {{Accioly}}}\ and\ \bibinfo {author} {\bibfnamefont {R.}~\bibnamefont
  {{Paszko}}},\ }\href {\doibase 10.1103/PhysRevD.69.107501} {\bibfield
  {journal} {\bibinfo  {journal} {\prd}\ }\textbf {\bibinfo {volume} {69}},\
  \bibinfo {eid} {107501} (\bibinfo {year} {2004})}\BibitemShut {NoStop}%
\bibitem [{\citenamefont {{Bhadra}}\ \emph {et~al.}(2007)\citenamefont
  {{Bhadra}}, \citenamefont {{Sarkar}},\ and\ \citenamefont
  {{Nandi}}}]{BSN2007}%
  \BibitemOpen
  \bibfield  {author} {\bibinfo {author} {\bibfnamefont {A.}~\bibnamefont
  {{Bhadra}}}, \bibinfo {author} {\bibfnamefont {K.}~\bibnamefont {{Sarkar}}},
  \ and\ \bibinfo {author} {\bibfnamefont {K.~K.}\ \bibnamefont {{Nandi}}},\
  }\href {\doibase 10.1103/PhysRevD.75.123004} {\bibfield  {journal} {\bibinfo
  {journal} {\prd}\ }\textbf {\bibinfo {volume} {75}},\ \bibinfo {eid} {123004}
  (\bibinfo {year} {2007})},\ \Eprint {http://arxiv.org/abs/gr-qc/0610089}
  {arXiv:gr-qc/0610089 [gr-qc]} \BibitemShut {NoStop}%
\bibitem [{\citenamefont {{Patla}}\ \emph
  {et~al.}(2014{\natexlab{a}})\citenamefont {{Patla}}, \citenamefont
  {{Nemiroff}}, \citenamefont {{Hoffmann}},\ and\ \citenamefont
  {{Zioutas}}}]{PNH2014}%
  \BibitemOpen
  \bibfield  {author} {\bibinfo {author} {\bibfnamefont {B.~R.}\ \bibnamefont
  {{Patla}}}, \bibinfo {author} {\bibfnamefont {R.~J.}\ \bibnamefont
  {{Nemiroff}}}, \bibinfo {author} {\bibfnamefont {D.~H.~H.}\ \bibnamefont
  {{Hoffmann}}}, \ and\ \bibinfo {author} {\bibfnamefont {K.}~\bibnamefont
  {{Zioutas}}},\ }\href {\doibase 10.1088/0004-637X/780/2/158} {\bibfield
  {journal} {\bibinfo  {journal} {\apj}\ }\textbf {\bibinfo {volume} {780}},\
  \bibinfo {eid} {158} (\bibinfo {year} {2014}{\natexlab{a}})},\ \Eprint
  {http://arxiv.org/abs/1305.2454} {arXiv:1305.2454 [astro-ph.EP]} \BibitemShut
  {NoStop}%
\bibitem [{\citenamefont {{Crisnejo}}\ and\ \citenamefont
  {{Gallo}}(2018)}]{CG2018}%
  \BibitemOpen
  \bibfield  {author} {\bibinfo {author} {\bibfnamefont {G.}~\bibnamefont
  {{Crisnejo}}}\ and\ \bibinfo {author} {\bibfnamefont {E.}~\bibnamefont
  {{Gallo}}},\ }\href {\doibase 10.1103/PhysRevD.97.124016} {\bibfield
  {journal} {\bibinfo  {journal} {\prd}\ }\textbf {\bibinfo {volume} {97}},\
  \bibinfo {eid} {124016} (\bibinfo {year} {2018})},\ \Eprint
  {http://arxiv.org/abs/1804.05473} {arXiv:1804.05473 [gr-qc]} \BibitemShut
  {NoStop}%
\bibitem [{\citenamefont {{Jusufi}}(2018)}]{J2018}%
  \BibitemOpen
  \bibfield  {author} {\bibinfo {author} {\bibfnamefont {K.}~\bibnamefont
  {{Jusufi}}},\ }\href {\doibase 10.1103/PhysRevD.98.064017} {\bibfield
  {journal} {\bibinfo  {journal} {\prd}\ }\textbf {\bibinfo {volume} {98}},\
  \bibinfo {eid} {064017} (\bibinfo {year} {2018})},\ \Eprint
  {http://arxiv.org/abs/1806.01256} {arXiv:1806.01256 [gr-qc]} \BibitemShut
  {NoStop}%
\bibitem [{\citenamefont {{Crisnejo}}\ \emph
  {et~al.}(2019{\natexlab{a}})\citenamefont {{Crisnejo}}, \citenamefont
  {{Gallo}},\ and\ \citenamefont {{Villanueva}}}]{CGV2019}%
  \BibitemOpen
  \bibfield  {author} {\bibinfo {author} {\bibfnamefont {G.}~\bibnamefont
  {{Crisnejo}}}, \bibinfo {author} {\bibfnamefont {E.}~\bibnamefont {{Gallo}}},
  \ and\ \bibinfo {author} {\bibfnamefont {J.~R.}\ \bibnamefont
  {{Villanueva}}},\ }\href {\doibase 10.1103/PhysRevD.100.044006} {\bibfield
  {journal} {\bibinfo  {journal} {\prd}\ }\textbf {\bibinfo {volume} {100}},\
  \bibinfo {eid} {044006} (\bibinfo {year} {2019}{\natexlab{a}})},\ \Eprint
  {http://arxiv.org/abs/1905.02125} {arXiv:1905.02125 [gr-qc]} \BibitemShut
  {NoStop}%
\bibitem [{\citenamefont {{Crisnejo}}\ \emph
  {et~al.}(2019{\natexlab{b}})\citenamefont {{Crisnejo}}, \citenamefont
  {{Gallo}},\ and\ \citenamefont {{Jusufi}}}]{CGJ2019}%
  \BibitemOpen
  \bibfield  {author} {\bibinfo {author} {\bibfnamefont {G.}~\bibnamefont
  {{Crisnejo}}}, \bibinfo {author} {\bibfnamefont {E.}~\bibnamefont {{Gallo}}},
  \ and\ \bibinfo {author} {\bibfnamefont {K.}~\bibnamefont {{Jusufi}}},\
  }\href {\doibase 10.1103/PhysRevD.100.104045} {\bibfield  {journal} {\bibinfo
   {journal} {\prd}\ }\textbf {\bibinfo {volume} {100}},\ \bibinfo {eid}
  {104045} (\bibinfo {year} {2019}{\natexlab{b}})},\ \Eprint
  {http://arxiv.org/abs/1910.02030} {arXiv:1910.02030 [gr-qc]} \BibitemShut
  {NoStop}%
\bibitem [{\citenamefont {{Jusufi}}\ \emph {et~al.}(2019)\citenamefont
  {{Jusufi}}, \citenamefont {{Banerjee}}, \citenamefont {{Gyulchev}},\ and\
  \citenamefont {{Amir}}}]{JBGA2019}%
  \BibitemOpen
  \bibfield  {author} {\bibinfo {author} {\bibfnamefont {K.}~\bibnamefont
  {{Jusufi}}}, \bibinfo {author} {\bibfnamefont {A.}~\bibnamefont
  {{Banerjee}}}, \bibinfo {author} {\bibfnamefont {G.}~\bibnamefont
  {{Gyulchev}}}, \ and\ \bibinfo {author} {\bibfnamefont {M.}~\bibnamefont
  {{Amir}}},\ }\href {\doibase 10.1140/epjc/s10052-019-6557-2} {\bibfield
  {journal} {\bibinfo  {journal} {Eur. Phys. J. C}\ }\textbf {\bibinfo {volume}
  {79}},\ \bibinfo {eid} {28} (\bibinfo {year} {2019})},\ \Eprint
  {http://arxiv.org/abs/1808.02751} {arXiv:1808.02751 [gr-qc]} \BibitemShut
  {NoStop}%
\bibitem [{\citenamefont {{Jia}}\ and\ \citenamefont {{Liu}}(2019)}]{JL2019}%
  \BibitemOpen
  \bibfield  {author} {\bibinfo {author} {\bibfnamefont {J.}~\bibnamefont
  {{Jia}}}\ and\ \bibinfo {author} {\bibfnamefont {H.}~\bibnamefont {{Liu}}},\
  }\href {\doibase 10.1103/PhysRevD.100.124050} {\bibfield  {journal} {\bibinfo
   {journal} {\prd}\ }\textbf {\bibinfo {volume} {100}},\ \bibinfo {eid}
  {124050} (\bibinfo {year} {2019})},\ \Eprint
  {http://arxiv.org/abs/1906.11833} {arXiv:1906.11833 [gr-qc]} \BibitemShut
  {NoStop}%
\bibitem [{\citenamefont {{Li}}\ \emph {et~al.}(2021)\citenamefont {{Li}},
  \citenamefont {{Liu}},\ and\ \citenamefont {{Jia}}}]{LLJ2021}%
  \BibitemOpen
  \bibfield  {author} {\bibinfo {author} {\bibfnamefont {Z.}~\bibnamefont
  {{Li}}}, \bibinfo {author} {\bibfnamefont {H.}~\bibnamefont {{Liu}}}, \ and\
  \bibinfo {author} {\bibfnamefont {J.}~\bibnamefont {{Jia}}},\ }\href
  {\doibase 10.1103/PhysRevD.104.084027} {\bibfield  {journal} {\bibinfo
  {journal} {\prd}\ }\textbf {\bibinfo {volume} {104}},\ \bibinfo {eid}
  {084027} (\bibinfo {year} {2021})},\ \Eprint
  {http://arxiv.org/abs/2107.11616} {arXiv:2107.11616 [gr-qc]} \BibitemShut
  {NoStop}%
\bibitem [{\citenamefont {{Huang}}\ and\ \citenamefont {{Cao}}(2023)}]{HC2023}%
  \BibitemOpen
  \bibfield  {author} {\bibinfo {author} {\bibfnamefont {Y.}~\bibnamefont
  {{Huang}}}\ and\ \bibinfo {author} {\bibfnamefont {Z.}~\bibnamefont
  {{Cao}}},\ }\href {\doibase 10.1140/epjc/s10052-023-11180-z} {\bibfield
  {journal} {\bibinfo  {journal} {Eur. Phys. J. C}\ }\textbf {\bibinfo {volume}
  {83}},\ \bibinfo {eid} {80} (\bibinfo {year} {2023})},\ \Eprint
  {http://arxiv.org/abs/2212.04254} {arXiv:2212.04254 [gr-qc]} \BibitemShut
  {NoStop}%
\bibitem [{\citenamefont {{Huang}}\ \emph {et~al.}(2024)\citenamefont
  {{Huang}}, \citenamefont {{Cao}},\ and\ \citenamefont {{Lu}}}]{HCL2024}%
  \BibitemOpen
  \bibfield  {author} {\bibinfo {author} {\bibfnamefont {Y.}~\bibnamefont
  {{Huang}}}, \bibinfo {author} {\bibfnamefont {Z.}~\bibnamefont {{Cao}}}, \
  and\ \bibinfo {author} {\bibfnamefont {Z.}~\bibnamefont {{Lu}}},\ }\href
  {\doibase 10.1088/1475-7516/2024/01/013} {\bibfield  {journal} {\bibinfo
  {journal} {J. Cosmol. Astropart. Phys.}\ }\textbf {\bibinfo {volume}
  {2024}},\ \bibinfo {eid} {013} (\bibinfo {year} {2024})},\ \Eprint
  {http://arxiv.org/abs/2306.04145} {arXiv:2306.04145 [gr-qc]} \BibitemShut
  {NoStop}%
\bibitem [{\citenamefont {{Tsupko}}(2014)}]{Tsupko2014}%
  \BibitemOpen
  \bibfield  {author} {\bibinfo {author} {\bibfnamefont {O.~Y.}\ \bibnamefont
  {{Tsupko}}},\ }\href {\doibase 10.1103/PhysRevD.89.084075} {\bibfield
  {journal} {\bibinfo  {journal} {\prd}\ }\textbf {\bibinfo {volume} {89}},\
  \bibinfo {eid} {084075} (\bibinfo {year} {2014})},\ \Eprint
  {http://arxiv.org/abs/1505.06481} {arXiv:1505.06481 [gr-qc]} \BibitemShut
  {NoStop}%
\bibitem [{\citenamefont {{Baker}}\ and\ \citenamefont
  {{Trodden}}(2017)}]{BT2017}%
  \BibitemOpen
  \bibfield  {author} {\bibinfo {author} {\bibfnamefont {T.}~\bibnamefont
  {{Baker}}}\ and\ \bibinfo {author} {\bibfnamefont {M.}~\bibnamefont
  {{Trodden}}},\ }\href {\doibase 10.1103/PhysRevD.95.063512} {\bibfield
  {journal} {\bibinfo  {journal} {\prd}\ }\textbf {\bibinfo {volume} {95}},\
  \bibinfo {eid} {063512} (\bibinfo {year} {2017})},\ \Eprint
  {http://arxiv.org/abs/1612.02004} {arXiv:1612.02004 [astro-ph.CO]}
  \BibitemShut {NoStop}%
\bibitem [{\citenamefont {{Huang}}\ \emph {et~al.}(2023)\citenamefont
  {{Huang}}, \citenamefont {{Sun}},\ and\ \citenamefont {{Cao}}}]{HSC2023}%
  \BibitemOpen
  \bibfield  {author} {\bibinfo {author} {\bibfnamefont {Y.}~\bibnamefont
  {{Huang}}}, \bibinfo {author} {\bibfnamefont {B.}~\bibnamefont {{Sun}}}, \
  and\ \bibinfo {author} {\bibfnamefont {Z.}~\bibnamefont {{Cao}}},\ }\href
  {\doibase 10.1103/PhysRevD.107.104046} {\bibfield  {journal} {\bibinfo
  {journal} {\prd}\ }\textbf {\bibinfo {volume} {107}},\ \bibinfo {eid}
  {104046} (\bibinfo {year} {2023})},\ \Eprint
  {http://arxiv.org/abs/2212.04251} {arXiv:2212.04251 [gr-qc]} \BibitemShut
  {NoStop}%
\bibitem [{\citenamefont {{Keivani}}\ \emph {et~al.}(2018)\citenamefont
  {{Keivani}} \emph {et~al.}}]{Keivani2018}%
  \BibitemOpen
  \bibfield  {author} {\bibinfo {author} {\bibfnamefont {A.}~\bibnamefont
  {{Keivani}}} \emph {et~al.},\ }\href {\doibase 10.3847/1538-4357/aad59a}
  {\bibfield  {journal} {\bibinfo  {journal} {\apj}\ }\textbf {\bibinfo
  {volume} {864}},\ \bibinfo {eid} {84} (\bibinfo {year} {2018})},\ \Eprint
  {http://arxiv.org/abs/1807.04537} {arXiv:1807.04537 [astro-ph.HE]}
  \BibitemShut {NoStop}%
\bibitem [{\citenamefont {{IceCube Collaboration}}\ \emph
  {et~al.}(2018)\citenamefont {{IceCube Collaboration}} \emph
  {et~al.}}]{IceCube2018}%
  \BibitemOpen
  \bibfield  {author} {\bibinfo {author} {\bibnamefont {{IceCube
  Collaboration}}} \emph {et~al.},\ }\href {\doibase 10.1126/science.aat1378}
  {\bibfield  {journal} {\bibinfo  {journal} {Science}\ }\textbf {\bibinfo
  {volume} {361}},\ \bibinfo {eid} {eaat1378} (\bibinfo {year} {2018})},\
  \Eprint {http://arxiv.org/abs/1807.08816} {arXiv:1807.08816 [astro-ph.HE]}
  \BibitemShut {NoStop}%
\bibitem [{\citenamefont {{Perryman}}\ \emph {et~al.}(2001)\citenamefont
  {{Perryman}} \emph {et~al.}}]{Perryman2001}%
  \BibitemOpen
  \bibfield  {author} {\bibinfo {author} {\bibfnamefont {M.~A.~C.}\
  \bibnamefont {{Perryman}}} \emph {et~al.},\ }\href {\doibase
  10.1051/0004-6361:20010085} {\bibfield  {journal} {\bibinfo  {journal}
  {Astron. Astrophys.}\ }\textbf {\bibinfo {volume} {369}},\ \bibinfo {pages}
  {339} (\bibinfo {year} {2001})},\ \Eprint
  {http://arxiv.org/abs/astro-ph/0101235} {arXiv:astro-ph/0101235 [astro-ph]}
  \BibitemShut {NoStop}%
\bibitem [{\citenamefont {{Shao}}\ and\ \citenamefont
  {{Nemati}}(2009)}]{SN2009}%
  \BibitemOpen
  \bibfield  {author} {\bibinfo {author} {\bibfnamefont {M.}~\bibnamefont
  {{Shao}}}\ and\ \bibinfo {author} {\bibfnamefont {B.}~\bibnamefont
  {{Nemati}}},\ }\href {\doibase 10.1086/596661} {\bibfield  {journal}
  {\bibinfo  {journal} {Publ. Astron. Soc. Pac.}\ }\textbf {\bibinfo {volume}
  {121}},\ \bibinfo {pages} {41} (\bibinfo {year} {2009})},\ \Eprint
  {http://arxiv.org/abs/0812.1530} {arXiv:0812.1530 [astro-ph]} \BibitemShut
  {NoStop}%
\bibitem [{\citenamefont {{Reid}}\ \emph {et~al.}(2009)\citenamefont {{Reid}}
  \emph {et~al.}}]{Reid2009}%
  \BibitemOpen
  \bibfield  {author} {\bibinfo {author} {\bibfnamefont {M.~J.}\ \bibnamefont
  {{Reid}}} \emph {et~al.},\ }\href {\doibase 10.1088/0004-637X/700/1/137}
  {\bibfield  {journal} {\bibinfo  {journal} {Astrophys. J.}\ }\textbf
  {\bibinfo {volume} {700}},\ \bibinfo {pages} {137} (\bibinfo {year}
  {2009})},\ \Eprint {http://arxiv.org/abs/0902.3913} {arXiv:0902.3913
  [astro-ph.GA]} \BibitemShut {NoStop}%
\bibitem [{\citenamefont {{Trippe}}\ \emph {et~al.}(2010)\citenamefont
  {{Trippe}}, \citenamefont {{Davies}}, \citenamefont {{Eisenhauer}},
  \citenamefont {{F{\"o}rster Schreiber}}, \citenamefont {{Fritz}},\ and\
  \citenamefont {{Genzel}}}]{Trippe2010}%
  \BibitemOpen
  \bibfield  {author} {\bibinfo {author} {\bibfnamefont {S.}~\bibnamefont
  {{Trippe}}}, \bibinfo {author} {\bibfnamefont {R.}~\bibnamefont {{Davies}}},
  \bibinfo {author} {\bibfnamefont {F.}~\bibnamefont {{Eisenhauer}}}, \bibinfo
  {author} {\bibfnamefont {N.~M.}\ \bibnamefont {{F{\"o}rster Schreiber}}},
  \bibinfo {author} {\bibfnamefont {T.~K.}\ \bibnamefont {{Fritz}}}, \ and\
  \bibinfo {author} {\bibfnamefont {R.}~\bibnamefont {{Genzel}}},\ }\href
  {\doibase 10.1111/j.1365-2966.2009.15940.x} {\bibfield  {journal} {\bibinfo
  {journal} {Mon. Not. R. Astron. Soc.}\ }\textbf {\bibinfo {volume} {402}},\
  \bibinfo {pages} {1126} (\bibinfo {year} {2010})},\ \Eprint
  {http://arxiv.org/abs/0910.5114} {arXiv:0910.5114 [astro-ph.IM]} \BibitemShut
  {NoStop}%
\bibitem [{\citenamefont {{Malbet}}\ \emph {et~al.}(2012)\citenamefont
  {{Malbet}} \emph {et~al.}}]{Malbet2012}%
  \BibitemOpen
  \bibfield  {author} {\bibinfo {author} {\bibfnamefont {F.}~\bibnamefont
  {{Malbet}}} \emph {et~al.},\ }\href {\doibase 10.1007/s10686-011-9246-1}
  {\bibfield  {journal} {\bibinfo  {journal} {Exp. Astron.}\ }\textbf {\bibinfo
  {volume} {34}},\ \bibinfo {pages} {385} (\bibinfo {year} {2012})},\ \Eprint
  {http://arxiv.org/abs/1107.3643} {arXiv:1107.3643 [astro-ph.EP]} \BibitemShut
  {NoStop}%
\bibitem [{\citenamefont {{Zhang}}\ \emph {et~al.}(2013)\citenamefont
  {{Zhang}}, \citenamefont {{Reid}}, \citenamefont {{Menten}}, \citenamefont
  {{Zheng}}, \citenamefont {{Brunthaler}}, \citenamefont {{Dame}},\ and\
  \citenamefont {{Xu}}}]{ZRMZBDX2013}%
  \BibitemOpen
  \bibfield  {author} {\bibinfo {author} {\bibfnamefont {B.}~\bibnamefont
  {{Zhang}}}, \bibinfo {author} {\bibfnamefont {M.~J.}\ \bibnamefont {{Reid}}},
  \bibinfo {author} {\bibfnamefont {K.~M.}\ \bibnamefont {{Menten}}}, \bibinfo
  {author} {\bibfnamefont {X.~W.}\ \bibnamefont {{Zheng}}}, \bibinfo {author}
  {\bibfnamefont {A.}~\bibnamefont {{Brunthaler}}}, \bibinfo {author}
  {\bibfnamefont {T.~M.}\ \bibnamefont {{Dame}}}, \ and\ \bibinfo {author}
  {\bibfnamefont {Y.}~\bibnamefont {{Xu}}},\ }\href {\doibase
  10.1088/0004-637X/775/1/79} {\bibfield  {journal} {\bibinfo  {journal}
  {\apj}\ }\textbf {\bibinfo {volume} {775}},\ \bibinfo {eid} {79} (\bibinfo
  {year} {2013})},\ \Eprint {http://arxiv.org/abs/1312.3856} {arXiv:1312.3856
  [astro-ph.GA]} \BibitemShut {NoStop}%
\bibitem [{\citenamefont {{Reid}}\ and\ \citenamefont
  {{Honma}}(2014)}]{RH2014}%
  \BibitemOpen
  \bibfield  {author} {\bibinfo {author} {\bibfnamefont {M.~J.}\ \bibnamefont
  {{Reid}}}\ and\ \bibinfo {author} {\bibfnamefont {M.}~\bibnamefont
  {{Honma}}},\ }\href {\doibase 10.1146/annurev-astro-081913-040006} {\bibfield
   {journal} {\bibinfo  {journal} {Ann. Rev. Astron. Astrophys.}\ }\textbf
  {\bibinfo {volume} {52}},\ \bibinfo {pages} {339} (\bibinfo {year} {2014})},\
  \Eprint {http://arxiv.org/abs/1312.2871} {arXiv:1312.2871 [astro-ph.IM]}
  \BibitemShut {NoStop}%
\bibitem [{\citenamefont {{Malbet}}\ \emph {et~al.}(2014)\citenamefont
  {{Malbet}}, \citenamefont {{Crouzier}}, \citenamefont {{L{\'e}ger}},
  \citenamefont {{Shao}},\ and\ \citenamefont {{Goullioud}}}]{Malbet2014}%
  \BibitemOpen
  \bibfield  {author} {\bibinfo {author} {\bibfnamefont {F.}~\bibnamefont
  {{Malbet}}}, \bibinfo {author} {\bibfnamefont {A.}~\bibnamefont
  {{Crouzier}}}, \bibinfo {author} {\bibfnamefont {A.}~\bibnamefont
  {{L{\'e}ger}}}, \bibinfo {author} {\bibfnamefont {M.}~\bibnamefont {{Shao}}},
  \ and\ \bibinfo {author} {\bibfnamefont {R.}~\bibnamefont {{Goullioud}}},\
  }in\ \href {\doibase 10.1117/12.2056628} {\emph {\bibinfo {booktitle} {Space
  Telescopes and Instrumentation 2014: Optical, Infrared, and Millimeter
  Wave}}},\ \bibinfo {series} {Society of Photo-Optical Instrumentation
  Engineers (SPIE) Conference Series}, Vol.\ \bibinfo {volume} {9143},\
  \bibinfo {editor} {edited by\ \bibinfo {editor} {\bibfnamefont
  {J.}~\bibnamefont {{Oschmann}}, \bibfnamefont {Jacobus~M.}}, \bibinfo
  {editor} {\bibfnamefont {M.}~\bibnamefont {{Clampin}}}, \bibinfo {editor}
  {\bibfnamefont {G.~G.}\ \bibnamefont {{Fazio}}}, \ and\ \bibinfo {editor}
  {\bibfnamefont {H.~A.}\ \bibnamefont {{MacEwen}}}}\ (\bibinfo {year} {2014})\
  p.\ \bibinfo {pages} {91432L}\BibitemShut {NoStop}%
\bibitem [{\citenamefont {{Prusti}}\ \emph {et~al.}(2016)\citenamefont
  {{Prusti}} \emph {et~al.}}]{Prusti2016}%
  \BibitemOpen
  \bibfield  {author} {\bibinfo {author} {\bibfnamefont {T.}~\bibnamefont
  {{Prusti}}} \emph {et~al.},\ }\href {\doibase 10.1051/0004-6361/201629272}
  {\bibfield  {journal} {\bibinfo  {journal} {Astron. Astrophys.}\ }\textbf
  {\bibinfo {volume} {595}},\ \bibinfo {eid} {A1} (\bibinfo {year} {2016})},\
  \Eprint {http://arxiv.org/abs/1609.04153} {arXiv:1609.04153 [astro-ph.IM]}
  \BibitemShut {NoStop}%
\bibitem [{\citenamefont {{Murphy}}\ \emph {et~al.}(2018)\citenamefont
  {{Murphy}} \emph {et~al.}}]{Murphy2018}%
  \BibitemOpen
  \bibfield  {author} {\bibinfo {author} {\bibfnamefont {E.~J.}\ \bibnamefont
  {{Murphy}}} \emph {et~al.},\ }in\ \href {\doibase 10.48550/arXiv.1810.07524}
  {\emph {\bibinfo {booktitle} {Science with a Next Generation Very Large
  Array}}},\ \bibinfo {series} {Astronomical Society of the Pacific Conference
  Series}, Vol.\ \bibinfo {volume} {517},\ \bibinfo {editor} {edited by\
  \bibinfo {editor} {\bibfnamefont {E.}~\bibnamefont {{Murphy}}}}\ (\bibinfo
  {year} {2018})\ p.~\bibinfo {pages} {3},\ \Eprint
  {http://arxiv.org/abs/1810.07524} {arXiv:1810.07524 [astro-ph.IM]}
  \BibitemShut {NoStop}%
\bibitem [{\citenamefont {{Rioja}}\ and\ \citenamefont
  {{Dodson}}(2020)}]{RD2020}%
  \BibitemOpen
  \bibfield  {author} {\bibinfo {author} {\bibfnamefont {M.~J.}\ \bibnamefont
  {{Rioja}}}\ and\ \bibinfo {author} {\bibfnamefont {R.}~\bibnamefont
  {{Dodson}}},\ }\href {\doibase 10.1007/s00159-020-00126-z} {\bibfield
  {journal} {\bibinfo  {journal} {Astron. Astrophys. Rev.}\ }\textbf {\bibinfo
  {volume} {28}},\ \bibinfo {eid} {6} (\bibinfo {year} {2020})},\ \Eprint
  {http://arxiv.org/abs/2010.02156} {arXiv:2010.02156 [astro-ph.IM]}
  \BibitemShut {NoStop}%
\bibitem [{\citenamefont {{Brown}}(2021)}]{Brown2021}%
  \BibitemOpen
  \bibfield  {author} {\bibinfo {author} {\bibfnamefont {A.~G.~A.}\
  \bibnamefont {{Brown}}},\ }\href {\doibase
  10.1146/annurev-astro-112320-035628} {\bibfield  {journal} {\bibinfo
  {journal} {Ann. Rev. Astron. Astrophys.}\ }\textbf {\bibinfo {volume} {59}},\
  \bibinfo {pages} {59} (\bibinfo {year} {2021})},\ \Eprint
  {http://arxiv.org/abs/2102.11712} {arXiv:2102.11712 [astro-ph.IM]}
  \BibitemShut {NoStop}%
\bibitem [{\citenamefont {{Li}}\ \emph
  {et~al.}(2022{\natexlab{a}})\citenamefont {{Li}}, \citenamefont {{Xu}},
  \citenamefont {{Li}}, \citenamefont {{Wu}}, \citenamefont {{Bian}},
  \citenamefont {{Lin}}, \citenamefont {{Yang}}, \citenamefont {{Hao}},\ and\
  \citenamefont {{Liu}}}]{LXLWBLYHL2022}%
  \BibitemOpen
  \bibfield  {author} {\bibinfo {author} {\bibfnamefont {Y.}~\bibnamefont
  {{Li}}}, \bibinfo {author} {\bibfnamefont {Y.}~\bibnamefont {{Xu}}}, \bibinfo
  {author} {\bibfnamefont {J.}~\bibnamefont {{Li}}}, \bibinfo {author}
  {\bibfnamefont {Y.}~\bibnamefont {{Wu}}}, \bibinfo {author} {\bibfnamefont
  {S.}~\bibnamefont {{Bian}}}, \bibinfo {author} {\bibfnamefont
  {Z.}~\bibnamefont {{Lin}}}, \bibinfo {author} {\bibfnamefont
  {W.}~\bibnamefont {{Yang}}}, \bibinfo {author} {\bibfnamefont
  {C.}~\bibnamefont {{Hao}}}, \ and\ \bibinfo {author} {\bibfnamefont
  {D.}~\bibnamefont {{Liu}}},\ }\href {\doibase 10.3847/1538-4357/ac3821}
  {\bibfield  {journal} {\bibinfo  {journal} {\apj}\ }\textbf {\bibinfo
  {volume} {925}},\ \bibinfo {eid} {47} (\bibinfo {year}
  {2022}{\natexlab{a}})},\ \Eprint {http://arxiv.org/abs/2111.14095}
  {arXiv:2111.14095 [gr-qc]} \BibitemShut {NoStop}%
\bibitem [{\citenamefont {{Li}}\ \emph
  {et~al.}(2022{\natexlab{b}})\citenamefont {{Li}}, \citenamefont {{Xu}},
  \citenamefont {{Bian}}, \citenamefont {{Lin}}, \citenamefont {{Li}},
  \citenamefont {{Liu}},\ and\ \citenamefont {{Hao}}}]{LXBLLLH2022}%
  \BibitemOpen
  \bibfield  {author} {\bibinfo {author} {\bibfnamefont {Y.}~\bibnamefont
  {{Li}}}, \bibinfo {author} {\bibfnamefont {Y.}~\bibnamefont {{Xu}}}, \bibinfo
  {author} {\bibfnamefont {S.}~\bibnamefont {{Bian}}}, \bibinfo {author}
  {\bibfnamefont {Z.}~\bibnamefont {{Lin}}}, \bibinfo {author} {\bibfnamefont
  {J.}~\bibnamefont {{Li}}}, \bibinfo {author} {\bibfnamefont {D.}~\bibnamefont
  {{Liu}}}, \ and\ \bibinfo {author} {\bibfnamefont {C.}~\bibnamefont
  {{Hao}}},\ }\href {\doibase 10.3847/1538-4357/ac8df8} {\bibfield  {journal}
  {\bibinfo  {journal} {\apj}\ }\textbf {\bibinfo {volume} {938}},\ \bibinfo
  {eid} {58} (\bibinfo {year} {2022}{\natexlab{b}})},\ \Eprint
  {http://arxiv.org/abs/2209.08702} {arXiv:2209.08702 [gr-qc]} \BibitemShut
  {NoStop}%
\bibitem [{\citenamefont {Braun}\ \emph {et~al.}(2015)\citenamefont {Braun},
  \citenamefont {Bourke}, \citenamefont {Green}, \citenamefont {Keane},\ and\
  \citenamefont {Wagg}}]{BBGKW2015}%
  \BibitemOpen
  \bibfield  {author} {\bibinfo {author} {\bibfnamefont {R.}~\bibnamefont
  {Braun}}, \bibinfo {author} {\bibfnamefont {T.}~\bibnamefont {Bourke}},
  \bibinfo {author} {\bibfnamefont {J.~A.}\ \bibnamefont {Green}}, \bibinfo
  {author} {\bibfnamefont {E.}~\bibnamefont {Keane}}, \ and\ \bibinfo {author}
  {\bibfnamefont {J.}~\bibnamefont {Wagg}},\ }\href {\doibase
  10.22323/1.215.0174} {\bibfield  {journal} {\bibinfo  {journal} {PoS}\
  }\textbf {\bibinfo {volume} {AASKA14}},\ \bibinfo {pages} {174} (\bibinfo
  {year} {2015})}\BibitemShut {NoStop}%
\bibitem [{\citenamefont {{Aab}}\ \emph {et~al.}(2014)\citenamefont {{Aab}}
  \emph {et~al.}}]{Aab2014}%
  \BibitemOpen
  \bibfield  {author} {\bibinfo {author} {\bibfnamefont {A.}~\bibnamefont
  {{Aab}}} \emph {et~al.},\ }\href {\doibase 10.1088/0004-637X/794/2/172}
  {\bibfield  {journal} {\bibinfo  {journal} {Astrophys. J.}\ }\textbf
  {\bibinfo {volume} {794}},\ \bibinfo {eid} {172} (\bibinfo {year} {2014})},\
  \Eprint {http://arxiv.org/abs/1409.3128} {arXiv:1409.3128 [astro-ph.HE]}
  \BibitemShut {NoStop}%
\bibitem [{\citenamefont {{Bartoli}}\ \emph {et~al.}(2019)\citenamefont
  {{Bartoli}} \emph {et~al.}}]{Bartoli2019}%
  \BibitemOpen
  \bibfield  {author} {\bibinfo {author} {\bibfnamefont {B.}~\bibnamefont
  {{Bartoli}}} \emph {et~al.},\ }\href {\doibase 10.3847/1538-4357/aafe06}
  {\bibfield  {journal} {\bibinfo  {journal} {Astrophys. J.}\ }\textbf
  {\bibinfo {volume} {872}},\ \bibinfo {eid} {143} (\bibinfo {year} {2019})},\
  \Eprint {http://arxiv.org/abs/1901.04201} {arXiv:1901.04201 [astro-ph.HE]}
  \BibitemShut {NoStop}%
\bibitem [{\citenamefont {{Albert}}\ \emph {et~al.}(2020)\citenamefont
  {{Albert}} \emph {et~al.}}]{Albert2020}%
  \BibitemOpen
  \bibfield  {author} {\bibinfo {author} {\bibfnamefont {A.}~\bibnamefont
  {{Albert}}} \emph {et~al.},\ }\href {\doibase 10.1103/PhysRevD.102.122007}
  {\bibfield  {journal} {\bibinfo  {journal} {\prd}\ }\textbf {\bibinfo
  {volume} {102}},\ \bibinfo {eid} {122007} (\bibinfo {year} {2020})},\ \Eprint
  {http://arxiv.org/abs/2007.00931} {arXiv:2007.00931 [astro-ph.HE]}
  \BibitemShut {NoStop}%
\bibitem [{\citenamefont {{Koch}}\ \emph {et~al.}(2010)\citenamefont {{Koch}}
  \emph {et~al.}}]{Koch2010}%
  \BibitemOpen
  \bibfield  {author} {\bibinfo {author} {\bibfnamefont {D.~G.}\ \bibnamefont
  {{Koch}}} \emph {et~al.},\ }\href {\doibase 10.1088/2041-8205/713/2/L79}
  {\bibfield  {journal} {\bibinfo  {journal} {Astrophys. J. Lett.}\ }\textbf
  {\bibinfo {volume} {713}},\ \bibinfo {pages} {L79} (\bibinfo {year}
  {2010})},\ \Eprint {http://arxiv.org/abs/1001.0268} {arXiv:1001.0268
  [astro-ph.EP]} \BibitemShut {NoStop}%
\bibitem [{\citenamefont {{Bowman}}\ and\ \citenamefont
  {{Kurtz}}(2018)}]{BK2018}%
  \BibitemOpen
  \bibfield  {author} {\bibinfo {author} {\bibfnamefont {D.~M.}\ \bibnamefont
  {{Bowman}}}\ and\ \bibinfo {author} {\bibfnamefont {D.~W.}\ \bibnamefont
  {{Kurtz}}},\ }\href {\doibase 10.1093/mnras/sty449} {\bibfield  {journal}
  {\bibinfo  {journal} {Mon. Not. R. Astron. Soc.}\ }\textbf {\bibinfo {volume}
  {476}},\ \bibinfo {pages} {3169} (\bibinfo {year} {2018})},\ \Eprint
  {http://arxiv.org/abs/1802.05433} {arXiv:1802.05433 [astro-ph.SR]}
  \BibitemShut {NoStop}%
\bibitem [{\citenamefont {{Kurtz}}\ \emph {et~al.}(2005)\citenamefont {{Kurtz}}
  \emph {et~al.}}]{Kurtz2005}%
  \BibitemOpen
  \bibfield  {author} {\bibinfo {author} {\bibfnamefont {D.~W.}\ \bibnamefont
  {{Kurtz}}} \emph {et~al.},\ }\href {\doibase
  10.1111/j.1365-2966.2005.08807.x} {\bibfield  {journal} {\bibinfo  {journal}
  {Mon. Not. R. Astron. Soc.}\ }\textbf {\bibinfo {volume} {358}},\ \bibinfo
  {pages} {651} (\bibinfo {year} {2005})}\BibitemShut {NoStop}%
\bibitem [{\citenamefont {{Vanderburg}}\ and\ \citenamefont
  {{Johnson}}(2014)}]{VJ2014}%
  \BibitemOpen
  \bibfield  {author} {\bibinfo {author} {\bibfnamefont {A.}~\bibnamefont
  {{Vanderburg}}}\ and\ \bibinfo {author} {\bibfnamefont {J.~A.}\ \bibnamefont
  {{Johnson}}},\ }\href {\doibase 10.1086/678764} {\bibfield  {journal}
  {\bibinfo  {journal} {Publ. Astron. Soc. Pac.}\ }\textbf {\bibinfo {volume}
  {126}},\ \bibinfo {pages} {948} (\bibinfo {year} {2014})},\ \Eprint
  {http://arxiv.org/abs/1408.3853} {arXiv:1408.3853 [astro-ph.IM]} \BibitemShut
  {NoStop}%
\bibitem [{\citenamefont {{Aigrain}}\ \emph {et~al.}(2015)\citenamefont
  {{Aigrain}}, \citenamefont {{Hodgkin}}, \citenamefont {{Irwin}},
  \citenamefont {{Lewis}},\ and\ \citenamefont {{Roberts}}}]{AHILR2015}%
  \BibitemOpen
  \bibfield  {author} {\bibinfo {author} {\bibfnamefont {S.}~\bibnamefont
  {{Aigrain}}}, \bibinfo {author} {\bibfnamefont {S.~T.}\ \bibnamefont
  {{Hodgkin}}}, \bibinfo {author} {\bibfnamefont {M.~J.}\ \bibnamefont
  {{Irwin}}}, \bibinfo {author} {\bibfnamefont {J.~R.}\ \bibnamefont
  {{Lewis}}}, \ and\ \bibinfo {author} {\bibfnamefont {S.~J.}\ \bibnamefont
  {{Roberts}}},\ }\href {\doibase 10.1093/mnras/stu2638} {\bibfield  {journal}
  {\bibinfo  {journal} {Mon. Not. R. Astron. Soc.}\ }\textbf {\bibinfo {volume}
  {447}},\ \bibinfo {pages} {2880} (\bibinfo {year} {2015})},\ \Eprint
  {http://arxiv.org/abs/1412.6304} {arXiv:1412.6304 [astro-ph.IM]} \BibitemShut
  {NoStop}%
\bibitem [{\citenamefont {{Huber}}\ \emph {et~al.}(2016)\citenamefont {{Huber}}
  \emph {et~al.}}]{Huber2016}%
  \BibitemOpen
  \bibfield  {author} {\bibinfo {author} {\bibfnamefont {D.}~\bibnamefont
  {{Huber}}} \emph {et~al.},\ }\href {\doibase 10.3847/0067-0049/224/1/2}
  {\bibfield  {journal} {\bibinfo  {journal} {Astrophys. J. Suppl. Ser.}\
  }\textbf {\bibinfo {volume} {224}},\ \bibinfo {eid} {2} (\bibinfo {year}
  {2016})},\ \Eprint {http://arxiv.org/abs/1512.02643} {arXiv:1512.02643
  [astro-ph.SR]} \BibitemShut {NoStop}%
\bibitem [{\citenamefont {{Qi}}\ \emph {et~al.}(2021)\citenamefont {{Qi}},
  \citenamefont {{Jin}}, \citenamefont {{Fan}}, \citenamefont {{Zhang}},\ and\
  \citenamefont {{Zhang}}}]{QJFZZ2021}%
  \BibitemOpen
  \bibfield  {author} {\bibinfo {author} {\bibfnamefont {J.-Z.}\ \bibnamefont
  {{Qi}}}, \bibinfo {author} {\bibfnamefont {S.-J.}\ \bibnamefont {{Jin}}},
  \bibinfo {author} {\bibfnamefont {X.-L.}\ \bibnamefont {{Fan}}}, \bibinfo
  {author} {\bibfnamefont {J.-F.}\ \bibnamefont {{Zhang}}}, \ and\ \bibinfo
  {author} {\bibfnamefont {X.}~\bibnamefont {{Zhang}}},\ }\href {\doibase
  10.1088/1475-7516/2021/12/042} {\bibfield  {journal} {\bibinfo  {journal} {J.
  Cosmol. Astropart. Phys.}\ }\textbf {\bibinfo {volume} {2021}},\ \bibinfo
  {eid} {042} (\bibinfo {year} {2021})},\ \Eprint
  {http://arxiv.org/abs/2102.01292} {arXiv:2102.01292 [astro-ph.CO]}
  \BibitemShut {NoStop}%
\bibitem [{\citenamefont {{Feleppa}}\ \emph {et~al.}(2025)\citenamefont
  {{Feleppa}}, \citenamefont {{Bozza}},\ and\ \citenamefont
  {{Tsupko}}}]{FBT2025}%
  \BibitemOpen
  \bibfield  {author} {\bibinfo {author} {\bibfnamefont {F.}~\bibnamefont
  {{Feleppa}}}, \bibinfo {author} {\bibfnamefont {V.}~\bibnamefont {{Bozza}}},
  \ and\ \bibinfo {author} {\bibfnamefont {O.~Y.}\ \bibnamefont {{Tsupko}}},\
  }\href {\doibase 10.1103/PhysRevD.111.044018} {\bibfield  {journal} {\bibinfo
   {journal} {\prd}\ }\textbf {\bibinfo {volume} {111}},\ \bibinfo {eid}
  {044018} (\bibinfo {year} {2025})},\ \Eprint
  {http://arxiv.org/abs/2412.16712} {arXiv:2412.16712 [gr-qc]} \BibitemShut
  {NoStop}%
\bibitem [{\citenamefont {{Eiroa}}\ \emph {et~al.}(2002)\citenamefont
  {{Eiroa}}, \citenamefont {{Romero}},\ and\ \citenamefont
  {{Torres}}}]{ERT2002}%
  \BibitemOpen
  \bibfield  {author} {\bibinfo {author} {\bibfnamefont {E.~F.}\ \bibnamefont
  {{Eiroa}}}, \bibinfo {author} {\bibfnamefont {G.~E.}\ \bibnamefont
  {{Romero}}}, \ and\ \bibinfo {author} {\bibfnamefont {D.~F.}\ \bibnamefont
  {{Torres}}},\ }\href {\doibase 10.1103/PhysRevD.66.024010} {\bibfield
  {journal} {\bibinfo  {journal} {\prd}\ }\textbf {\bibinfo {volume} {66}},\
  \bibinfo {eid} {024010} (\bibinfo {year} {2002})},\ \Eprint
  {http://arxiv.org/abs/gr-qc/0203049} {arXiv:gr-qc/0203049 [gr-qc]}
  \BibitemShut {NoStop}%
\bibitem [{\citenamefont {{Virbhadra}}(2009)}]{Virbha2009}%
  \BibitemOpen
  \bibfield  {author} {\bibinfo {author} {\bibfnamefont {K.~S.}\ \bibnamefont
  {{Virbhadra}}},\ }\href {\doibase 10.1103/PhysRevD.79.083004} {\bibfield
  {journal} {\bibinfo  {journal} {\prd}\ }\textbf {\bibinfo {volume} {79}},\
  \bibinfo {eid} {083004} (\bibinfo {year} {2009})},\ \Eprint
  {http://arxiv.org/abs/0810.2109} {arXiv:0810.2109 [gr-qc]} \BibitemShut
  {NoStop}%
\bibitem [{\citenamefont {{Virbhadra}}(2022)}]{Virbhadra2022}%
  \BibitemOpen
  \bibfield  {author} {\bibinfo {author} {\bibfnamefont {K.~S.}\ \bibnamefont
  {{Virbhadra}}},\ }\href {\doibase 10.1103/PhysRevD.106.064038} {\bibfield
  {journal} {\bibinfo  {journal} {\prd}\ }\textbf {\bibinfo {volume} {106}},\
  \bibinfo {eid} {064038} (\bibinfo {year} {2022})},\ \Eprint
  {http://arxiv.org/abs/2204.01879} {arXiv:2204.01879 [gr-qc]} \BibitemShut
  {NoStop}%
\bibitem [{\citenamefont {{Keeton}}\ and\ \citenamefont
  {{Petters}}(2006{\natexlab{a}})}]{KP2006a}%
  \BibitemOpen
  \bibfield  {author} {\bibinfo {author} {\bibfnamefont {C.~R.}\ \bibnamefont
  {{Keeton}}}\ and\ \bibinfo {author} {\bibfnamefont {A.~O.}\ \bibnamefont
  {{Petters}}},\ }\href {\doibase 10.1103/PhysRevD.73.044024} {\bibfield
  {journal} {\bibinfo  {journal} {\prd}\ }\textbf {\bibinfo {volume} {73}},\
  \bibinfo {eid} {044024} (\bibinfo {year} {2006}{\natexlab{a}})},\ \Eprint
  {http://arxiv.org/abs/gr-qc/0601053} {arXiv:gr-qc/0601053 [gr-qc]}
  \BibitemShut {NoStop}%
\bibitem [{\citenamefont {{Keeton}}\ and\ \citenamefont
  {{Petters}}(2006{\natexlab{b}})}]{KP2006b}%
  \BibitemOpen
  \bibfield  {author} {\bibinfo {author} {\bibfnamefont {C.~R.}\ \bibnamefont
  {{Keeton}}}\ and\ \bibinfo {author} {\bibfnamefont {A.~O.}\ \bibnamefont
  {{Petters}}},\ }\href {\doibase 10.1103/PhysRevD.73.104032} {\bibfield
  {journal} {\bibinfo  {journal} {\prd}\ }\textbf {\bibinfo {volume} {73}},\
  \bibinfo {eid} {104032} (\bibinfo {year} {2006}{\natexlab{b}})},\ \Eprint
  {http://arxiv.org/abs/gr-qc/0603061} {arXiv:gr-qc/0603061 [gr-qc]}
  \BibitemShut {NoStop}%
\bibitem [{\citenamefont {{Simpson}}\ \emph {et~al.}(2019)\citenamefont
  {{Simpson}}, \citenamefont {{Mart{\'\i}n-Moruno}},\ and\ \citenamefont
  {{Visser}}}]{SMV2019}%
  \BibitemOpen
  \bibfield  {author} {\bibinfo {author} {\bibfnamefont {A.}~\bibnamefont
  {{Simpson}}}, \bibinfo {author} {\bibfnamefont {P.}~\bibnamefont
  {{Mart{\'\i}n-Moruno}}}, \ and\ \bibinfo {author} {\bibfnamefont
  {M.}~\bibnamefont {{Visser}}},\ }\href {\doibase 10.1088/1361-6382/ab28a5}
  {\bibfield  {journal} {\bibinfo  {journal} {Classical and Quantum Gravity}\
  }\textbf {\bibinfo {volume} {36}},\ \bibinfo {eid} {145007} (\bibinfo {year}
  {2019})},\ \Eprint {http://arxiv.org/abs/1902.04232} {arXiv:1902.04232
  [gr-qc]} \BibitemShut {NoStop}%
\bibitem [{\citenamefont {{Lobo}}\ \emph {et~al.}(2021)\citenamefont {{Lobo}},
  \citenamefont {{Rodrigues}}, \citenamefont {{Silva}}, \citenamefont
  {{Simpson}},\ and\ \citenamefont {{Visser}}}]{LRSSV2021}%
  \BibitemOpen
  \bibfield  {author} {\bibinfo {author} {\bibfnamefont {F.~S.~N.}\
  \bibnamefont {{Lobo}}}, \bibinfo {author} {\bibfnamefont {M.~E.}\
  \bibnamefont {{Rodrigues}}}, \bibinfo {author} {\bibfnamefont {M.~V. d.~S.}\
  \bibnamefont {{Silva}}}, \bibinfo {author} {\bibfnamefont {A.}~\bibnamefont
  {{Simpson}}}, \ and\ \bibinfo {author} {\bibfnamefont {M.}~\bibnamefont
  {{Visser}}},\ }\href {\doibase 10.1103/PhysRevD.103.084052} {\bibfield
  {journal} {\bibinfo  {journal} {\prd}\ }\textbf {\bibinfo {volume} {103}},\
  \bibinfo {eid} {084052} (\bibinfo {year} {2021})},\ \Eprint
  {http://arxiv.org/abs/2009.12057} {arXiv:2009.12057 [gr-qc]} \BibitemShut
  {NoStop}%
\bibitem [{\citenamefont {{Straumann}}(1984)}]{St1984}%
  \BibitemOpen
  \bibfield  {author} {\bibinfo {author} {\bibfnamefont {N.}~\bibnamefont
  {{Straumann}}},\ }\href@noop {} {\emph {\bibinfo {title} {{General Relativity
  and Relativistic Astrophysics}}}}\ (\bibinfo  {publisher} {Springer},\
  \bibinfo {address} {Berlin},\ \bibinfo {year} {1984})\BibitemShut {NoStop}%
\bibitem [{\citenamefont {Islam}\ and\ \citenamefont {Rahaman}(2023)}]{2023IR}%
  \BibitemOpen
  \bibfield  {author} {\bibinfo {author} {\bibfnamefont {S.}~\bibnamefont
  {Islam}}\ and\ \bibinfo {author} {\bibfnamefont {F.}~\bibnamefont
  {Rahaman}},\ }\href {\doibase 10.3390/axioms12040364} {\bibfield  {journal}
  {\bibinfo  {journal} {Axioms}\ }\textbf {\bibinfo {volume} {12}},\ \bibinfo
  {pages} {364} (\bibinfo {year} {2023})}\BibitemShut {NoStop}%
\bibitem [{\citenamefont {{Misner}}\ \emph {et~al.}(1973)\citenamefont
  {{Misner}}, \citenamefont {{Thorne}},\ and\ \citenamefont
  {{Wheeler}}}]{MTW1973}%
  \BibitemOpen
  \bibfield  {author} {\bibinfo {author} {\bibfnamefont {C.~W.}\ \bibnamefont
  {{Misner}}}, \bibinfo {author} {\bibfnamefont {K.~S.}\ \bibnamefont
  {{Thorne}}}, \ and\ \bibinfo {author} {\bibfnamefont {J.~A.}\ \bibnamefont
  {{Wheeler}}},\ }\href@noop {} {\emph {\bibinfo {title} {{Gravitation}}}}\
  (\bibinfo  {publisher} {W. H. Freeman},\ \bibinfo {address} {San Francisco},\
  \bibinfo {year} {1973})\BibitemShut {NoStop}%
\bibitem [{\citenamefont {{GRAVITY Collaboration}}\ \emph
  {et~al.}(2020)\citenamefont {{GRAVITY Collaboration}} \emph
  {et~al.}}]{GRAVITY2020}%
  \BibitemOpen
  \bibfield  {author} {\bibinfo {author} {\bibnamefont {{GRAVITY
  Collaboration}}} \emph {et~al.},\ }\href {\doibase
  10.1051/0004-6361/202037813} {\bibfield  {journal} {\bibinfo  {journal}
  {Astron. Astrophys.}\ }\textbf {\bibinfo {volume} {636}},\ \bibinfo {eid}
  {L5} (\bibinfo {year} {2020})},\ \Eprint {http://arxiv.org/abs/2004.07187}
  {arXiv:2004.07187 [astro-ph.GA]} \BibitemShut {NoStop}%
\bibitem [{\citenamefont {{Deng}}(2020)}]{Deng2020}%
  \BibitemOpen
  \bibfield  {author} {\bibinfo {author} {\bibfnamefont {X.-M.}\ \bibnamefont
  {{Deng}}},\ }\href {\doibase 10.1016/j.dark.2020.100629} {\bibfield
  {journal} {\bibinfo  {journal} {Phys. Dark Univ.}\ }\textbf {\bibinfo
  {volume} {30}},\ \bibinfo {eid} {100629} (\bibinfo {year}
  {2020})}\BibitemShut {NoStop}%
\bibitem [{\citenamefont {{Lin}}\ and\ \citenamefont {{Deng}}(2022)}]{LDE2022}%
  \BibitemOpen
  \bibfield  {author} {\bibinfo {author} {\bibfnamefont {H.-Y.}\ \bibnamefont
  {{Lin}}}\ and\ \bibinfo {author} {\bibfnamefont {X.-M.}\ \bibnamefont
  {{Deng}}},\ }\href {\doibase 10.1140/epjp/s13360-022-02391-6} {\bibfield
  {journal} {\bibinfo  {journal} {Eur. Phys. J. Plus}\ }\textbf {\bibinfo
  {volume} {137}},\ \bibinfo {eid} {176} (\bibinfo {year} {2022})}\BibitemShut
  {NoStop}%
\bibitem [{\citenamefont {{Lin}}\ and\ \citenamefont
  {{Deng}}(2023{\natexlab{a}})}]{LDE2023}%
  \BibitemOpen
  \bibfield  {author} {\bibinfo {author} {\bibfnamefont {H.-Y.}\ \bibnamefont
  {{Lin}}}\ and\ \bibinfo {author} {\bibfnamefont {X.-M.}\ \bibnamefont
  {{Deng}}},\ }\href {\doibase 10.1140/epjc/s10052-023-11487-x} {\bibfield
  {journal} {\bibinfo  {journal} {Eur. Phys. J. C}\ }\textbf {\bibinfo {volume}
  {83}},\ \bibinfo {eid} {311} (\bibinfo {year}
  {2023}{\natexlab{a}})}\BibitemShut {NoStop}%
\bibitem [{\citenamefont {{Liu}}\ \emph {et~al.}(2023)\citenamefont {{Liu}},
  \citenamefont {{Mustafa}}, \citenamefont {{Maurya}},\ and\ \citenamefont
  {{Javed}}}]{LMMJ2023}%
  \BibitemOpen
  \bibfield  {author} {\bibinfo {author} {\bibfnamefont {Y.}~\bibnamefont
  {{Liu}}}, \bibinfo {author} {\bibfnamefont {G.}~\bibnamefont {{Mustafa}}},
  \bibinfo {author} {\bibfnamefont {S.~K.}\ \bibnamefont {{Maurya}}}, \ and\
  \bibinfo {author} {\bibfnamefont {F.}~\bibnamefont {{Javed}}},\ }\href
  {\doibase 10.1140/epjc/s10052-023-11702-9} {\bibfield  {journal} {\bibinfo
  {journal} {Eur. Phys. J. C}\ }\textbf {\bibinfo {volume} {83}},\ \bibinfo
  {eid} {584} (\bibinfo {year} {2023})}\BibitemShut {NoStop}%
\bibitem [{\citenamefont {{Lin}}\ and\ \citenamefont
  {{Deng}}(2023{\natexlab{b}})}]{LDA2023}%
  \BibitemOpen
  \bibfield  {author} {\bibinfo {author} {\bibfnamefont {H.-Y.}\ \bibnamefont
  {{Lin}}}\ and\ \bibinfo {author} {\bibfnamefont {X.-M.}\ \bibnamefont
  {{Deng}}},\ }\href {\doibase 10.1016/j.aop.2023.169360} {\bibfield  {journal}
  {\bibinfo  {journal} {Ann. Phys.}\ }\textbf {\bibinfo {volume} {455}},\
  \bibinfo {eid} {169360} (\bibinfo {year} {2023}{\natexlab{b}})}\BibitemShut
  {NoStop}%
\bibitem [{\citenamefont {{Huang}}\ and\ \citenamefont
  {{Deng}}(2024{\natexlab{a}})}]{HDE2024}%
  \BibitemOpen
  \bibfield  {author} {\bibinfo {author} {\bibfnamefont {L.}~\bibnamefont
  {{Huang}}}\ and\ \bibinfo {author} {\bibfnamefont {X.-M.}\ \bibnamefont
  {{Deng}}},\ }\href {\doibase 10.1140/epjc/s10052-024-12969-2} {\bibfield
  {journal} {\bibinfo  {journal} {Eur. Phys. J. C}\ }\textbf {\bibinfo {volume}
  {84}},\ \bibinfo {eid} {615} (\bibinfo {year}
  {2024}{\natexlab{a}})}\BibitemShut {NoStop}%
\bibitem [{\citenamefont {{Huang}}\ and\ \citenamefont
  {{Deng}}(2024{\natexlab{b}})}]{HDP2024}%
  \BibitemOpen
  \bibfield  {author} {\bibinfo {author} {\bibfnamefont {L.}~\bibnamefont
  {{Huang}}}\ and\ \bibinfo {author} {\bibfnamefont {X.-M.}\ \bibnamefont
  {{Deng}}},\ }\href {\doibase 10.1103/PhysRevD.109.124005} {\bibfield
  {journal} {\bibinfo  {journal} {\prd}\ }\textbf {\bibinfo {volume} {109}},\
  \bibinfo {eid} {124005} (\bibinfo {year} {2024}{\natexlab{b}})}\BibitemShut
  {NoStop}%
\bibitem [{\citenamefont {{Li}}\ and\ \citenamefont {{Deng}}(2025)}]{LD2025P}%
  \BibitemOpen
  \bibfield  {author} {\bibinfo {author} {\bibfnamefont {C.-H.}\ \bibnamefont
  {{Li}}}\ and\ \bibinfo {author} {\bibfnamefont {X.-M.}\ \bibnamefont
  {{Deng}}},\ }\href {\doibase 10.1103/q4c6-hw51} {\bibfield  {journal}
  {\bibinfo  {journal} {\prd}\ }\textbf {\bibinfo {volume} {111}},\ \bibinfo
  {eid} {124051} (\bibinfo {year} {2025})}\BibitemShut {NoStop}%
\bibitem [{\citenamefont {{Zakharov}}(1994)}]{Zakh1994}%
  \BibitemOpen
  \bibfield  {author} {\bibinfo {author} {\bibfnamefont {A.~F.}\ \bibnamefont
  {{Zakharov}}},\ }\href {\doibase 10.1088/0264-9381/11/4/018} {\bibfield
  {journal} {\bibinfo  {journal} {Classical and Quantum Gravity}\ }\textbf
  {\bibinfo {volume} {11}},\ \bibinfo {pages} {1027} (\bibinfo {year}
  {1994})}\BibitemShut {NoStop}%
\bibitem [{\citenamefont {{Zakharov}}(2018)}]{Zakh2018}%
  \BibitemOpen
  \bibfield  {author} {\bibinfo {author} {\bibfnamefont {A.~F.}\ \bibnamefont
  {{Zakharov}}},\ }\href {\doibase 10.1088/1361-6382/aa964e} {\bibfield
  {journal} {\bibinfo  {journal} {Classical and Quantum Gravity}\ }\textbf
  {\bibinfo {volume} {35}},\ \bibinfo {eid} {028001} (\bibinfo {year}
  {2018})}\BibitemShut {NoStop}%
\bibitem [{\citenamefont {{Cardoso}}\ \emph {et~al.}(2009)\citenamefont
  {{Cardoso}}, \citenamefont {{Miranda}}, \citenamefont {{Berti}},
  \citenamefont {{Witek}},\ and\ \citenamefont {{Zanchin}}}]{CMBWZ2009}%
  \BibitemOpen
  \bibfield  {author} {\bibinfo {author} {\bibfnamefont {V.}~\bibnamefont
  {{Cardoso}}}, \bibinfo {author} {\bibfnamefont {A.~S.}\ \bibnamefont
  {{Miranda}}}, \bibinfo {author} {\bibfnamefont {E.}~\bibnamefont {{Berti}}},
  \bibinfo {author} {\bibfnamefont {H.}~\bibnamefont {{Witek}}}, \ and\
  \bibinfo {author} {\bibfnamefont {V.~T.}\ \bibnamefont {{Zanchin}}},\ }\href
  {\doibase 10.1103/PhysRevD.79.064016} {\bibfield  {journal} {\bibinfo
  {journal} {\prd}\ }\textbf {\bibinfo {volume} {79}},\ \bibinfo {eid} {064016}
  (\bibinfo {year} {2009})},\ \Eprint {http://arxiv.org/abs/0812.1806}
  {arXiv:0812.1806 [hep-th]} \BibitemShut {NoStop}%
\bibitem [{\citenamefont {{Jefremov}}\ \emph {et~al.}(2015)\citenamefont
  {{Jefremov}}, \citenamefont {{Tsupko}},\ and\ \citenamefont
  {{Bisnovatyi-Kogan}}}]{JTB2015}%
  \BibitemOpen
  \bibfield  {author} {\bibinfo {author} {\bibfnamefont {P.~I.}\ \bibnamefont
  {{Jefremov}}}, \bibinfo {author} {\bibfnamefont {O.~Y.}\ \bibnamefont
  {{Tsupko}}}, \ and\ \bibinfo {author} {\bibfnamefont {G.~S.}\ \bibnamefont
  {{Bisnovatyi-Kogan}}},\ }\href {\doibase 10.1103/PhysRevD.91.124030}
  {\bibfield  {journal} {\bibinfo  {journal} {\prd}\ }\textbf {\bibinfo
  {volume} {91}},\ \bibinfo {eid} {124030} (\bibinfo {year} {2015})},\ \Eprint
  {http://arxiv.org/abs/1503.07060} {arXiv:1503.07060 [gr-qc]} \BibitemShut
  {NoStop}%
\bibitem [{\citenamefont {{Bland-Hawthorn}}\ and\ \citenamefont
  {{Gerhard}}(2016)}]{BG2016}%
  \BibitemOpen
  \bibfield  {author} {\bibinfo {author} {\bibfnamefont {J.}~\bibnamefont
  {{Bland-Hawthorn}}}\ and\ \bibinfo {author} {\bibfnamefont {O.}~\bibnamefont
  {{Gerhard}}},\ }\href {\doibase 10.1146/annurev-astro-081915-023441}
  {\bibfield  {journal} {\bibinfo  {journal} {Annu. Rev. Astron. Astrophys.}\
  }\textbf {\bibinfo {volume} {54}},\ \bibinfo {pages} {529} (\bibinfo {year}
  {2016})},\ \Eprint {http://arxiv.org/abs/1602.07702} {arXiv:1602.07702
  [astro-ph.GA]} \BibitemShut {NoStop}%
\bibitem [{\citenamefont {{Parsa}}\ \emph {et~al.}(2017)\citenamefont
  {{Parsa}}, \citenamefont {{Eckart}}, \citenamefont {{Shahzamanian}},
  \citenamefont {{Karas}}, \citenamefont {{Zaja{\v{c}}ek}}, \citenamefont
  {{Zensus}},\ and\ \citenamefont {{Straubmeier}}}]{Parsa2017}%
  \BibitemOpen
  \bibfield  {author} {\bibinfo {author} {\bibfnamefont {M.}~\bibnamefont
  {{Parsa}}}, \bibinfo {author} {\bibfnamefont {A.}~\bibnamefont {{Eckart}}},
  \bibinfo {author} {\bibfnamefont {B.}~\bibnamefont {{Shahzamanian}}},
  \bibinfo {author} {\bibfnamefont {V.}~\bibnamefont {{Karas}}}, \bibinfo
  {author} {\bibfnamefont {M.}~\bibnamefont {{Zaja{\v{c}}ek}}}, \bibinfo
  {author} {\bibfnamefont {J.~A.}\ \bibnamefont {{Zensus}}}, \ and\ \bibinfo
  {author} {\bibfnamefont {C.}~\bibnamefont {{Straubmeier}}},\ }\href {\doibase
  10.3847/1538-4357/aa7bf0} {\bibfield  {journal} {\bibinfo  {journal} {\apj}\
  }\textbf {\bibinfo {volume} {845}},\ \bibinfo {eid} {22} (\bibinfo {year}
  {2017})},\ \Eprint {http://arxiv.org/abs/1708.03507} {arXiv:1708.03507
  [astro-ph.GA]} \BibitemShut {NoStop}%
\bibitem [{\citenamefont {{Patla}}\ \emph
  {et~al.}(2014{\natexlab{b}})\citenamefont {{Patla}}, \citenamefont
  {{Nemiroff}}, \citenamefont {{Hoffmann}},\ and\ \citenamefont
  {{Zioutas}}}]{PNHZ2014}%
  \BibitemOpen
  \bibfield  {author} {\bibinfo {author} {\bibfnamefont {B.~R.}\ \bibnamefont
  {{Patla}}}, \bibinfo {author} {\bibfnamefont {R.~J.}\ \bibnamefont
  {{Nemiroff}}}, \bibinfo {author} {\bibfnamefont {D.~H.~H.}\ \bibnamefont
  {{Hoffmann}}}, \ and\ \bibinfo {author} {\bibfnamefont {K.}~\bibnamefont
  {{Zioutas}}},\ }\href {\doibase 10.1088/0004-637X/780/2/158} {\bibfield
  {journal} {\bibinfo  {journal} {\apj}\ }\textbf {\bibinfo {volume} {780}},\
  \bibinfo {eid} {158} (\bibinfo {year} {2014}{\natexlab{b}})},\ \Eprint
  {http://arxiv.org/abs/1305.2454} {arXiv:1305.2454 [astro-ph.EP]} \BibitemShut
  {NoStop}%
\bibitem [{\citenamefont {{Werner}}\ and\ \citenamefont
  {{Petters}}(2007)}]{WP2007}%
  \BibitemOpen
  \bibfield  {author} {\bibinfo {author} {\bibfnamefont {M.~C.}\ \bibnamefont
  {{Werner}}}\ and\ \bibinfo {author} {\bibfnamefont {A.~O.}\ \bibnamefont
  {{Petters}}},\ }\href {\doibase 10.1103/PhysRevD.76.064024} {\bibfield
  {journal} {\bibinfo  {journal} {\prd}\ }\textbf {\bibinfo {volume} {76}},\
  \bibinfo {eid} {064024} (\bibinfo {year} {2007})},\ \Eprint
  {http://arxiv.org/abs/0706.0132} {arXiv:0706.0132 [gr-qc]} \BibitemShut
  {NoStop}%
\bibitem [{\citenamefont {{Ibrahim}}\ \emph {et~al.}(2018)\citenamefont
  {{Ibrahim}}, \citenamefont {{Malasan}}, \citenamefont {{Kunjaya}},
  \citenamefont {{Timur Jaelani}}, \citenamefont {{Puannandra Putri}},\ and\
  \citenamefont {{Djamal}}}]{IMKTP2018}%
  \BibitemOpen
  \bibfield  {author} {\bibinfo {author} {\bibfnamefont {I.}~\bibnamefont
  {{Ibrahim}}}, \bibinfo {author} {\bibfnamefont {H.~L.}\ \bibnamefont
  {{Malasan}}}, \bibinfo {author} {\bibfnamefont {C.}~\bibnamefont
  {{Kunjaya}}}, \bibinfo {author} {\bibfnamefont {A.}~\bibnamefont {{Timur
  Jaelani}}}, \bibinfo {author} {\bibfnamefont {G.}~\bibnamefont {{Puannandra
  Putri}}}, \ and\ \bibinfo {author} {\bibfnamefont {M.}~\bibnamefont
  {{Djamal}}},\ }\href {\doibase 10.1088/1674-4527/18/4/41} {\bibfield
  {journal} {\bibinfo  {journal} {Res. Astron. Astrophys.}\ }\textbf {\bibinfo
  {volume} {18}},\ \bibinfo {eid} {041} (\bibinfo {year} {2018})}\BibitemShut
  {NoStop}%
\bibitem [{\citenamefont {{Tsukamoto}}(2021{\natexlab{c}})}]{Tsukamoto2021}%
  \BibitemOpen
  \bibfield  {author} {\bibinfo {author} {\bibfnamefont {N.}~\bibnamefont
  {{Tsukamoto}}},\ }\href {\doibase 10.1103/PhysRevD.103.024033} {\bibfield
  {journal} {\bibinfo  {journal} {\prd}\ }\textbf {\bibinfo {volume} {103}},\
  \bibinfo {eid} {024033} (\bibinfo {year} {2021}{\natexlab{c}})},\ \Eprint
  {http://arxiv.org/abs/2011.03932} {arXiv:2011.03932 [gr-qc]} \BibitemShut
  {NoStop}%
\bibitem [{\citenamefont {{Frost}}(2023)}]{Frost2023}%
  \BibitemOpen
  \bibfield  {author} {\bibinfo {author} {\bibfnamefont {T.~C.}\ \bibnamefont
  {{Frost}}},\ }\href {\doibase 10.1103/PhysRevD.108.124019} {\bibfield
  {journal} {\bibinfo  {journal} {\prd}\ }\textbf {\bibinfo {volume} {108}},\
  \bibinfo {eid} {124019} (\bibinfo {year} {2023})},\ \Eprint
  {http://arxiv.org/abs/2304.12563} {arXiv:2304.12563 [gr-qc]} \BibitemShut
  {NoStop}%
\bibitem [{\citenamefont {{IceCube Collaboration}}\ \emph
  {et~al.}(2006)\citenamefont {{IceCube Collaboration}} \emph
  {et~al.}}]{IceCube2006}%
  \BibitemOpen
  \bibfield  {author} {\bibinfo {author} {\bibnamefont {{IceCube
  Collaboration}}} \emph {et~al.},\ }\href {\doibase
  10.1016/j.astropartphys.2006.06.007} {\bibfield  {journal} {\bibinfo
  {journal} {Astropart. Phys.}\ }\textbf {\bibinfo {volume} {26}},\ \bibinfo
  {pages} {155} (\bibinfo {year} {2006})},\ \Eprint
  {http://arxiv.org/abs/astro-ph/0604450} {arXiv:astro-ph/0604450 [astro-ph]}
  \BibitemShut {NoStop}%
\bibitem [{\citenamefont {{Albert}}\ \emph {et~al.}(2017)\citenamefont
  {{Albert}} \emph {et~al.}}]{Albert2017}%
  \BibitemOpen
  \bibfield  {author} {\bibinfo {author} {\bibfnamefont {A.}~\bibnamefont
  {{Albert}}} \emph {et~al.},\ }\href {\doibase 10.3847/2041-8213/aa9aed}
  {\bibfield  {journal} {\bibinfo  {journal} {Astrophys. J. Lett.}\ }\textbf
  {\bibinfo {volume} {850}},\ \bibinfo {eid} {L35} (\bibinfo {year} {2017})},\
  \Eprint {http://arxiv.org/abs/1710.05839} {arXiv:1710.05839 [astro-ph.HE]}
  \BibitemShut {NoStop}%
\bibitem [{\citenamefont {{Kampert}}\ \emph {et~al.}(2019)\citenamefont
  {{Kampert}}, \citenamefont {{Alejandro Mostafa}}, \citenamefont {{Zas}},\
  and\ \citenamefont {{Pierre Auger Collaboration}}}]{KAZP2019}%
  \BibitemOpen
  \bibfield  {author} {\bibinfo {author} {\bibfnamefont {K.-H.}\ \bibnamefont
  {{Kampert}}}, \bibinfo {author} {\bibfnamefont {M.}~\bibnamefont {{Alejandro
  Mostafa}}}, \bibinfo {author} {\bibfnamefont {E.}~\bibnamefont {{Zas}}}, \
  and\ \bibinfo {author} {\bibnamefont {{Pierre Auger Collaboration}}},\ }\href
  {\doibase 10.3389/fspas.2019.00024} {\bibfield  {journal} {\bibinfo
  {journal} {Front. Astron. Space Sci.}\ }\textbf {\bibinfo {volume} {6}},\
  \bibinfo {eid} {24} (\bibinfo {year} {2019})}\BibitemShut {NoStop}%
\bibitem [{\citenamefont {{Dundovic}}\ \emph {et~al.}(2021)\citenamefont
  {{Dundovic}}, \citenamefont {{Evoli}}, \citenamefont {{Gaggero}},\ and\
  \citenamefont {{Grasso}}}]{DEGG2021}%
  \BibitemOpen
  \bibfield  {author} {\bibinfo {author} {\bibfnamefont {A.}~\bibnamefont
  {{Dundovic}}}, \bibinfo {author} {\bibfnamefont {C.}~\bibnamefont {{Evoli}}},
  \bibinfo {author} {\bibfnamefont {D.}~\bibnamefont {{Gaggero}}}, \ and\
  \bibinfo {author} {\bibfnamefont {D.}~\bibnamefont {{Grasso}}},\ }\href
  {\doibase 10.1051/0004-6361/202140801} {\bibfield  {journal} {\bibinfo
  {journal} {Astron. Astrophys.}\ }\textbf {\bibinfo {volume} {653}},\ \bibinfo
  {eid} {A18} (\bibinfo {year} {2021})},\ \Eprint
  {http://arxiv.org/abs/2105.13165} {arXiv:2105.13165 [astro-ph.HE]}
  \BibitemShut {NoStop}%
\bibitem [{\citenamefont {{Sharma}}(2024)}]{Sharma2024}%
  \BibitemOpen
  \bibfield  {author} {\bibinfo {author} {\bibfnamefont {A.}~\bibnamefont
  {{Sharma}}},\ }\href {\doibase 10.3390/universe10080326} {\bibfield
  {journal} {\bibinfo  {journal} {Universe}\ }\textbf {\bibinfo {volume}
  {10}},\ \bibinfo {eid} {326} (\bibinfo {year} {2024})},\ \Eprint
  {http://arxiv.org/abs/2408.11353} {arXiv:2408.11353 [astro-ph.HE]}
  \BibitemShut {NoStop}%
\bibitem [{\citenamefont {Acciari}\ \emph {et~al.}(2023)\citenamefont {Acciari}
  \emph {et~al.}}]{MAGIC:2022fww}%
  \BibitemOpen
  \bibfield  {author} {\bibinfo {author} {\bibfnamefont {V.~A.}\ \bibnamefont
  {Acciari}} \emph {et~al.} (\bibinfo {collaboration} {MAGIC}),\ }\href
  {\doibase 10.1051/0004-6361/202244477} {\bibfield  {journal} {\bibinfo
  {journal} {Astron. Astrophys.}\ }\textbf {\bibinfo {volume} {670}},\ \bibinfo
  {pages} {A49} (\bibinfo {year} {2023})},\ \Eprint
  {http://arxiv.org/abs/2211.13268} {arXiv:2211.13268 [astro-ph.HE]}
  \BibitemShut {NoStop}%
\bibitem [{\citenamefont {{Lico}}\ \emph {et~al.}(2023)\citenamefont {{Lico}}
  \emph {et~al.}}]{Lico2023}%
  \BibitemOpen
  \bibfield  {author} {\bibinfo {author} {\bibfnamefont {R.}~\bibnamefont
  {{Lico}}} \emph {et~al.},\ }\href {\doibase 10.3390/galaxies11010017}
  {\bibfield  {journal} {\bibinfo  {journal} {Galaxies}\ }\textbf {\bibinfo
  {volume} {11}},\ \bibinfo {eid} {17} (\bibinfo {year} {2023})},\ \Eprint
  {http://arxiv.org/abs/2301.05699} {arXiv:2301.05699 [astro-ph.HE]}
  \BibitemShut {NoStop}%
\bibitem [{\citenamefont {Akiyama}\ \emph
  {et~al.}(2019{\natexlab{a}})\citenamefont {Akiyama} \emph
  {et~al.}}]{2019ApJ...875L...6E}%
  \BibitemOpen
  \bibfield  {author} {\bibinfo {author} {\bibfnamefont {K.}~\bibnamefont
  {Akiyama}} \emph {et~al.} (\bibinfo {collaboration} {Event Horizon Telescope
  Collaboration}),\ }\href {\doibase 10.3847/2041-8213/ab1141} {\bibfield
  {journal} {\bibinfo  {journal} {Astrophys. J. Lett.}\ }\textbf {\bibinfo
  {volume} {875}},\ \bibinfo {eid} {L6} (\bibinfo {year}
  {2019}{\natexlab{a}})},\ \Eprint {http://arxiv.org/abs/1906.11243}
  {arXiv:1906.11243 [astro-ph.GA]} \BibitemShut {NoStop}%
\bibitem [{\citenamefont {{Agafonova}}\ \emph {et~al.}(2012)\citenamefont
  {{Agafonova}} \emph {et~al.}}]{Agafon2012}%
  \BibitemOpen
  \bibfield  {author} {\bibinfo {author} {\bibfnamefont {N.~Y.}\ \bibnamefont
  {{Agafonova}}} \emph {et~al.},\ }\href {\doibase
  10.1103/PhysRevLett.109.070801} {\bibfield  {journal} {\bibinfo  {journal}
  {\prl}\ }\textbf {\bibinfo {volume} {109}},\ \bibinfo {eid} {070801}
  (\bibinfo {year} {2012})},\ \Eprint {http://arxiv.org/abs/1208.1392}
  {arXiv:1208.1392 [hep-ex]} \BibitemShut {NoStop}%
\bibitem [{\citenamefont {{Adam}}\ \emph {et~al.}(2012)\citenamefont {{Adam}}
  \emph {et~al.}}]{Adam2012}%
  \BibitemOpen
  \bibfield  {author} {\bibinfo {author} {\bibfnamefont {T.}~\bibnamefont
  {{Adam}}} \emph {et~al.},\ }\href {\doibase 10.1007/JHEP10(2012)093}
  {\bibfield  {journal} {\bibinfo  {journal} {J. High Energy Phys.}\ }\textbf
  {\bibinfo {volume} {2012}},\ \bibinfo {eid} {93} (\bibinfo {year} {2012})},\
  \Eprint {http://arxiv.org/abs/1109.4897} {arXiv:1109.4897 [hep-ex]}
  \BibitemShut {NoStop}%
\bibitem [{\citenamefont {{Adamson}}\ \emph {et~al.}(2015)\citenamefont
  {{Adamson}} \emph {et~al.}}]{Adamson2015}%
  \BibitemOpen
  \bibfield  {author} {\bibinfo {author} {\bibfnamefont {P.}~\bibnamefont
  {{Adamson}}} \emph {et~al.},\ }\href {\doibase 10.1103/PhysRevD.92.052005}
  {\bibfield  {journal} {\bibinfo  {journal} {\prd}\ }\textbf {\bibinfo
  {volume} {92}},\ \bibinfo {eid} {052005} (\bibinfo {year}
  {2015})}\BibitemShut {NoStop}%
\bibitem [{\citenamefont {Akiyama}\ \emph
  {et~al.}(2019{\natexlab{b}})\citenamefont {Akiyama} \emph
  {et~al.}}]{2019ApJ875L1E}%
  \BibitemOpen
  \bibfield  {author} {\bibinfo {author} {\bibfnamefont {K.}~\bibnamefont
  {Akiyama}} \emph {et~al.} (\bibinfo {collaboration} {Event Horizon Telescope
  Collaboration}),\ }\href {\doibase 10.3847/2041-8213/ab0ec7} {\bibfield
  {journal} {\bibinfo  {journal} {Astrophys. J. Lett.}\ }\textbf {\bibinfo
  {volume} {875}},\ \bibinfo {eid} {L1} (\bibinfo {year}
  {2019}{\natexlab{b}})},\ \Eprint {http://arxiv.org/abs/1906.11238}
  {arXiv:1906.11238 [astro-ph.GA]} \BibitemShut {NoStop}%
\bibitem [{\citenamefont {{Gralla}}(2021)}]{Gralla2021}%
  \BibitemOpen
  \bibfield  {author} {\bibinfo {author} {\bibfnamefont {S.~E.}\ \bibnamefont
  {{Gralla}}},\ }\href {\doibase 10.1103/PhysRevD.103.024023} {\bibfield
  {journal} {\bibinfo  {journal} {\prd}\ }\textbf {\bibinfo {volume} {103}},\
  \bibinfo {eid} {024023} (\bibinfo {year} {2021})},\ \Eprint
  {http://arxiv.org/abs/2010.08557} {arXiv:2010.08557 [astro-ph.HE]}
  \BibitemShut {NoStop}%
\bibitem [{\citenamefont {{Kulsrud}}\ and\ \citenamefont
  {{Loeb}}(1992)}]{KL1992}%
  \BibitemOpen
  \bibfield  {author} {\bibinfo {author} {\bibfnamefont {R.}~\bibnamefont
  {{Kulsrud}}}\ and\ \bibinfo {author} {\bibfnamefont {A.}~\bibnamefont
  {{Loeb}}},\ }\href {\doibase 10.1103/PhysRevD.45.525} {\bibfield  {journal}
  {\bibinfo  {journal} {\prd}\ }\textbf {\bibinfo {volume} {45}},\ \bibinfo
  {pages} {525} (\bibinfo {year} {1992})}\BibitemShut {NoStop}%
\bibitem [{\citenamefont {{Virbhadra}}\ and\ \citenamefont
  {{Keeton}}(2008)}]{VK2008}%
  \BibitemOpen
  \bibfield  {author} {\bibinfo {author} {\bibfnamefont {K.~S.}\ \bibnamefont
  {{Virbhadra}}}\ and\ \bibinfo {author} {\bibfnamefont {C.~R.}\ \bibnamefont
  {{Keeton}}},\ }\href {\doibase 10.1103/PhysRevD.77.124014} {\bibfield
  {journal} {\bibinfo  {journal} {\prd}\ }\textbf {\bibinfo {volume} {77}},\
  \bibinfo {eid} {124014} (\bibinfo {year} {2008})},\ \Eprint
  {http://arxiv.org/abs/0710.2333} {arXiv:0710.2333 [gr-qc]} \BibitemShut
  {NoStop}%
\bibitem [{\citenamefont {{Wei}}\ \emph {et~al.}(2012)\citenamefont {{Wei}},
  \citenamefont {{Liu}}, \citenamefont {{Fu}},\ and\ \citenamefont
  {{Yang}}}]{WLFY2012}%
  \BibitemOpen
  \bibfield  {author} {\bibinfo {author} {\bibfnamefont {S.-W.}\ \bibnamefont
  {{Wei}}}, \bibinfo {author} {\bibfnamefont {Y.-X.}\ \bibnamefont {{Liu}}},
  \bibinfo {author} {\bibfnamefont {C.-E.}\ \bibnamefont {{Fu}}}, \ and\
  \bibinfo {author} {\bibfnamefont {K.}~\bibnamefont {{Yang}}},\ }\href
  {\doibase 10.1088/1475-7516/2012/10/053} {\bibfield  {journal} {\bibinfo
  {journal} {J. Cosmol. Astropart. Phys.}\ }\textbf {\bibinfo {volume}
  {2012}},\ \bibinfo {eid} {053} (\bibinfo {year} {2012})},\ \Eprint
  {http://arxiv.org/abs/1104.0776} {arXiv:1104.0776 [hep-th]} \BibitemShut
  {NoStop}%
\bibitem [{\citenamefont {{Chen}}\ and\ \citenamefont
  {{Jing}}(2015)}]{CJ2015J}%
  \BibitemOpen
  \bibfield  {author} {\bibinfo {author} {\bibfnamefont {S.}~\bibnamefont
  {{Chen}}}\ and\ \bibinfo {author} {\bibfnamefont {J.}~\bibnamefont
  {{Jing}}},\ }\href {\doibase 10.1088/1475-7516/2015/10/002} {\bibfield
  {journal} {\bibinfo  {journal} {J. Cosmol. Astropart. Phys.}\ }\textbf
  {\bibinfo {volume} {2015}},\ \bibinfo {pages} {002} (\bibinfo {year}
  {2015})},\ \Eprint {http://arxiv.org/abs/1502.01088} {arXiv:1502.01088
  [gr-qc]} \BibitemShut {NoStop}%
\bibitem [{\citenamefont {{Liu}}\ \emph
  {et~al.}(2016{\natexlab{b}})\citenamefont {{Liu}} \emph {et~al.}}]{Liu2016}%
  \BibitemOpen
  \bibfield  {author} {\bibinfo {author} {\bibfnamefont {X.}~\bibnamefont
  {{Liu}}} \emph {et~al.},\ }\href {\doibase 10.1103/PhysRevLett.117.051101}
  {\bibfield  {journal} {\bibinfo  {journal} {\prl}\ }\textbf {\bibinfo
  {volume} {117}},\ \bibinfo {eid} {051101} (\bibinfo {year}
  {2016}{\natexlab{b}})},\ \Eprint {http://arxiv.org/abs/1607.00184}
  {arXiv:1607.00184 [astro-ph.CO]} \BibitemShut {NoStop}%
\bibitem [{\citenamefont {{Tsukamoto}}\ and\ \citenamefont
  {{Gong}}(2017)}]{TG2017}%
  \BibitemOpen
  \bibfield  {author} {\bibinfo {author} {\bibfnamefont {N.}~\bibnamefont
  {{Tsukamoto}}}\ and\ \bibinfo {author} {\bibfnamefont {Y.}~\bibnamefont
  {{Gong}}},\ }\href {\doibase 10.1103/PhysRevD.95.064034} {\bibfield
  {journal} {\bibinfo  {journal} {\prd}\ }\textbf {\bibinfo {volume} {95}},\
  \bibinfo {eid} {064034} (\bibinfo {year} {2017})},\ \Eprint
  {http://arxiv.org/abs/1612.08250} {arXiv:1612.08250 [gr-qc]} \BibitemShut
  {NoStop}%
\bibitem [{\citenamefont {{He}}\ and\ \citenamefont
  {{Lin}}(2017{\natexlab{b}})}]{HL2017a}%
  \BibitemOpen
  \bibfield  {author} {\bibinfo {author} {\bibfnamefont {G.}~\bibnamefont
  {{He}}}\ and\ \bibinfo {author} {\bibfnamefont {W.}~\bibnamefont {{Lin}}},\
  }\href {\doibase 10.1088/1361-6382/aa5203} {\bibfield  {journal} {\bibinfo
  {journal} {Classical and Quantum Gravity}\ }\textbf {\bibinfo {volume}
  {34}},\ \bibinfo {eid} {029401} (\bibinfo {year}
  {2017}{\natexlab{b}})}\BibitemShut {NoStop}%
\bibitem [{\citenamefont {{Zhao}}\ and\ \citenamefont {{Xie}}(2017)}]{ZX2017E}%
  \BibitemOpen
  \bibfield  {author} {\bibinfo {author} {\bibfnamefont {S.-S.}\ \bibnamefont
  {{Zhao}}}\ and\ \bibinfo {author} {\bibfnamefont {Y.}~\bibnamefont {{Xie}}},\
  }\href {\doibase 10.1140/epjc/s10052-017-4850-5} {\bibfield  {journal}
  {\bibinfo  {journal} {Eur. Phys. J. C}\ }\textbf {\bibinfo {volume} {77}},\
  \bibinfo {eid} {272} (\bibinfo {year} {2017})},\ \Eprint
  {http://arxiv.org/abs/1704.02434} {arXiv:1704.02434 [gr-qc]} \BibitemShut
  {NoStop}%
\bibitem [{\citenamefont {{Cao}}\ and\ \citenamefont {{Xie}}(2018)}]{CX2018}%
  \BibitemOpen
  \bibfield  {author} {\bibinfo {author} {\bibfnamefont {W.-G.}\ \bibnamefont
  {{Cao}}}\ and\ \bibinfo {author} {\bibfnamefont {Y.}~\bibnamefont {{Xie}}},\
  }\href {\doibase 10.1140/epjc/s10052-018-5684-5} {\bibfield  {journal}
  {\bibinfo  {journal} {Eur. Phys. J. C}\ }\textbf {\bibinfo {volume} {78}},\
  \bibinfo {eid} {191} (\bibinfo {year} {2018})}\BibitemShut {NoStop}%
\bibitem [{\citenamefont {{Tsukamoto}}\ and\ \citenamefont
  {{Gong}}(2018)}]{TG2018}%
  \BibitemOpen
  \bibfield  {author} {\bibinfo {author} {\bibfnamefont {N.}~\bibnamefont
  {{Tsukamoto}}}\ and\ \bibinfo {author} {\bibfnamefont {Y.}~\bibnamefont
  {{Gong}}},\ }\href {\doibase 10.1103/PhysRevD.97.084051} {\bibfield
  {journal} {\bibinfo  {journal} {\prd}\ }\textbf {\bibinfo {volume} {97}},\
  \bibinfo {eid} {084051} (\bibinfo {year} {2018})},\ \Eprint
  {http://arxiv.org/abs/1711.04560} {arXiv:1711.04560 [gr-qc]} \BibitemShut
  {NoStop}%
\bibitem [{\citenamefont {{Marques}}\ \emph {et~al.}(2019)\citenamefont
  {{Marques}}, \citenamefont {{Liu}}, \citenamefont {{Zorrilla Matilla}},
  \citenamefont {{Haiman}}, \citenamefont {{Bernui}},\ and\ \citenamefont
  {{Novaes}}}]{MLMHBN2019}%
  \BibitemOpen
  \bibfield  {author} {\bibinfo {author} {\bibfnamefont {G.~A.}\ \bibnamefont
  {{Marques}}}, \bibinfo {author} {\bibfnamefont {J.}~\bibnamefont {{Liu}}},
  \bibinfo {author} {\bibfnamefont {J.~M.}\ \bibnamefont {{Zorrilla Matilla}}},
  \bibinfo {author} {\bibfnamefont {Z.}~\bibnamefont {{Haiman}}}, \bibinfo
  {author} {\bibfnamefont {A.}~\bibnamefont {{Bernui}}}, \ and\ \bibinfo
  {author} {\bibfnamefont {C.~P.}\ \bibnamefont {{Novaes}}},\ }\href {\doibase
  10.1088/1475-7516/2019/06/019} {\bibfield  {journal} {\bibinfo  {journal} {J.
  Cosmol. Astropart. Phys.}\ }\textbf {\bibinfo {volume} {2019}},\ \bibinfo
  {eid} {019} (\bibinfo {year} {2019})},\ \Eprint
  {http://arxiv.org/abs/1812.08206} {arXiv:1812.08206 [astro-ph.CO]}
  \BibitemShut {NoStop}%
\bibitem [{\citenamefont {{Javed}}\ \emph {et~al.}(2019)\citenamefont
  {{Javed}}, \citenamefont {{Babar}},\ and\ \citenamefont
  {{{\"O}vg{\"u}n}}}]{JBO2019}%
  \BibitemOpen
  \bibfield  {author} {\bibinfo {author} {\bibfnamefont {W.}~\bibnamefont
  {{Javed}}}, \bibinfo {author} {\bibfnamefont {R.}~\bibnamefont {{Babar}}}, \
  and\ \bibinfo {author} {\bibfnamefont {A.}~\bibnamefont {{{\"O}vg{\"u}n}}},\
  }\href {\doibase 10.1103/PhysRevD.99.084012} {\bibfield  {journal} {\bibinfo
  {journal} {\prd}\ }\textbf {\bibinfo {volume} {99}},\ \bibinfo {eid} {084012}
  (\bibinfo {year} {2019})},\ \Eprint {http://arxiv.org/abs/1903.11657}
  {arXiv:1903.11657 [gr-qc]} \BibitemShut {NoStop}%
\bibitem [{\citenamefont {{Lu}}\ and\ \citenamefont {{Xie}}(2019)}]{LX2019E}%
  \BibitemOpen
  \bibfield  {author} {\bibinfo {author} {\bibfnamefont {X.}~\bibnamefont
  {{Lu}}}\ and\ \bibinfo {author} {\bibfnamefont {Y.}~\bibnamefont {{Xie}}},\
  }\href {\doibase 10.1140/epjc/s10052-019-7537-2} {\bibfield  {journal}
  {\bibinfo  {journal} {Eur. Phys. J. C}\ }\textbf {\bibinfo {volume} {79}},\
  \bibinfo {eid} {1016} (\bibinfo {year} {2019})}\BibitemShut {NoStop}%
\bibitem [{\citenamefont {{Zhu}}\ and\ \citenamefont {{Xie}}(2020)}]{ZX2020a}%
  \BibitemOpen
  \bibfield  {author} {\bibinfo {author} {\bibfnamefont {X.-Y.}\ \bibnamefont
  {{Zhu}}}\ and\ \bibinfo {author} {\bibfnamefont {Y.}~\bibnamefont {{Xie}}},\
  }\href {\doibase 10.1140/epjc/s10052-020-8021-8} {\bibfield  {journal}
  {\bibinfo  {journal} {Eur. Phys. J. C}\ }\textbf {\bibinfo {volume} {80}},\
  \bibinfo {eid} {444} (\bibinfo {year} {2020})}\BibitemShut {NoStop}%
\bibitem [{\citenamefont {{Kumar}}\ \emph
  {et~al.}(2020{\natexlab{b}})\citenamefont {{Kumar}}, \citenamefont
  {{Islam}},\ and\ \citenamefont {{Ghosh}}}]{KIG2020a}%
  \BibitemOpen
  \bibfield  {author} {\bibinfo {author} {\bibfnamefont {R.}~\bibnamefont
  {{Kumar}}}, \bibinfo {author} {\bibfnamefont {S.~U.}\ \bibnamefont
  {{Islam}}}, \ and\ \bibinfo {author} {\bibfnamefont {S.~G.}\ \bibnamefont
  {{Ghosh}}},\ }\href {\doibase 10.1140/epjc/s10052-020-08606-3} {\bibfield
  {journal} {\bibinfo  {journal} {Eur. Phys. J. C}\ }\textbf {\bibinfo {volume}
  {80}},\ \bibinfo {eid} {1128} (\bibinfo {year} {2020}{\natexlab{b}})},\
  \Eprint {http://arxiv.org/abs/2004.12970} {arXiv:2004.12970 [gr-qc]}
  \BibitemShut {NoStop}%
\bibitem [{\citenamefont {{Lu}}\ and\ \citenamefont {{Xie}}(2021)}]{LX2021}%
  \BibitemOpen
  \bibfield  {author} {\bibinfo {author} {\bibfnamefont {X.}~\bibnamefont
  {{Lu}}}\ and\ \bibinfo {author} {\bibfnamefont {Y.}~\bibnamefont {{Xie}}},\
  }\href {\doibase 10.1140/epjc/s10052-021-09440-x} {\bibfield  {journal}
  {\bibinfo  {journal} {Eur. Phys. J. C}\ }\textbf {\bibinfo {volume} {81}},\
  \bibinfo {eid} {627} (\bibinfo {year} {2021})}\BibitemShut {NoStop}%
\bibitem [{\citenamefont {{Hsieh}}\ \emph {et~al.}(2021)\citenamefont
  {{Hsieh}}, \citenamefont {{Lee}},\ and\ \citenamefont {{Lin}}}]{HLL2021a}%
  \BibitemOpen
  \bibfield  {author} {\bibinfo {author} {\bibfnamefont {T.}~\bibnamefont
  {{Hsieh}}}, \bibinfo {author} {\bibfnamefont {D.-S.}\ \bibnamefont {{Lee}}},
  \ and\ \bibinfo {author} {\bibfnamefont {C.-Y.}\ \bibnamefont {{Lin}}},\
  }\href {\doibase 10.1103/PhysRevD.103.104063} {\bibfield  {journal} {\bibinfo
   {journal} {\prd}\ }\textbf {\bibinfo {volume} {103}},\ \bibinfo {eid}
  {104063} (\bibinfo {year} {2021})},\ \Eprint
  {http://arxiv.org/abs/2101.09008} {arXiv:2101.09008 [gr-qc]} \BibitemShut
  {NoStop}%
\bibitem [{\citenamefont {{Javed}}\ \emph {et~al.}(2022)\citenamefont
  {{Javed}}, \citenamefont {{Riaz}}, \citenamefont {{Pantig}},\ and\
  \citenamefont {{{\"O}vg{\"u}n}}}]{JRPO2022}%
  \BibitemOpen
  \bibfield  {author} {\bibinfo {author} {\bibfnamefont {W.}~\bibnamefont
  {{Javed}}}, \bibinfo {author} {\bibfnamefont {S.}~\bibnamefont {{Riaz}}},
  \bibinfo {author} {\bibfnamefont {R.~C.}\ \bibnamefont {{Pantig}}}, \ and\
  \bibinfo {author} {\bibfnamefont {A.}~\bibnamefont {{{\"O}vg{\"u}n}}},\
  }\href {\doibase 10.1140/epjc/s10052-022-11030-4} {\bibfield  {journal}
  {\bibinfo  {journal} {Eur. Phys. J. C}\ }\textbf {\bibinfo {volume} {82}},\
  \bibinfo {eid} {1057} (\bibinfo {year} {2022})},\ \Eprint
  {http://arxiv.org/abs/2212.00804} {arXiv:2212.00804 [gr-qc]} \BibitemShut
  {NoStop}%
\bibitem [{\citenamefont {{Gao}}\ and\ \citenamefont {{Xie}}(2022)}]{GX2022}%
  \BibitemOpen
  \bibfield  {author} {\bibinfo {author} {\bibfnamefont {Y.-X.}\ \bibnamefont
  {{Gao}}}\ and\ \bibinfo {author} {\bibfnamefont {Y.}~\bibnamefont {{Xie}}},\
  }\href {\doibase 10.1140/epjc/s10052-022-10128-z} {\bibfield  {journal}
  {\bibinfo  {journal} {Eur. Phys. J. C}\ }\textbf {\bibinfo {volume} {82}},\
  \bibinfo {eid} {162} (\bibinfo {year} {2022})}\BibitemShut {NoStop}%
\bibitem [{\citenamefont {{Soares}}\ \emph {et~al.}(2023)\citenamefont
  {{Soares}}, \citenamefont {{Vit{\'o}ria}},\ and\ \citenamefont
  {{Pereira}}}]{SVP2023}%
  \BibitemOpen
  \bibfield  {author} {\bibinfo {author} {\bibfnamefont {A.~R.}\ \bibnamefont
  {{Soares}}}, \bibinfo {author} {\bibfnamefont {R.~L.~L.}\ \bibnamefont
  {{Vit{\'o}ria}}}, \ and\ \bibinfo {author} {\bibfnamefont {C.~F.~S.}\
  \bibnamefont {{Pereira}}},\ }\href {\doibase 10.1140/epjc/s10052-023-12071-z}
  {\bibfield  {journal} {\bibinfo  {journal} {Eur. Phys. J. C}\ }\textbf
  {\bibinfo {volume} {83}},\ \bibinfo {eid} {903} (\bibinfo {year} {2023})},\
  \Eprint {http://arxiv.org/abs/2305.11105} {arXiv:2305.11105 [gr-qc]}
  \BibitemShut {NoStop}%
\bibitem [{\citenamefont {{Gao}}\ \emph {et~al.}(2023)\citenamefont {{Gao}},
  \citenamefont {{Yan}}, \citenamefont {{Yin}},\ and\ \citenamefont
  {{Hu}}}]{GYYH2023}%
  \BibitemOpen
  \bibfield  {author} {\bibinfo {author} {\bibfnamefont {X.-J.}\ \bibnamefont
  {{Gao}}}, \bibinfo {author} {\bibfnamefont {X.-k.}\ \bibnamefont {{Yan}}},
  \bibinfo {author} {\bibfnamefont {Y.}~\bibnamefont {{Yin}}}, \ and\ \bibinfo
  {author} {\bibfnamefont {Y.-P.}\ \bibnamefont {{Hu}}},\ }\href {\doibase
  10.1140/epjc/s10052-023-11414-0} {\bibfield  {journal} {\bibinfo  {journal}
  {Eur. Phys. J. C}\ }\textbf {\bibinfo {volume} {83}},\ \bibinfo {eid} {281}
  (\bibinfo {year} {2023})},\ \Eprint {http://arxiv.org/abs/2303.00190}
  {arXiv:2303.00190 [gr-qc]} \BibitemShut {NoStop}%
\bibitem [{\citenamefont {{Puli{\c{c}}e}}\ \emph {et~al.}(2023)\citenamefont
  {{Puli{\c{c}}e}}, \citenamefont {{Pantig}}, \citenamefont {{{\"O}vg{\"u}n}},\
  and\ \citenamefont {{Demir}}}]{PPOD2023}%
  \BibitemOpen
  \bibfield  {author} {\bibinfo {author} {\bibfnamefont {B.}~\bibnamefont
  {{Puli{\c{c}}e}}}, \bibinfo {author} {\bibfnamefont {R.~C.}\ \bibnamefont
  {{Pantig}}}, \bibinfo {author} {\bibfnamefont {A.}~\bibnamefont
  {{{\"O}vg{\"u}n}}}, \ and\ \bibinfo {author} {\bibfnamefont {D.}~\bibnamefont
  {{Demir}}},\ }\href {\doibase 10.1088/1361-6382/acf08c} {\bibfield  {journal}
  {\bibinfo  {journal} {Classical and Quantum Gravity}\ }\textbf {\bibinfo
  {volume} {40}},\ \bibinfo {eid} {195003} (\bibinfo {year} {2023})},\ \Eprint
  {http://arxiv.org/abs/2308.08415} {arXiv:2308.08415 [gr-qc]} \BibitemShut
  {NoStop}%
\bibitem [{\citenamefont {{Atamurotov}}\ \emph {et~al.}(2023)\citenamefont
  {{Atamurotov}}, \citenamefont {{Alibekov}}, \citenamefont {{Abdujabbarov}},
  \citenamefont {{Mustafa}},\ and\ \citenamefont {{Aripov}}}]{aaaa2023}%
  \BibitemOpen
  \bibfield  {author} {\bibinfo {author} {\bibfnamefont {F.}~\bibnamefont
  {{Atamurotov}}}, \bibinfo {author} {\bibfnamefont {H.}~\bibnamefont
  {{Alibekov}}}, \bibinfo {author} {\bibfnamefont {A.}~\bibnamefont
  {{Abdujabbarov}}}, \bibinfo {author} {\bibfnamefont {G.}~\bibnamefont
  {{Mustafa}}}, \ and\ \bibinfo {author} {\bibfnamefont {M.~M.}\ \bibnamefont
  {{Aripov}}},\ }\href {\doibase 10.3390/sym15040848} {\bibfield  {journal}
  {\bibinfo  {journal} {Symmetry}\ }\textbf {\bibinfo {volume} {15}},\ \bibinfo
  {eid} {848} (\bibinfo {year} {2023})}\BibitemShut {NoStop}%
\bibitem [{\citenamefont {{Soares}}\ \emph {et~al.}(2024)\citenamefont
  {{Soares}}, \citenamefont {{Vit{\'o}ria}},\ and\ \citenamefont
  {{Pereira}}}]{SVP2024}%
  \BibitemOpen
  \bibfield  {author} {\bibinfo {author} {\bibfnamefont {A.~R.}\ \bibnamefont
  {{Soares}}}, \bibinfo {author} {\bibfnamefont {R.~L.~L.}\ \bibnamefont
  {{Vit{\'o}ria}}}, \ and\ \bibinfo {author} {\bibfnamefont {C.~F.~S.}\
  \bibnamefont {{Pereira}}},\ }\href {\doibase 10.1103/PhysRevD.110.084004}
  {\bibfield  {journal} {\bibinfo  {journal} {\prd}\ }\textbf {\bibinfo
  {volume} {110}},\ \bibinfo {eid} {084004} (\bibinfo {year} {2024})},\ \Eprint
  {http://arxiv.org/abs/2408.03217} {arXiv:2408.03217 [gr-qc]} \BibitemShut
  {NoStop}%
\bibitem [{\citenamefont {{Zhang}}\ and\ \citenamefont {{Xie}}(2024)}]{ZX2024}%
  \BibitemOpen
  \bibfield  {author} {\bibinfo {author} {\bibfnamefont {J.}~\bibnamefont
  {{Zhang}}}\ and\ \bibinfo {author} {\bibfnamefont {Y.}~\bibnamefont
  {{Xie}}},\ }\href {\doibase 10.1103/PhysRevD.109.043032} {\bibfield
  {journal} {\bibinfo  {journal} {\prd}\ }\textbf {\bibinfo {volume} {109}},\
  \bibinfo {eid} {043032} (\bibinfo {year} {2024})}\BibitemShut {NoStop}%
\bibitem [{\citenamefont {{Shan}}\ \emph {et~al.}(2024)\citenamefont {{Shan}},
  \citenamefont {{Chen}}, \citenamefont {{Hu}},\ and\ \citenamefont
  {{Li}}}]{SCHL2024}%
  \BibitemOpen
  \bibfield  {author} {\bibinfo {author} {\bibfnamefont {X.}~\bibnamefont
  {{Shan}}}, \bibinfo {author} {\bibfnamefont {X.}~\bibnamefont {{Chen}}},
  \bibinfo {author} {\bibfnamefont {B.}~\bibnamefont {{Hu}}}, \ and\ \bibinfo
  {author} {\bibfnamefont {G.}~\bibnamefont {{Li}}},\ }\href {\doibase
  10.1007/s11433-023-2334-9} {\bibfield  {journal} {\bibinfo  {journal} {Sci.
  China-Phys. Mech. Astron.}\ }\textbf {\bibinfo {volume} {67}},\ \bibinfo
  {eid} {269511} (\bibinfo {year} {2024})},\ \Eprint
  {http://arxiv.org/abs/2306.14796} {arXiv:2306.14796 [astro-ph.CO]}
  \BibitemShut {NoStop}%
\bibitem [{\citenamefont {{Saketh}}\ \emph {et~al.}(2025)\citenamefont
  {{Saketh}}, \citenamefont {{Ghosh}},\ and\ \citenamefont
  {{Mishra}}}]{SGM2025}%
  \BibitemOpen
  \bibfield  {author} {\bibinfo {author} {\bibfnamefont {M.~V.~S.}\
  \bibnamefont {{Saketh}}}, \bibinfo {author} {\bibfnamefont {R.}~\bibnamefont
  {{Ghosh}}}, \ and\ \bibinfo {author} {\bibfnamefont {A.}~\bibnamefont
  {{Mishra}}},\ }\href {\doibase 10.48550/arXiv.2511.23110} {\bibfield
  {journal} {\bibinfo  {journal} {arXiv e-prints}\ ,\ \bibinfo {eid}
  {arXiv:2511.23110}} (\bibinfo {year} {2025})},\ \Eprint
  {http://arxiv.org/abs/2511.23110} {arXiv:2511.23110 [gr-qc]} \BibitemShut
  {NoStop}%
\bibitem [{\citenamefont {{Igata}}(2026)}]{Igata2026}%
  \BibitemOpen
  \bibfield  {author} {\bibinfo {author} {\bibfnamefont {T.}~\bibnamefont
  {{Igata}}},\ }\href {\doibase 10.1103/ylrj-rm9j} {\bibfield  {journal}
  {\bibinfo  {journal} {\prd}\ }\textbf {\bibinfo {volume} {113}},\ \bibinfo
  {eid} {024036} (\bibinfo {year} {2026})},\ \Eprint
  {http://arxiv.org/abs/2504.07906} {arXiv:2504.07906 [gr-qc]} \BibitemShut
  {NoStop}%
\bibitem [{\citenamefont {{Boos}}\ and\ \citenamefont {{Hu}}(2026)}]{BH2026}%
  \BibitemOpen
  \bibfield  {author} {\bibinfo {author} {\bibfnamefont {J.}~\bibnamefont
  {{Boos}}}\ and\ \bibinfo {author} {\bibfnamefont {H.}~\bibnamefont {{Hu}}},\
  }\href {\doibase 10.1103/dv28-xm9x} {\bibfield  {journal} {\bibinfo
  {journal} {\prd}\ }\textbf {\bibinfo {volume} {113}},\ \bibinfo {eid}
  {024065} (\bibinfo {year} {2026})},\ \Eprint
  {http://arxiv.org/abs/2510.10282} {arXiv:2510.10282 [gr-qc]} \BibitemShut
  {NoStop}%
\bibitem [{\citenamefont {{Tamborra}}(2025)}]{Tambo2025}%
  \BibitemOpen
  \bibfield  {author} {\bibinfo {author} {\bibfnamefont {I.}~\bibnamefont
  {{Tamborra}}},\ }\href {\doibase 10.1038/s42254-025-00828-2} {\bibfield
  {journal} {\bibinfo  {journal} {Nat. Rev. Phys.}\ }\textbf {\bibinfo {volume}
  {7}},\ \bibinfo {pages} {285} (\bibinfo {year} {2025})},\ \Eprint
  {http://arxiv.org/abs/2412.09699} {arXiv:2412.09699 [astro-ph.HE]}
  \BibitemShut {NoStop}%
\end{thebibliography}
\end{document}